\documentclass[useAMS,usenatbib]{mnras}
\usepackage{graphicx}
\usepackage{amsmath}
\usepackage{amssymb}
\usepackage{color}
\usepackage{hyperref}

\title[Discs, bulges and stellar haloes in simulated galaxies]{NIHAO XVI: The properties and evolution of kinematically selected discs, bulges and stellar haloes}
 \author[Obreja et al]{Aura Obreja$^{1,2}$\thanks{E-mail: obreja@usm.lmu.de}, Aaron A. Dutton$^{2}$, Andrea V. Macci\`{o}$^{2,3}$, Benjamin Moster$^{1,4}$,
 \and Tobias Buck$^{5}$, Glenn van den Ven$^{6,7}$, Liang Wang$^{8}$, Gregory S. Stinson and 
 \and Ling Zhu$^{9}$\\ 
 $^{1}$University Observatory Munich, Scheinerstra\ss e 1, D-81679 Munich, Germany\\
 $^{2}$New York University Abu Dhabi, PO Box 129188, Saadiyat Island, Abu Dhabi, United Arab Emirates\\ 
 $^{3}$Max-Planck-Institut f\"{u}r Astronomie, K\"{o}nigstuhl 17, 69117 Heidelberg, Germany\\
 $^{4}$Max-Planck-Institut f\"{u}r Astrophysik, Karl-Schwarzschild Stra\ss e 1, 85748 Garching, Germany\\
 $^{5}$Leibniz-Institut für Astrophysik Potsdam (AIP), An der Sternwarte 16, D-14482, Potsdam, Germany\\
 $^{6}$European Southern Observatory, Karl-Schwarzschild-Str. 2, D-85748 Garching bei M\"{u}nchen, Germany\\
 $^{7}$Department of Astrophysics, University of Vienna, T\"urkenschanzstrasse 17, 1180, Vienna, Austria\\
 $^{8}$University of Western Australia, Crawley, WA 6009, Australia\\
 $^{9}$Shanghai Astronomical Observatory, Chinese Academy of Sciences, 80 Nandan Road, Shanghai 200030, China}
\begin{document}

\maketitle

\label{firstpage}

\begin{abstract}
We use 25 simulated galaxies from the NIHAO project to define and characterize a variety of kinematic stellar structures:
thin and thick discs, large scale single discs, classical and pseudo bulges, spheroids, inner discs, and stellar haloes. 
These structures have masses, spins, shapes and rotational support in good agreement with theoretical expectations and observational data. 
Above a dark matter halo mass of $2.5\times10^{\rm~11}M_{\rm\odot}$, all galaxies have a classical bulge and 70\% have a thin and thick disc.
The kinematic (thin) discs follow a power-law relation between angular momentum and stellar mass 
$J_{\rm *}=3.4M_{\rm *}^{\rm1.26\pm0.06}$, 
in very good agreement with the prediction based on the empirical stellar-to-halo mass relation in the same mass range,
and show a strong correlation between maximum `observed' rotation velocity and dark matter halo circular velocity
$v_{\rm c}=6.4v_{\rm max}^{0.64\pm0.04}$.
Tracing back in time these structures' progenitors, we find all to lose a fraction $1-f_j$ of their maximum angular momentum.
Thin discs are significantly better at retaining their high-redshift spins ($f_j\sim0.70$) than thick ones ($f_j\sim0.40$).           
Stellar haloes have their progenitor baryons assembled the latest ($z_{\rm~1/2}\sim1.1$) and over the longest timescales ($\tau\sim6.2$~Gyr), 
and have the smallest fraction of stars born in-situ ($f_{\rm in-situ}=0.35\pm0.14$). 
All other structures have $1.5\lesssim z_{\rm1/2}\lesssim3$, $\tau=4\pm2$~Gyr and $f_{\rm in-situ}\gtrsim0.9$. 
\end{abstract}

\begin{keywords}
galaxies: stellar content - galaxies: structure - galaxies: kinematics and dynamics - galaxies: fundamental parameters - methods: numerical
\end{keywords}

\section{Introduction}

From the pioneering work of Hubble in the 1920s, a large effort has been put into the discovery and classification of galaxies, 
which constitute the luminous tracers of the Universe's elementary blocks, the dark matter haloes. 
For both historical and practical reasons, the most widely used galaxy classifications rely on photometry. 
Among the first, \citet{deVaucouleurs:1959}, \citet{vanHouten:1961}, \citet{Sersic:1963}, \citet{Freeman:1970}, \citet{Yoshizawa:1975}, \citet{Simien:1989} used fits of azimuthally 
averaged luminosity profiles of galaxies to distinguish between highly centrally concentrated ellipticals and extended fainter spirals \citep{Sandage:1961}.
With the recent observations with ever increasing spatial resolution, 
it became feasible to use the complete two-dimensional photometric data to derive structural parameters of galaxies, both at high and low redshifts \citep[e.g.][]{vanderWel:2012,Salo:2015,MendezAbreu:2017}.   

The Hubble classification typically relies on 2D or 1D two-components fits: one function describing the inner region (a de Vaucouleurs or a more general S\'{ersic})
and the other characterizing the outer region (an exponential, a low index S\'{ersic} or a truncated disc). 
The integral luminosity of the inner function divided by the total galaxy luminosity is known as bulge-to-total ratio ($B/T$), and is the main parameter defining the location of a galaxy in the Hubble diagram. 
With the development of new instruments able to simultaneously observe the morphological and kinematic properties of a galaxy, 
this classification started to become insufficient, as galaxies were discovered to sometimes host a large number of subcomponents.   
\citet{Dalcanton:2002}, \citet{Yoachim:2006}, \citet{Comeron:2011,Comeron:2014,Comeron:2018}, \citet{Elmegreen:2017}, for example, showed that many nearby galaxies have an additional thick disc component. 
Regarding the inner galactic regions, it has been known for a long time that galaxies can have bulges ranging from highly spherically symmetric to highly flattened configurations 
\citep{Andredakis:1994,Andredakis:1995}. 
Interestingly, some studies have also found various types of bulges and bars to coexist 
\citep[e.g.][]{Erwin:2003,Athanassoula:2005,Gadotti:2009,Aguerri:2009,Nowak:2010,Kormendy:2010,Mendez-Abreu:2014}. 
The stars of the Milky Way are also thought to form several components: (i) a thin and (ii) a thick disc \citep{Gilmore:1983}, (iii) a boxy/peanut bulge \citep{Okuda:1977}, 
(iv) a nuclear star cluster \citep{Becklin:1968}, (v) a bar \citep{Hammersley:2000} and (vi) a stellar halo \citep{Searle:1978}.  
Whether the Milky Way has a classical bulge is still uncertain \citep[see review by][and references therein]{Bland-Hawthorn:2016}.

One of the first classification schemes that goes beyond photometry has been proposed by \citet{Fall:1983}. This method places galaxies 
in the stellar mass ($M_{\rm *}$) - specific stellar angular momentum ($j_{\rm *}$) plane, where late type galaxies follow a relatively tight correlation $j_{\rm *}\sim M_{\rm *}^{\rm \alpha}$ 
with $\alpha\simeq2/3$. For a fixed stellar mass, elliptical galaxies appear on this diagram at much lower $j_{\rm *}$ than late types. 
The $M_{\rm *}-j_{\rm *}$ diagram has two important implications. First of all, the $\rm\Lambda$CDM cosmology predicts that dark matter haloes follow a similar power-law with the same 
exponent $j_{\rm halo}\sim M_{\rm halo}^{\rm 3/2}$ \citep{Shaya:1984}, therefore suggesting that there might also be a relation between the stellar and the dark matter specific angular momentum, 
and not only between their masses \citep[e.g.][]{Moster:2013,Behroozi:2013}. On the other hand, the normalization in the $M_{\rm *}-j_{\rm *}$ encodes information on the dynamical type of a galaxy 
\citep[e.g.][]{Romanowsky:2012,Fall:2013,Obreschkow:2014,Fall:2018}.

From a dynamical point of view, stellar orbit classification provides the most robust way to disentangle the various galactic stellar structures \citep{Binney:2013}.
Observationally, this approach can be used for the Milky Way, whose stars have been targeted in great numbers by recent surveys \citep[e.g.][]{Steinmetz:2006,Gaia:2016,Majewski:2017}. 
For extragalactic observations \citet{Zhu:2018a,Zhu:2018b} have just made the first attempt to use reconstructed stellar orbits on {\tt CALIFA} 
galaxies \citep{Sanchez:2012} in order to obtain a dynamical Hubble classification. 
\citet{Zhu:2018b} found that the dark matter content does not bias the orbit distribution itself.  
However, some of the most recent dark matter profiles found in hydrodynamical simulations \citep[e.g.][]{DiCintio:2014,Tollet:2016,Schaller:2015,Dutton:2016}, 
can differ considerably in the inner regions from the typical NFW one \citep{Navarro:1997}.
Therefore, the dynamical masses and hence the dark matter masses inferred from dynamical modeling can be significantly biased.   

In simulations, an approximate equivalent of orbit classification has been proposed by \citet{Abadi:2003}, 
who used the circularity parameter for separating the disc from the spheroid particles, by assuming that the latter has a circularity distribution symmetric around zero. 
This method has been later employed, with additional gravitational binding energy information, to advocate for the presence of 
thick discs \citep[e.g.][]{Brook:2004,Brooks:2008} and stellar haloes \citep[e.g.][]{Scannapieco:2010,Tissera:2012} in cosmological simulations of late-type galaxies. 
\citet{Domenech:2012} went a step further and employed a statistical cluster finding algorithm in a 3D stellar kinematic space to classify particles of 
a small sample of simulated galaxies as belonging to thin discs, thick discs or spheroids. 
\citet{Obreja:2016} have generalized the method proposed by \citet{Domenech:2012}, by switching to Gaussian Mixture Models as cluster finding algorithm.
They used Gaussian Mixture Models to recover observational trends for kinematic stellar discs and spheroids in a large sample of galaxies simulated in a cosmological 
context, which cover almost three decades in stellar mass \citep{Wang:2015}. 
In a previous paper \citep[][, hereafter Paper I]{Obreja:2018}, we significantly extended the method of \citet{Obreja:2016} such that more than two stellar components can now be separated.
The OMP parallel Python/Fortran code {\tt gsf} used for the decomposition has been made publicly available at \url{https://github.com/aobr/gsf}. 

In the present study we use the method and code of Paper I, to kinematically find/define the multitude of galaxy stellar structures in simulations 
(thin and thicks discs, classical and pseudo bulges, spheroids, inner discs and stellar haloes) that have equivalents in extragalactic and galactic observations. 
We study both their observable-like properties as well as their intrinsic ones, focusing on the kinematics. 
The analysis of the various structures' morphologies is defered for a future study in which we will compare photometric decomposition of these simulations with observations, 
and with the kinematic classification presented here. 

This work is structured as follows. The simulation sample is described in Section~\ref{sim_section}, while the classification method is briefly given in Section~\ref{methods_section}.
The results of {\tt gsf} applied to the complete galaxy sample are presented in Section~\ref{sec5}. 
The observational and intrinsic properties of the various stellar structures are discussed in Sections~\ref{obs_prop} and \ref{int_prop}, respectively.
Section~\ref{glabal_evolution} presents the study of the differences in the evolutionary patterns of the stellar kinematic components. 
The average properties of all types of kinematic stellar structure are presented in Section~\ref{summary}.
Finally, Section~\ref{conclusions} summarizes our conclusions and highlights some of the possible future lines of study.

\begin{figure*} 
\includegraphics[width=1.0\textwidth]{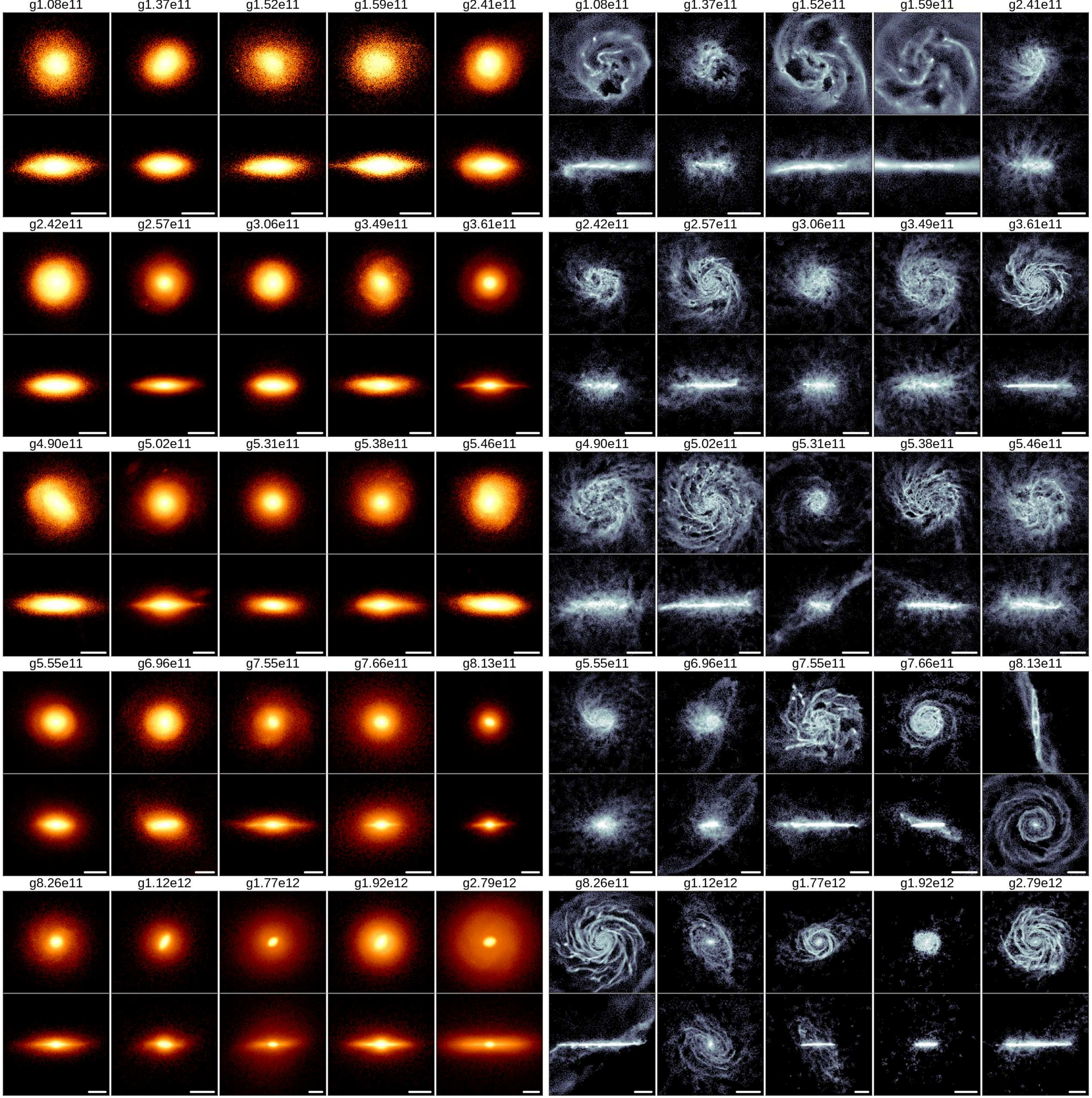}
\caption{The surface mass density maps for the stars (left) and the cold gas (right) for the 25 galaxy subsample of NIHAO.
The white horizontal lines represent the physical scale of 10 kpc. The dark matter halo masses increase from left to right, and top to bottom.}
\label{fig:sunrise}
\end{figure*}

\section{Simulations}
\label{sim_section}

This study uses galaxy zoom-in simulations from the Numerical Investigation of a Hundred Astrophysical Objects \citep[NIHAO Project,][]{Wang:2015}. 
This simulation suite assumes the Planck cosmology \citep{Planck:2014}, 
and has been run with an improved version \citep{Wadsley:2017} of the N-body SPH code {\tt GASOLINE} \citep{Wadsley:2004}. 

The new version of the code alleviates the old SPH problem of artificial cold blobs following \citet{Ritchie:2001}
and includes the treatment of artificial viscosity of \citet{Price:2008}. 
Also, the code now uses the Dehnen SPH kernel \citep{Dehnen:2012} with 50 neighbors, 
and the \citet{Saitoh:2009} time step limiter. 
The implementation of metal diffusion is described in \citet{Wadsley:2008}. 
We assume gas is heated through photoionization and photoheating produced by a redshift dependent UV background \citep{Haardt:2012},
while it is cooled through metal lines and Compton scattering \citep{Shen:2010}. 

Star formation is governed by a Kennicutt-Schmidt relation, where only the gas with temperatures lower than 15000 K and 
densities higher than 10.3 cm$^{\rm -3}$ is eligible to form stars. 
The stellar feedback includes the effect of SNe II blast-waves following \citet{Stinson:2006} and the 
heating of gas by massive stars prior to their SN phase, also known as ``early stellar feedback'' \citep{Stinson:2013a}.  
This implementation of the supernova feedback leads to galaxies that follow the stellar-to-halo mass relation determined with subhalo abundance matching 
\citep[e.g.][]{Moster:2013,Behroozi:2013} at all redshifts, with a fixed set of parameters that have been chosen based on only one Milky Way-like galaxy at $z=0$ \citep{Stinson:2013a}. 
The code uses a Chabrier Initial Mass Function (IMF) \citep{Chabrier:2003} and traces the heavy elements based on the SNe Ia yields of \citet{Thielemann:1986} 
and SNe II of \citet{Woosley:1995}.

\subsection{Galaxy sample}
\label{sample}

We use a sub-sample of NIHAO, where the selection criteria is that galaxies should not show signs of disturbances in stellar surface mass density
and have a stellar mass higher than 10$^{\rm 8}$M$_{\rm\odot}$. 
Therefore, the selected galaxies belong to either the resolution level 60.1 of \citet{Wang:2015} with gas and dark matter particle masses 
of 3.2$\rm\times$10$^{\rm 5}$M$_{\rm\odot}$ and 1.7$\rm\times$10$^{\rm 6}$M$_{\rm\odot}$, and force softenings of 400~pc and 931~pc, 
or to the level 60.2 with masses of 4$\rm\times$10$^{\rm 4}$M$_{\rm\odot}$ and 2.2$\rm\times$10$^{\rm 5}$M$_{\rm\odot}$, and softenings of 200~pc and 466~pc, 
for the gas and dark matter respectively.

The dark matter haloes are identified in the simulations using {\tt AHF} \citep{Gill:2004,Knollmann:2009}. 
The virial radius $r_{\rm vir}$ is defined as the radius enclosing a region whose average density is 200 times the cosmic critical matter density. 
Figure~\ref{fig:sunrise} gives the face-on and edge-on projections for the stellar (left) and cold gas (right) surface mass densities of all selected galaxies.
The gas panels on the right illustrate that some of the galaxies have experienced recent mergers or close encounters with satellites. 
Throughout this study \textit{cold gas} means gas with temperatures below 15000~K. 
This value is the same as one of the criteria to tag  gas particles eligible for star formation. 
Thus, the final sample consists of 25 galaxies, whose structural parameters are given in Table \ref{table1}.

By comparing the stellar and cold gas mass surface densities it is noticeable that some galaxies, 
like g1.59e11, g3.61e11, g7.55e11 and g2.79e12, have the cold gas and stellar discs very well aligned.
For these galaxies, no disturbances are visible in the edge-on gas surface mass density maps. 
On the other hand, the less massive galaxies (top left), show much thicker discs both in stars and gas. 
At the massive end of the sample, there are a few examples of very interesting galaxies. 
Simulations g7.66e11, g8.26e11 and g1.77e12 show imprints of recent minor mergers in the edge-on cold gas mass surface densities. 
A more extreme case is g8.13e11, which suffered a transversal major merger that left the cold gas disc in a plane perpendicular to the (older) stellar disc, 
which appears undisturbed at $z=0$. 

In this study we are interested in both the properties of the various stellar kinematic components of galaxies at $z=0$, as well as their evolutionary patterns. 
The next section briefly describes how {\tt gsf} works to split a simulated galaxy's stellar particles into multiple components.

\begin{table}
\centering
\begin{tabular}{ccccccc}
\hline
Sim & M$_{\rm h}$ & M$_{\rm *}$ & M$_{\rm cold}$ & f$_{\rm cold}$ & r$_{\rm vir}$ & v$_{\rm c}$\\
  & 10$^{\rm 12}$M$_{\odot}$ & 10$^{\rm 10}$M$_{\odot}$ & 10$^{\rm 10}$M$_{\odot}$ & & kpc & km/s\\
\hline
g1.08e11 & 0.11 & 0.09 & 0.32 & 0.53 & 105 & 92\\
g1.37e11 & 0.14 & 0.20 & 0.18 & 0.46 & 112 & 100\\
g1.52e11 & 0.14 & 0.09 & 0.65 & 0.47 & 117 & 96\\
g1.59e11 & 0.15 & 0.07 & 1.16 & 0.67 & 117 & 91\\
g2.41e11 & 0.24 & 0.42 & 0.61 & 0.50 & 135 & 116\\
g2.42e11 & 0.26 & 0.56 & 0.36 & 0.49 & 136 & 124\\
g2.57e11 & 0.27 & 1.09 & 0.76 & 0.49 & 140 & 131\\
g3.06e11 & 0.29 & 0.75 & 0.69 & 0.42 & 144 & 126\\
g3.49e11 & 0.39 & 0.44 & 1.71 & 0.49 & 174 & 120\\
g3.61e11 & 0.39 & 2.16 & 0.78 & 0.53 & 160 & 157\\
g4.90e11 & 0.30 & 0.35 & 1.15 & 0.47 & 146 & 119\\
g5.02e11 & 0.52 & 1.48 & 1.76 & 0.47 & 176 & 153\\
g5.31e11 & 0.48 & 1.66 & 1.10 & 0.33 & 172 & 151\\
g5.38e11 & 0.59 & 1.87 & 1.52 & 0.42 & 184 & 157\\
g5.46e11 & 0.30 & 0.38 & 1.13 & 0.49 & 146 & 119\\
g5.55e11 & 0.48 & 1.73 & 1.01 & 0.39 & 170 & 149\\
g6.96e11 & 0.72 & 3.40 & 2.02 & 0.41 & 197 & 174\\
g7.55e11 & 0.79 & 3.15 & 3.47 & 0.49 & 204 & 181\\
g7.66e11 & 0.83 & 5.96 & 1.03 & 0.25 & 207 & 193\\
g8.13e11 & 0.88 & 6.72 & 1.79 & 0.38 & 211 & 206\\
g8.26e11 & 0.90 & 4.74 & 4.20 & 0.57 & 213 & 204\\
g1.12e12 & 0.98 & 7.93 & 0.86 & 0.15 & 220 & 208\\
g1.77e12 & 1.97 & 13.83 & 2.29 & 0.20 & 276 & 243\\
g1.92e12 & 2.08 & 15.90 & 0.76 & 0.07 & 281 & 266\\
g2.79e12 & 3.12 & 20.06 & 4.25 & 0.21 & 322 & 290\\
\hline
\end{tabular}
\caption{Simulation name, dark matter halo mass ($M_{\rm h}$), stellar mass ($M_{\rm *}$), mass of cold gas ($M_{\rm cold}$),
mass fraction of cold gas with respect to the total virial gas mass ($f_{\rm cold}$),
virial radius ($r_{\rm vir}$) and circular velocity at $0.15r_{\rm vir}$ ($v_{\rm c}$) for the sample of 25 galaxies. 
Cold gas is the gas colder than 15000~K. All masses are computed inside the virial sphere.}
\label{table1}
\end{table}

\section{Methods}
\label{methods_section}

\begin{figure*}
\begin{center}
\includegraphics[width=0.48\textwidth]{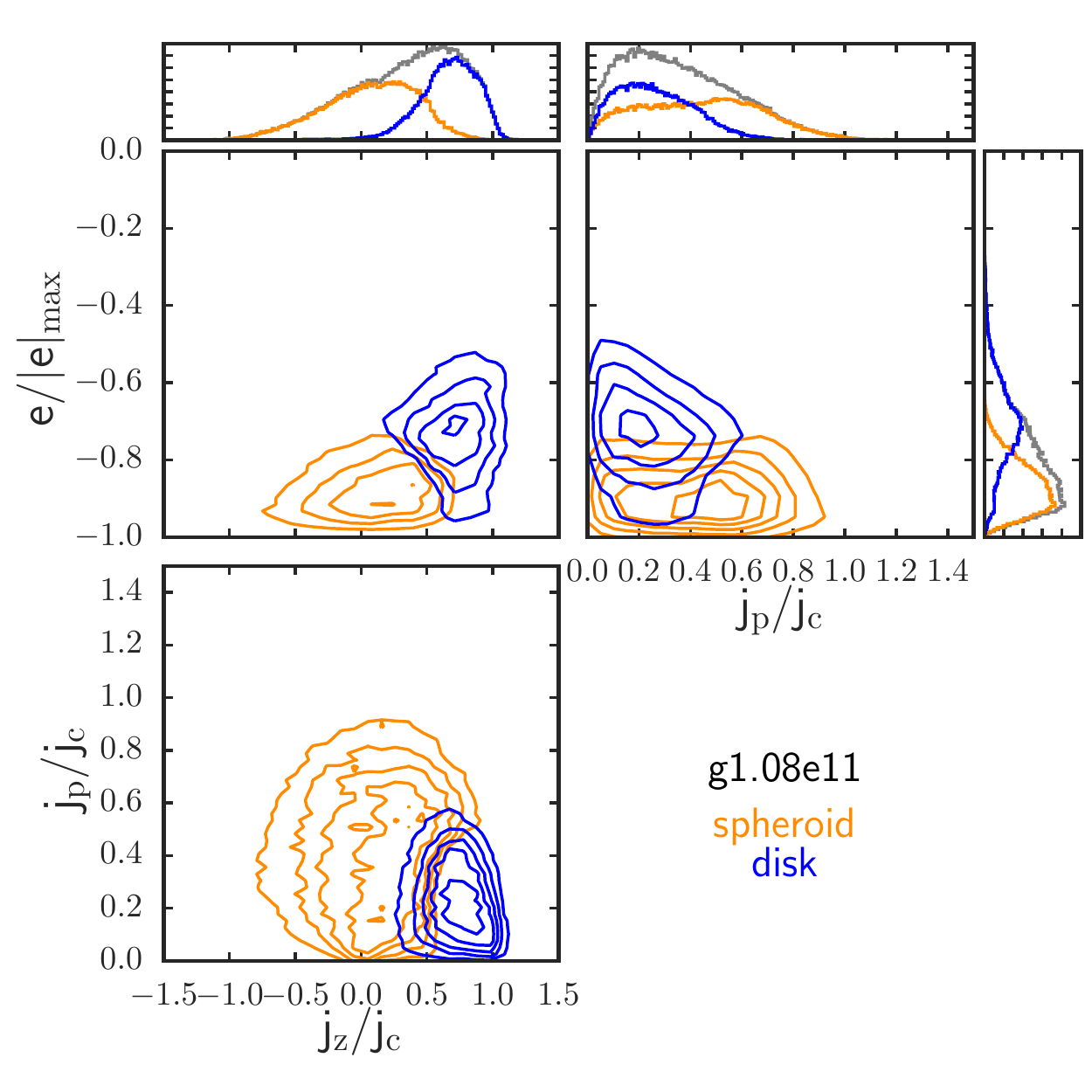}
\includegraphics[width=0.48\textwidth]{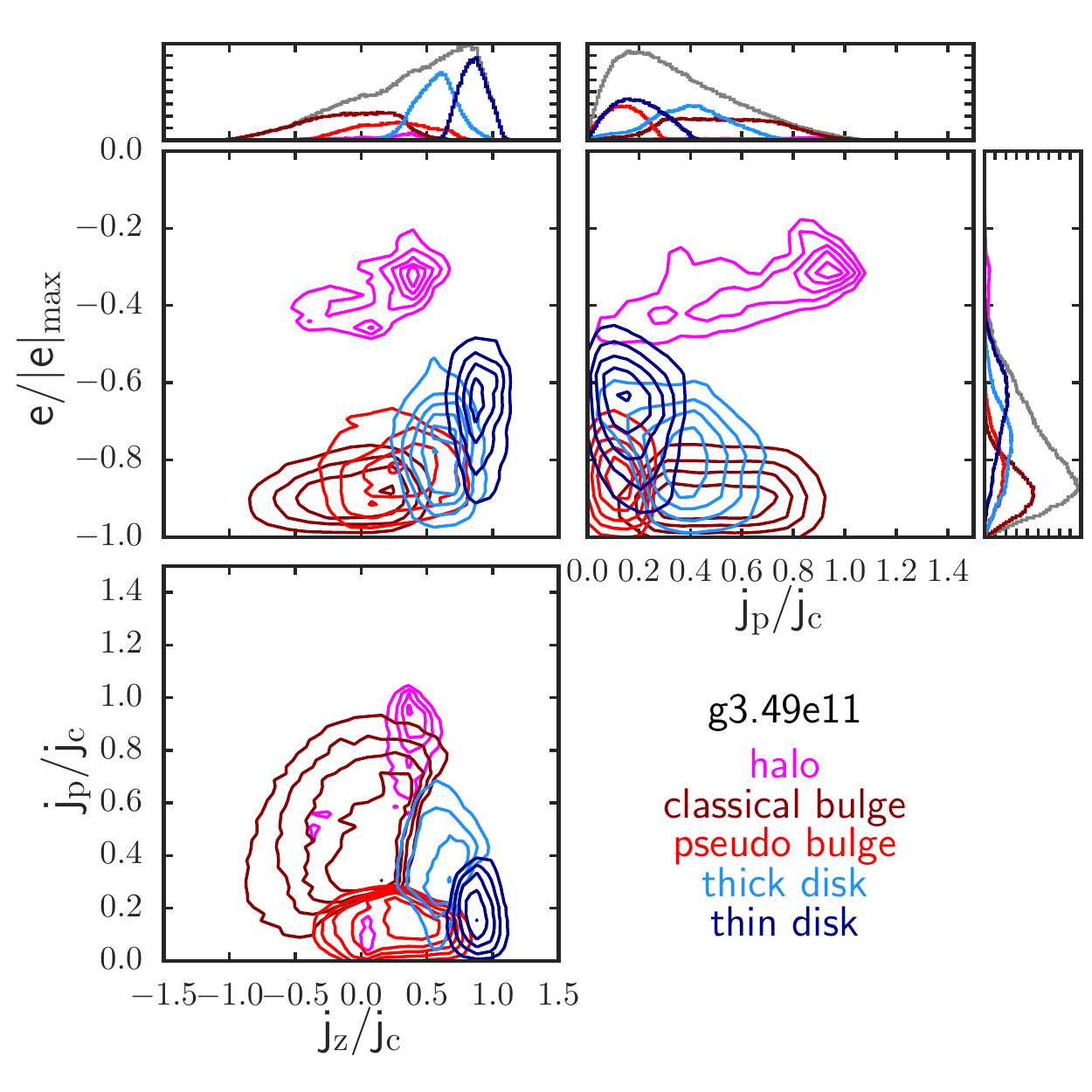}\\
\caption{Examples of two galaxies, g1.08e11 (left) and g3.49e11 (right), separated by {\tt gsf} in multiple stellar kinematic components. 
The large panels give the 2D projections of the 3D input space ($j_z/j_c$,$j_p/j_c$, $e/|e|_{\rm max}$) for each component, 
while the small ones show the corresponding 1D histograms. The names of the components are shown in their corresponding colors in the lower right corners.}
\label{fig_gmm_example}
\end{center}
\end{figure*}

The {\tt gsf} code, described in Paper I, recalculates accurately the gravitational force felt by each stellar particle in a given dark matter halo, assuming the halo in isolation. 
In a second step it computes the mapping between specific binding energy $e$ and the specific angular momentum of particles on purely circular orbits $j_{\rm c}$ in the equatorial plane. 
Considering that the galaxy is already oriented with its total angular momentum along the $z$-axis, the next step is to compute 
the specific angular momentum for each stellar particle $\vec{j}$ and decompose it as $\vec{j}=\vec{j_z}+\vec{j_p}$, where 
$j_z$ is the azimuthal component, quantifying the rotation, and $j_{p}$ is the in (equatorial)-plane component, quantifying the movement outside of the equatorial plane. 
By interpolating on the $j_c$-$e$ mapping the accurate values of the stellar binding energies, {\tt gsf} finds the $j_c$ normalization value for each particle and thus
computes the stellar circularities $j_z/j_c$ and $j_p/j_c$. 
The input space of ($j_z/j_c$,$j_p/j_c$, $e/|e|_{\rm max}$) is subsequently fed to the Gaussian Mixture Model algorithm together with 
the number of structures $nk$ one wishes to look for. The binding energies are normalized to the most bound stellar particle in the halo, $|e|_{\rm max}$.   
The output of the code is a mapping between each stellar particle and $nk$ groups defined by an index between $0$ and $nk-1$. 

As explained in Paper I, the $nk$ parameter needed by {\tt gsf} depends on the problem one wants to study. 
For the present study, we are mainly interested in two aspects of galaxies. 
The first one is how frequently galaxies have a single large scale dynamical disc instead of having two such discs (a thin and a thick one). 
The second aspect driving this work is related to stellar haloes. 
Stellar haloes are expected to have only a few per cent of the total stellar mass of a galaxy. 
Given these aspects and the particular resolution of NIHAO, we ran {\tt gsf} with $nk$ up to five in order to disentangle the stellar haloes. 

Figure~\ref{fig_gmm_example} shows two examples of how {\tt gsf} works. The galaxy on the left is one of the least massive in our sample of 25, 
and can be meaningfully separated in only two kinematic components: a spheroid (orange) and a disc (blue). The three large panels for each galaxy 
show the number density in linear scale of stellar particles in each of the 2D projections of the 3D input space ($j_z/j_c$,$j_p/j_c$, $e/|e|_{\rm max}$). 
The smaller panels give the corresponding 1D histograms, with the grey curves representing all the stellar particles of the galaxy. 
The disc of g1.08e11 is made of the stars with high circularities $j_z/j_c$ and low binding energies $e/|e|_{\rm max}$, while the spheroid has lower circularities and 
contains the particles with the highest binding energies. However, there is a considerable amount of overlap in this 3D space between the two components, most visible 
in the 2D projection $e/|e|_{\rm max}$ vs $j_p/j_c$.

The galaxy on the right side of Figure~\ref{fig_gmm_example}, g3.49e11, is almost five times more massive than g1.08e11 in terms of stellar mass, and has more than 
two kinematically distinct components. In particular, using {\tt gsf} we were able to distinguish five components for g3.49e11: a thin disc (navy), a thick disc (light blue), 
a classical bulge (dark red), a pseudo bulge (red) and a stellar halo (magenta). The separation among the components is best illustrated by the 2D projection $e/|e|_{\rm max}$ 
vs $j_z/j_c$, where the circularities increase from the particles of the classical bulge to those pseudo bulge, followed by those of the thick disc, and reaching the maximum for 
the thin disc, while the binding energies decrease in the same sequence. While there is considerable overlap among these four components, in the $e/|e|_{\rm max}$ vs $j_z/j_c$ 
projection the stellar halo of g3.49e11 appears well separated from the rest, in the region of the lowest binding energies $e/|e|_{\rm max}>-0.5$ and small circularities $-0.5<j_z/j_c<0.5$. 
However, the stellar halo overlaps in circularities with all other components but the thin disc, while in binding energies it overlaps precisely with the thin disc. 
The case of this particular galaxy is therefore a good example of why it is important to go beyond simple cuts in these parameters \citep[e.g.][]{Brook:2004,Scannapieco:2010} 
to distinguish between the various galactic components.

Finally, we should mention that in the context of this work the \textit{pseudo bulge} is a type of inner component with a relatively flatter surface mass density distribution 
than the classical bulge and sometimes small traces of coherent rotation, though at much smaller velocities than the discs. The two bulge types, classical and pseudo, are not 
distinguished based on the S\'{ersic} index \citep[e.g.][]{Kormendy:2004,Fisher:2016}. 
In the case of g3.49e11 in particular, the pseudo bulge makes the transition in circularities and binding energies from classical bulge to thick disc, 
while its deviation from spherical symmetry is mirrored by the small values of $j_p/j_c$, more common of the thin disc.

All galaxy components studied in this work are identified based on the regions they occupy in the 3D space ($j_z/j_c$,$j_p/j_c$, $e/|e|_{\rm max}$) and nicknamed based on their general 
appearances in the edge-on surface mass density and line-of-sight velocity maps. We do not use the S\'{ersic} index or the scaleheights to distinguish between the components. 
In this work we focus only on the dynamical properties of the galaxy components identified with {\tt gsf}, while in a follow-up paper
we will study the connection between their morphology as encoded by parameters such as S\'{ersic} index, scaleheight and scalelength, and their dynamical properties.

\begin{table*}
\centering
\begin{tabular}{ccccccccc}
\hline
Sim & disc & thin disc & thick disc & inner disc & spheroid & classical b & pseudo b & halo\\
\hline
g1.08e11 & 0.45 & - & - & - & 0.55 & - & - & -\\
g1.37e11 & 0.33 & - & - & - & 0.67 & - & - & -\\
g1.52e11 & 0.58 & - & - & - & 0.42 & - & - & -\\
g1.59e11 & 0.41 & - & - & - & 0.59 & - & - & -\\
g2.41e11 & 0.28 & - & - & - & 0.72 & - & - & -\\
g2.42e11 & - & 0.27 & 0.26 & - & - & 0.27 & 0.16 & 0.04\\
g2.57e11 & - & 0.27 & 0.24 & - & - & 0.32 & 0.17 & -\\
g3.06e11 & - & 0.28 & 0.26 & - & - & 0.24 & 0.17 & 0.05\\
g3.49e11 & - & 0.26 & 0.28 & - & - & 0.28 & 0.15 & 0.03\\
g3.61e11 & - & 0.19 & 0.15 & 0.17 & - & 0.31 & 0.19 & -\\
g4.90e11 & - & 0.22 & 0.29 & - & - & 0.29 & 0.17 & 0.03\\
g5.02e11 & - & 0.23 & 0.26 & - & - & 0.29 & 0.18 & 0.04\\
g5.31e11 & 0.35 & - & - & - & - & 0.38 & 0.27 & -\\
g5.38e11 & - & 0.30 & 0.25 & - & - & 0.28 & 0.15 & 0.02\\
g5.46e11 & - & 0.22 & 0.26 & - & - & 0.27 & 0.23 & 0.02\\
g5.55e11 & 0.15 & - & - & 0.22 & 0.31 & 0.28 & - & 0.03\\
g6.96e11 & 0.37 & - & - & - & 0.23 & 0.14 & 0.20 & 0.06\\
g7.55e11 & - & 0.18 & 0.32 & 0.10 & 0.26 & 0.14 & - & -\\
g7.66e11 & - & 0.19 & 0.21 & - & 0.16 & 0.30 & 0.14 & -\\
g8.13e11 & 0.18 & - & - & 0.39 & - & 0.23 & 0.20 & -\\
g8.26e11 & - & 0.22 & 0.33 & - & - & 0.24 & 0.14 & 0.06\\
g1.12e12 & 0.05 & - & - & 0.20 & - & 0.42 & 0.34 & -\\
g1.77e12 & 0.21 & - & - & 0.25 & - & 0.26 & 0.15 & 0.13\\
g1.92e12 & - & 0.18 & 0.24 & - & - & 0.49 & 0.10 & -\\
g2.79e12 & - & 0.30 & 0.10 & 0.31 & - & 0.08 & 0.20 & -\\
\hline
\end{tabular}
\centering\caption{The stellar mass fractions of the various kinematic components for the sample of 25 simulated galaxies.}
\label{table_massfractions}
\end{table*}

\section{The variety of stellar structures in simulated galaxies}
\label{sec5}

The 25 NIHAO galaxies in the current sample range from dwarfs to objects a few times more massive than the Milky Way. As for observed galaxies, 
it is not expected that all will exemplify the same combination of stellar substructures. 
For each one of them we ran {\tt gsf} with $nk=2, 3, 4, 5$, constructed the edge-on stellar mass surface density and line-of-sight velocity maps, 
and chose by visual inspection the \textit{optimal} $nk$. In the case of this study, \textit{optimal} $nk$ means the $nk$ which results in either 
clearly separated thin/thick discs, clearly separated classical/pseudo bulges, and/or the presence of a stellar halo. 
Figure~\ref{figure_allsampledeco} of the Appendix shows the edge-on surface mass density and mass-weighted line-of-sight velocities for all the \textit{optimal} components of the 25 galaxies.

\begin{figure*}
\begin{center}
\includegraphics[width=0.98\textwidth]{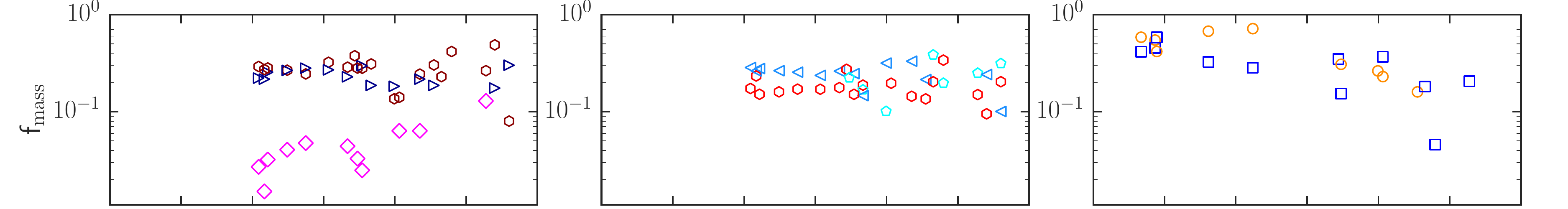}\\
\includegraphics[width=0.98\textwidth]{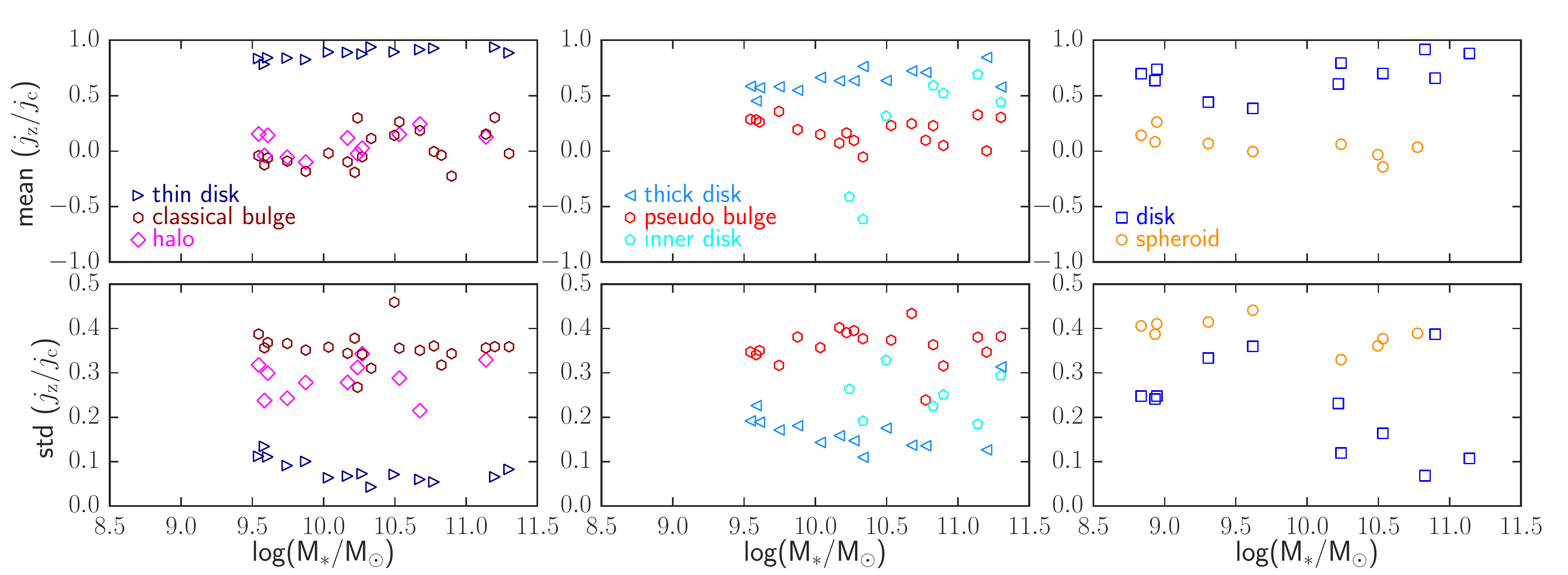}\\
\includegraphics[width=0.98\textwidth]{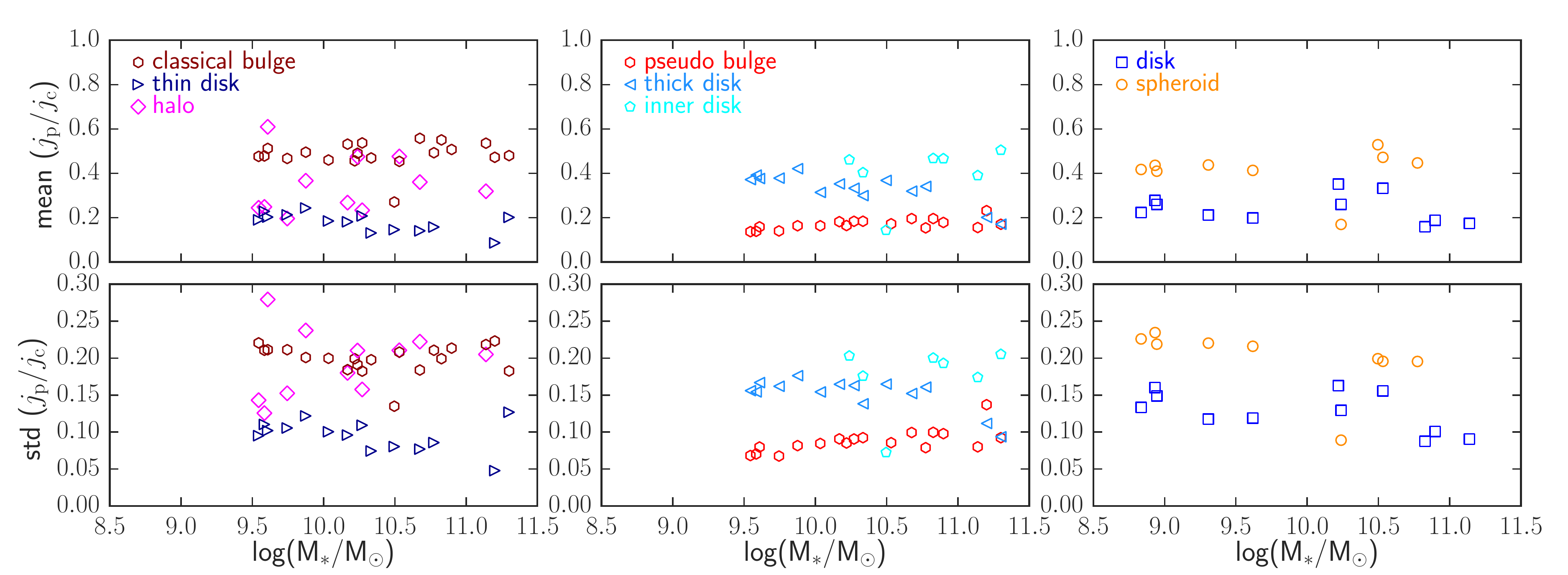}\\
\includegraphics[width=0.98\textwidth]{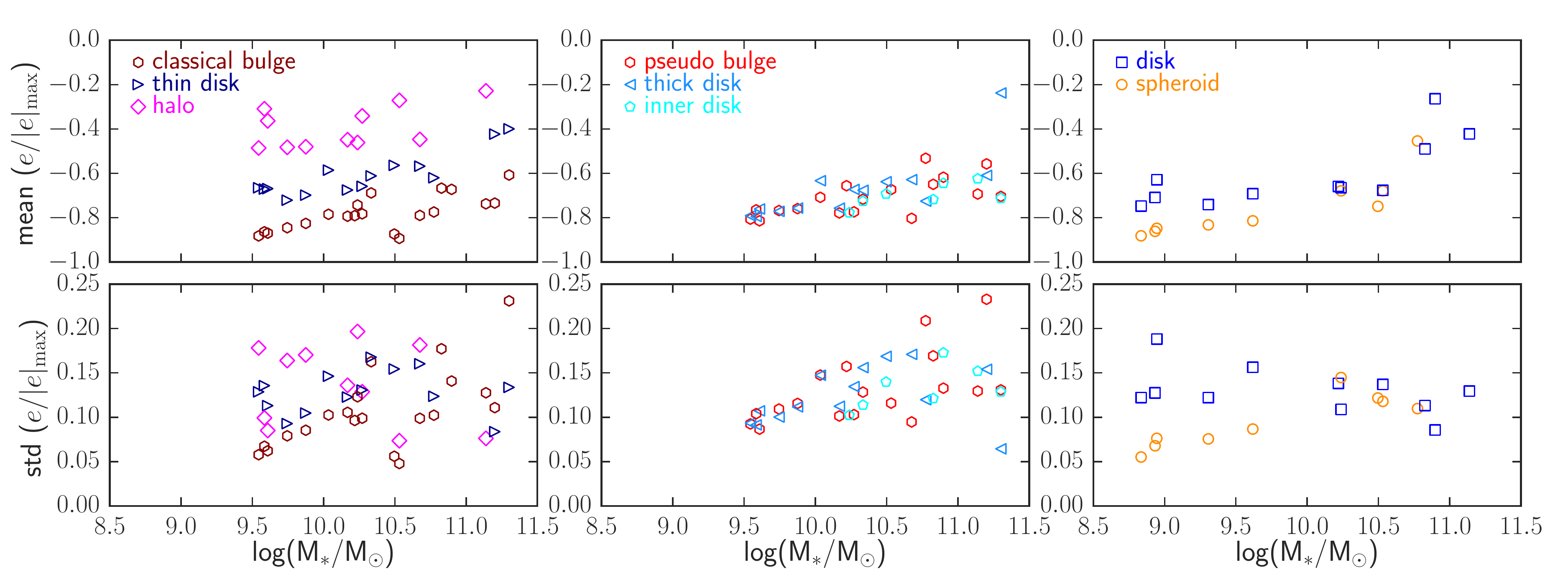}\\
\caption{The means and standard deviations of the distribution functions in the input clustering variables $j_z/j_c$ (2$^{\rm nd}$ and 3$^{\rm rd}$ rows), 
$j_p/j_c$ (4$^{\rm th}$ and 5$^{\rm th}$ rows) and $e/|e|_{\rm max}$ (6$^{\rm th}$ and 7$^{\rm th}$ rows) for all the components found in the sample of 25 simulated NIHAO galaxies, 
as a function of total stellar mass. The first row gives the corresponding stellar mass fractions $f_{\rm mass}$.}
\label{fig_gmm}
\end{center}
\end{figure*}

The five least massive galaxies have only two clearly distinct kinematic components. 
The first component comprises the part of stellar material least gravitationally bound and with eccentricities confined to a large extent under the left winged peak at $j_z/j_c=1$.
It displays a highly flattened surface mass density and a spider velocity pattern, and is therefore a \textit{disc}. 
The other component is made of the most gravitationally bound material and has a wide $j_z/j_c$ distribution, roughly centered on zero. 
These properties translate into it having a surface mass density more spherically symmetric and showing almost no signature of a coherent rotation pattern. 
Therefore, this component is a \textit{spheroid}. 

At higher stellar masses, {\tt gsf} starts finding two kinematically distinct components with disc properties. 
One component is much flatter, has higher rotation speeds and is slightly less bound than the other. 
The first one has a $j_z/j_c$ distribution sharply peaked at $1$, while the other shows a broader distribution, peaked at smaller eccentricities, $j_z/j_c\sim0.75$. 
Therefore, these two kinematic components correspond to a \textit{thin disc} and a \textit{thick disc}, respectively. 
However, not all of the 20 most massive galaxies in our sample have two distinct large scale discs. Six galaxies out of 20 have only one large scale disc. 
Therefore, we will denote by \textit{disc} both the rotation dominated component of dwarfs as well as the single large scale spinning component of more massive galaxies.

Another type of rotating component appears in a few of our galaxies. 
It is characterized by a small spatial extent and a large degree of flattening in the edge-on surface mass density map, 
and a coherent velocity pattern with maximum speeds comparable to those of thin discs. This component corresponds to an \textit{inner disc}. 
In a few cases, this component has a very boxy appearance in the surface mass density map, and as such one can be tempted to call it a bar. 
However, our analysis does not distinguish components based on orbit classification, which would be the appropriate diagnostic to identify bars,  
and as such we prefer to refer to these components as inner discs.

For the most massive 20 of 25 galaxies, {\tt gsf} also distinguishes various types of velocity dispersion dominated components. 
All of these 20 galaxies contain a highly compact, spherically symmetric, very slowly rotating (or non-rotating) component that comprises the most gravitationally bound material, 
which corresponds to a \textit{classical bulge}. 
Some of these 20 galaxies also show a second dispersion dominated component in the innermost region, which is flattened to some degree on the 
$z$-axis and shows signs of coherent rotation, albeit with relatively small velocities. When compared to the \textit{classical bulge} it is typically also less gravitationally bound. 
Based on these properties, this secondary central component is called \textit{pseudo bulge}. 

In 11 out of 25 galaxies, {\tt gsf} unveiled a kinematic component occupying the low binding energy tails of the galaxy distributions with a large scale diffuse appearance in the 
surface mass density maps. Only in some of the galaxies this component shows rotation signatures. This component is the \textit{stellar halo}. 
Though stellar haloes and spheroids might have misleading similar appearances in surface mass density, it is important to stress that the two occupy the opposite ends of the 
binding energy range. The global surface mass density of stellar haloes is much lower than that of spheroids.
On the other hand, classical bulges differentiate themselves from spheroids by having not only larger binding energies, but also by being much more compact and 
confined to the very inner galactic regions.

Figure~\ref{fig_gmm} gives the means and standard deviations of the distribution functions in the input space of $j_z/j_c$ (top), $j_p/j_c$ (centre) and $e/|e|_{\rm max}$ (bottom), 
and the stellar mass fractions of all the components found in the galaxy sample, as a function of the total stellar mass. 
The only types of structures that seem to have a slight correlation between $f_{\rm mass}$ and the total stellar mass are the single discs, spheroids and the stellar haloes. 

The means of the circularity parameter distributions in Figure~\ref{fig_gmm} show little to no stellar mass dependency
for all kinematic components, each type occupying different ranges of circularities.  
The disc components have significantly smaller standard deviations than the dispersion dominated ones, decreasing from single large scale discs, to thick and finally to thin discs. 
Overall, the standard deviations of the disc components show a decrease with increasing stellar mass. 
The standard deviations of the dispersion dominated structures do not correlate with the stellar mass.

In the forth row from the top of Figure~\ref{fig_gmm} the means of the $j_p/j_c$ distributions show also little to no correlations with the stellar mass. 
The thin discs, large scale single discs and pseudo bulges have very similar low $mean(j_p/j_c)\lesssim0.2$. 
The only correlation of the $j_p/j_c$ mean with the stellar mass occures for thick discs, 
which decreases from $\sim0.4$ at $M_{\rm *}=10^{\rm9.5}M_{\rm\odot}$ to  $\sim0.2$ at $M_{\rm *}=10^{\rm11.3}M_{\rm\odot}$. 
The spheroids have $mean(j_p/j_c)\sim0.4$, while the classical bulges have $mean(j_p/j_c)\sim0.5$. Most of the inner discs also have the mean between $0.4$ and $0.5$. 
The stellar haloes, on the other hand, appear all over the place, with values as low as $0.2$ and as high as $0.6$. 

In the specific binding energy panels, sixth row from top down, the mean of the $e/|e|_{\rm max}$ increases with the stellar mass for all types of kinematic structures.
At fixed $M_{\rm *}$, the least bound structures are the stellar haloes, followed by the thin discs, at their turn followed by the large scale single discs. 
All other types of structures (thick discs, inner discs, pseudo bulges, classical bulges and spheroids) share mainly the same region in the $mean(e/|e|_{\rm max})$-$M_{\rm *}$ plane, 
with a tendency of spheroids and classical bulges towards the largest absolute values of the specific binding energy.

In summary, in this sample of 25 NIHAO simulated galaxies, {\tt gsf} is able to identify eight types of stellar kinematic structures: discs, thin and thick discs, inner discs, spheroids, 
classical and pseudo bulges, and stellar haloes. The mass fractions in these components of all 25 galaxies are given in Table~\ref{table_massfractions}
and their edge-on surface mass densities and edge-on line-of-sight velocities are given in Figure~\ref{figure_allsampledeco} of the Appendix.
The means and standard deviations of $j_z/j_c$, $j_p/j_c$ and $e/|e|_{\rm max}$ for all stellar structures in the sample are given in Table~\ref{table_aux} of the Appendix,
together with all the other intrinsic properties analysed in this study.  

Most of the structures we identify dynamically, namely the thin/thick discs, classical bulges, spheroids and stellar haloes, have obvious counterparts in observations.
However, no such clear correspondence can be made for what we call \textit{pseudo bulges} and \textit{inner discs}. In observations, it is known that 
the inner regions of galaxies can simultaneously host a large variety of stellar substructures like nuclear rings, bars and/or various types of bulges 
\citep[e.g.][]{Erwin:2003,Athanassoula:2005,Gadotti:2009,Aguerri:2009,Nowak:2010,Kormendy:2010,Mendez-Abreu:2014,Fisher:2016}, and there is yet no consensus on a unified 
classification scheme and formation scenario for them \citep[e.g.][]{Graham:2016}. 
For example, ``observational pseudo bulges'' are typically taken to be the central components that have light distributions with S\'{ersic} indices $n<2$, 
while ``observational classical bulges'' are assumed to have $n>2$ \citep[e.g][]{Kormendy:2004}. 
However, it is still currently debated whether a hard limit of $n\sim2$ separating classical from pseudo bulges has any physical meaning \citep[e.g.][]{Andredakis:1994,Andredakis:1995,Graham:2016}.

In the context of our dynamical classification of components, the main difference between what we call \textit{pseudo bulge} and \textit{inner disc} 
is that the former has only small traces of coherent rotation, while the latter is a central structure that rotates with velocities similar to the large scale discs. 
This difference is nicely exemplified by the $mean(j_z/j_c)$ shown in Figure~\ref{fig_gmm}, where inner discs / pseudo bulges have absolute mean circularities larger/smaller than $\sim0.5$.
The intrinsic flattening of both \textit{pseudo bulges} and \textit{inner discs} have intermediate values between those of spheroids/classical bulges and those of (thin/thick) discs.
The most likely observational counterparts to what we call \textit{inner discs} are the bars and/or the boxy/peanut bulges \citep[e.g.][]{Athanassoula:2005}.

For the sake of simplicity we decided to group all the resulting stellar structures in these eight types, 
even though in a few cases the assignment to one of these groups is not obvious. The most ambiguous assignments refer to the pseudo bulges of g5.46e11 and g6.96e11. 
The pseudo bulge of g5.46e11 looks like a much slower rotating, secondary thick disc.  
The pseudo bulge of g6.96e11 also looks like a thick disc, but if we would have considered it as such the 
difference between the large scale single disc and the pseudo bulge / thick disc would have been given only by the velocity pattern and not by the vertical thickness. 

In this study a few very interesting stellar structure types were revealed, which will be the subject of future studies.
A few simulated galaxies have stellar structures which rotate in the opposite direction as their large scale disc(s): 
the classical bulges of g3.06e11, g5.31e11 and g1.12e12, the inner disc of g5.55e11, the spheroid of g6.96e11 and the stellar haloes of g2.42e11 and g3.06e11.
The counter-rotating structures are marked in column eight of Table~\ref{table_aux}.
Another interesting fact is that the classical bulge of g5.55e11 displays an X-shaped feature in the line-of-sight velocity maps. 

Finally, we should mention that the current version of {\tt gsf} available online does not suggest any names for the components, leaving this step for the users. 
The main reason we choose to do so is the lack of consensus on the definitions of the inner components of galaxies.

\begin{figure*}
 \includegraphics[width=0.48\textwidth]{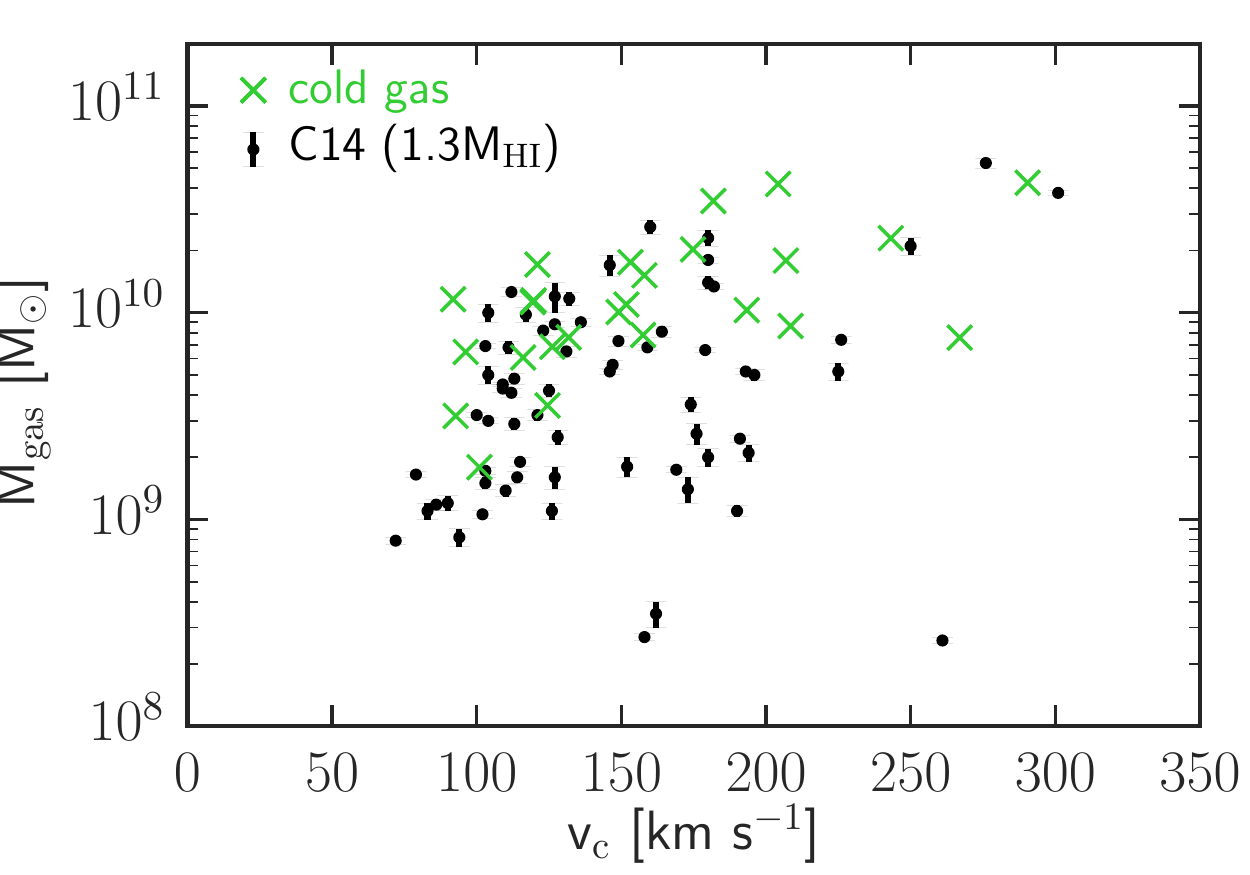}
 \includegraphics[width=0.48\textwidth]{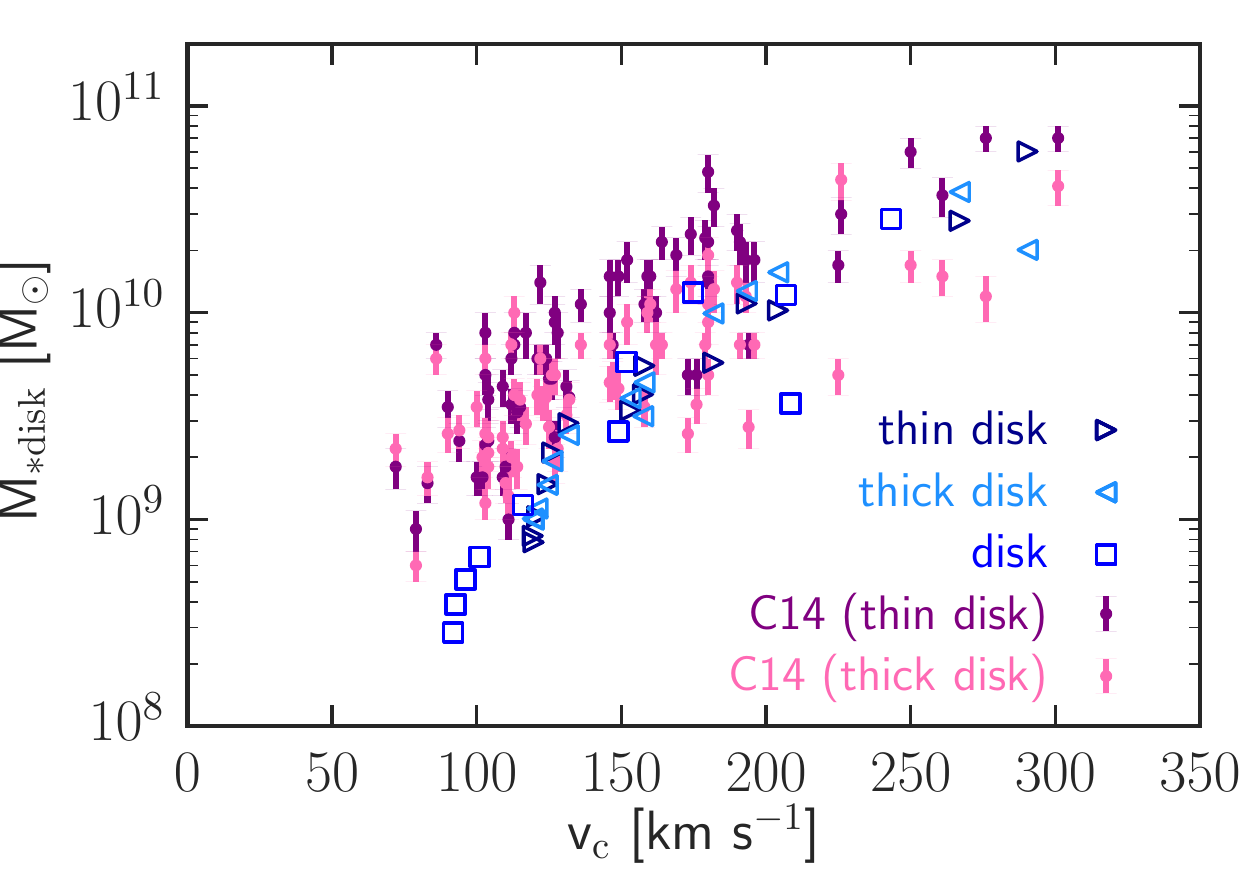}\\
 \includegraphics[width=0.48\textwidth]{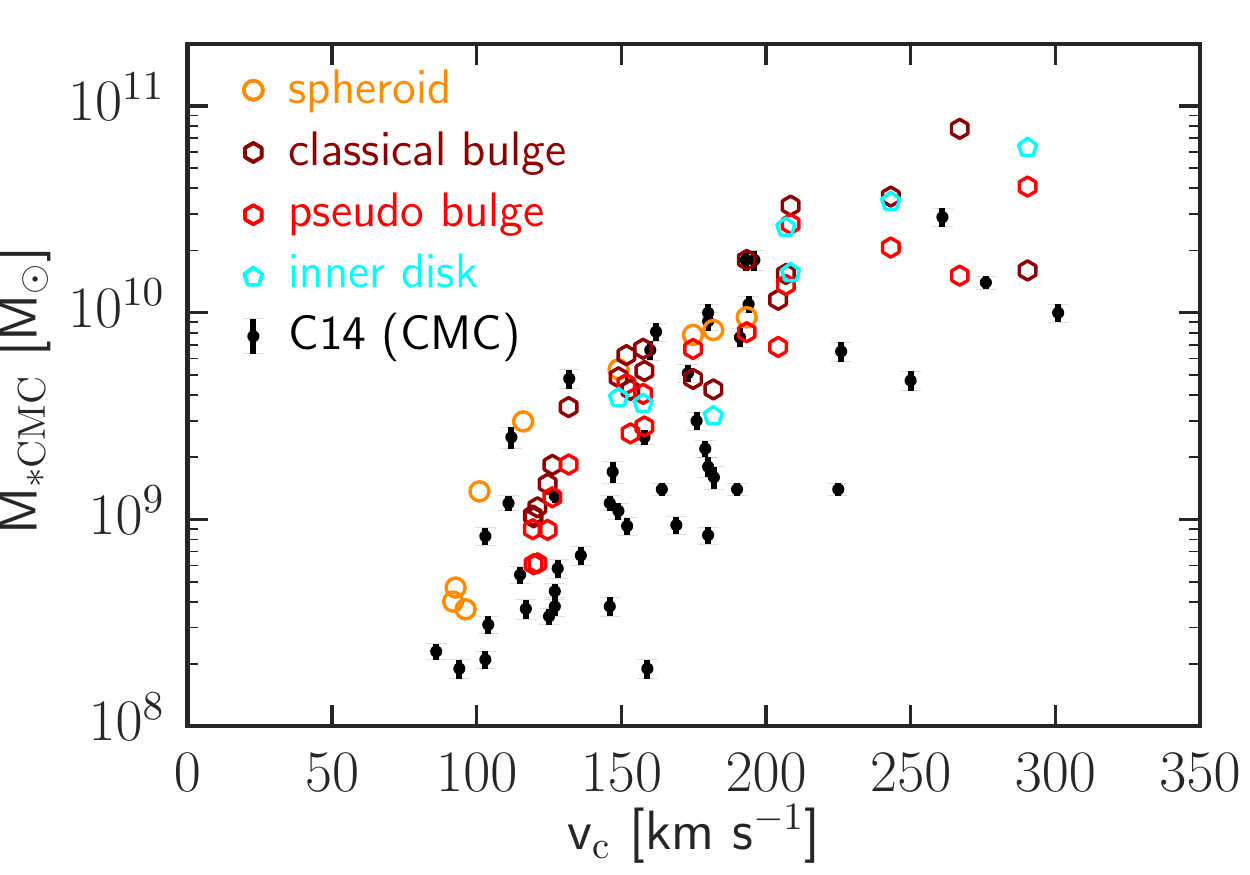}
 \includegraphics[width=0.48\textwidth]{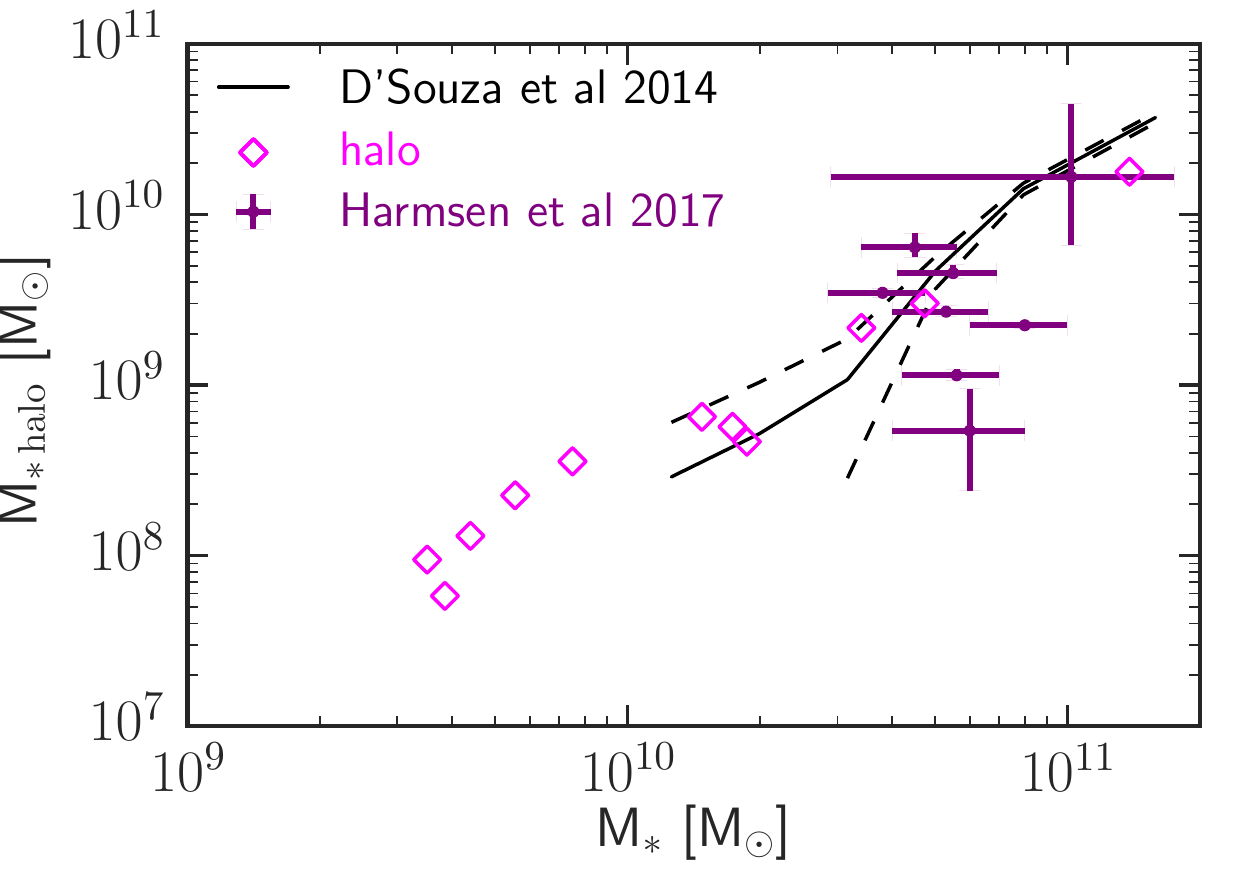}\\
\caption{The mass in the various components of the simulated galaxies as a function of the circular velocity $v_{\rm c}$ at $0.15r_{\rm vir}$ 
(top left and right, and bottom left panels), or as a function of the total stellar mass $M_{\rm *}$ (bottom right panel). 
All kinematic components but the stellar haloes are compared with the corresponding observational data of \citet{Comeron:2014} (C14), represented by the black, purple or pink symbols.
CMC stands for Central Mass Component. 
The bottom right panel shows the comparison between the kinematic stellar haloes and the SDSS stacking results of \citet{DSouza:2014} for low concentration galaxies (solid black curve). 
The purple data points are the observations of \citet{Harmsen:2017} for a small sample of nearby galaxies.}
 \label{figure_com2}
\end{figure*}

\subsection{Dynamical disc-to-total ratios}
\label{dt_ratios}

The fraction of light produced by the central, velocity dispersion dominated stellar component of galaxies, with respect to the total stellar light 
is a photometrically defined quantity, also known as bulge-to-total ratio $B/T$. 
The value of an observed galaxy $B/T$ determines its Hubble type, 
and is derived by fitting combinations of functions, like for example a S\'{ersic} plus an exponential, to either galaxy images or to azimuthally averaged galaxy luminosity profiles. 

In this work we are concerned with the dynamical state of galaxies, hence our derived $B/T$ or the equivalent disc-to-total mass ratios, $D/T=1-B/T$,
represent the fraction of a galaxy's stellar mass supported mainly by velocity dispersion or rotation, respectively. 
Almost all observational studies still assume that the dynamical and photometric $B/T$ are the same, 
e.g. the stellar mass in the exponential part of the profile is supported by rotation, while the material of the S\'{ersic} profile is mainly supported by velocity dispersion. 
However, this assumption is not necessarily true \citep[e.g.][]{Scannapieco:2010,Martig:2012,Obreja:2016,Sokolowska:2017,Grand:2017}.

In this study, we take disc mass $D$ to be the sum of all rotation dominated kinematic components, be they discs, thin or thick discs, or inner discs. 
The disc-to-total $D/T$ dynamical mass ratios calculated in this manner show no correlation with the total stellar mass $M_{\rm *}$. 
Across the sample of 25 galaxies $\langle D/T\rangle=0.47\pm0.11$ (see Table~\ref{table_massfractions} for the individual $D/T$s.)
Giving that we do not perform any kind of profile fitting in this study, all stellar masses we quote for these simulated galaxies are \textit{dynamical} masses. 
On the contrary, all observationally stellar masses from the literature that we use to compare our simulations with are intrinsically \textit{photometric} masses, 
e.g. derived from light profile fitting under different assumptions for $M/L$..  
 
For the galaxies in the simulated sample that have two distinct large scale kinematic discs, the mass is distributed roughly equal between them, 
and there is only a slight tendency for smaller thin/thick disc mass ratios with increasing $M_{\rm *}$. 
The average values for the 14 galaxies with a thin and a thick discs are: $D_{\rm thin}/T=0.23\pm0.04$ and $D_{\rm thick}/T=0.25\pm0.06$.

A thin/thick disc mass ratio of $\sim1$ might also be true for the Milky Way \citep[e.g.][]{Fuhrmann:2012}.
The most massive galaxy, g2.79e12 has a high total $D/T=0.71$ due to the presence of a relatively high mass inner disc component, 
identified as a bar by \citet{Buck:2018}.

\section{Observational properties}
\label{obs_prop}

One important parameter of galaxies is the total dynamical mass in the inner region of the dark matter halo. 
This parameter is sometimes cast as a velocity called \textit{circular velocity}: $v_{\rm c}=\sqrt{\rm GM(<r_{\rm c})/r_{\rm c}}$, with G the gravitational constant. 
For the current simulation sample we use $r_{\rm c}=0.15r_{\rm vir}$, which contains almost all the stellar mass. 
This radius roughly corresponds to the limit after which the radial profile of the circular velocity stays constant. 

We use as independent variable either the circular velocity $v_{\rm c}$ or the total stellar mass $M_{\rm *}$ to check whether 
the masses of the various kinematic components in the simulated galaxy sample agree with observational data. 
The two top and the left bottom panels of Figure~\ref{figure_com2} show the comparison between the masses of the various kinematic components in the simulations 
and the corresponding masses derived from photometric fits to a subsample of the objects from the Spitzer Survey of Stellar Structures in Galaxies, S$^{\rm 4}$G \citep{Comeron:2014}.
The independent variable in this case is the circular velocity.
These authors use a sample of 69 edge-on galaxies from S$^{\rm 4}$G to fit 3.6$\rm\mu$ and 4.5$\rm\mu$ radial profiles with combinations of a S\'{ersic} 
(for the central mass concentrations or CMCs) and up to four exponentials and/or truncated discs \citep{Erwin:2008}. 
Due to the limitations of their fitting, they exclude the galaxies with large CMCs, and 
as such their sample is strongly biased towards late Hubble types ($\langle M_{\rm CMC}/M_{\rm total}\rangle=0.09$).

There is a general good agreement between our simulations and the observations of \citet{Comeron:2014} 
for the central stellar components (bottom left panel), as well as for the gas masses (top left panel). 
The gas masses of \citet{Comeron:2014} are $1.3\times M_{\rm HI}$, such that they account for He and metals in atomic form as well, 
while cold gas in the simulations means gas with temperatures below 15000~K. Since our simulations do not resolve the molecular phase,
the green points in the top left panel represent $\simeq M_{\rm H2}+1.3\times M_{\rm HI}$, 
and thus are expected to be between $30\%$ and $73\%$ higher \citep[e.g.][]{Saintonge:2011} than the observations of \citet{Comeron:2014} 
at a fixed circular velocity / stellar mass.

When comparing the rotation dominated structures, the masses of the photometrically derived discs from observations 
tend to be higher on average than the masses of the kinematically defined discs from simulations at $v_{\rm c}\lesssim$150km~s$^{\rm -1}$ 
(top right panel of Figure~\ref{figure_com2}). At higher $v_{\rm c}$, however, the observations and simulations agree very well. 
Also, while the kinematic stellar discs of simulated galaxies form a tight sequence between the disc mass and circular velocity, 
the observations show a larger dispersion in masses for a given $v_{\rm c}$. 
This scatter comes partly from the fitting procedure (large number of parameters, some of which can be correlated). 
However, the fact that the scatter is higher for small values of $v_{\rm c}$ is most likely due to the fact that circular velocities are estimated from HI kinematics \citep{Courtois:2009}
under assumptions that hold for large disc-like galaxies, but not for less massive objects \citep[e.g.][]{Maccio:2016}. 

The bottom right panel of Figure~\ref{figure_com2} compares the masses of the stellar haloes in the simulated sample with those derived by 
\citet{DSouza:2014} using stacked SDSS data, as a function of the total stellar mass. 
The purple points in the same panel give the observations of \citet{Harmsen:2017} for a small sample of nearby galaxies. 
In the total stellar mass range where the simulations and observations overlap, there is a very good agreement between the two in terms of stellar halo masses.
Our simulations predict stellar haloes to sometimes exist in galaxies with total stellar masses below 10$^{\rm 10}$M$_{\rm\odot}$.

\subsection{Thick/thin disc ratios vs circular velocity}

\begin{figure}
 \includegraphics[width=0.48\textwidth]{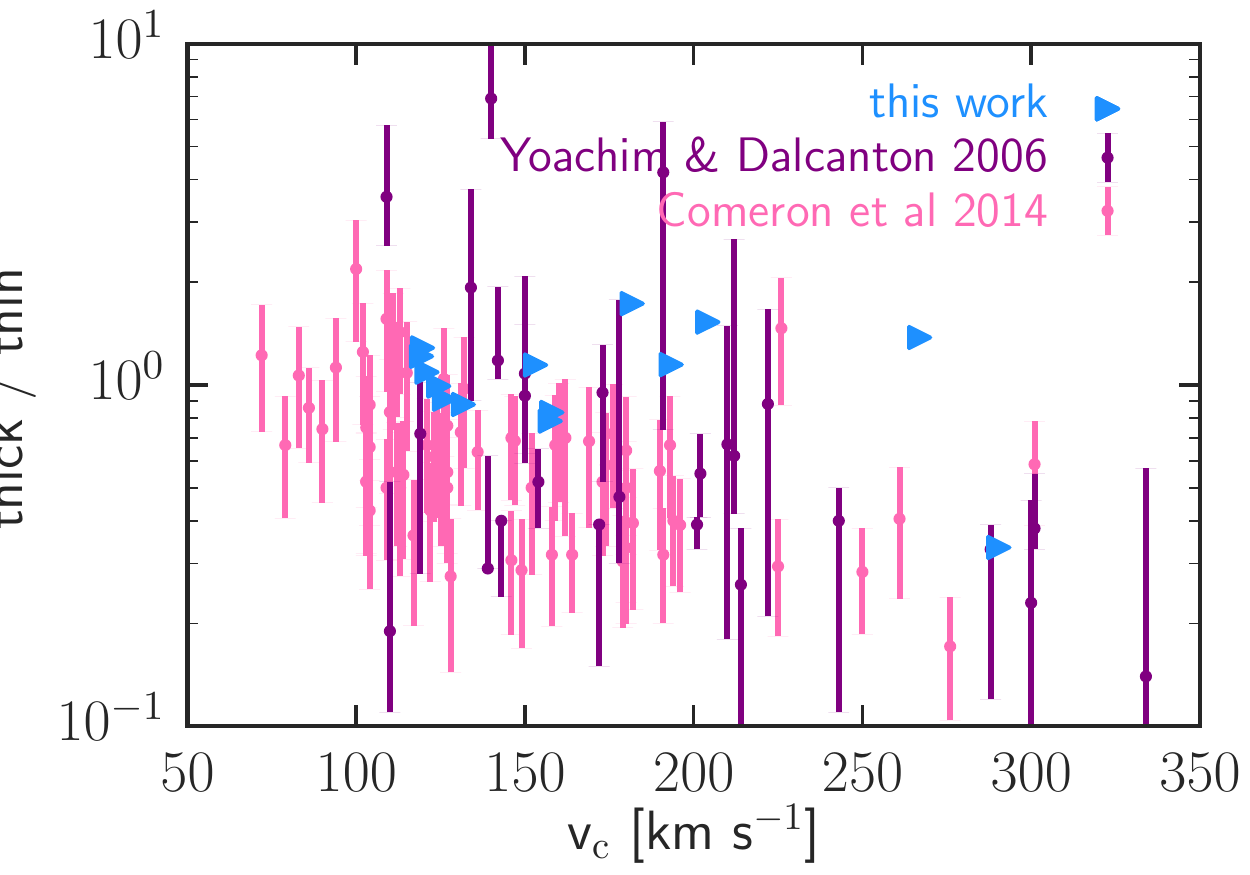}
\caption{The thick-to-thin ratio for the kinematic discs of the simulated galaxies (light blue triangles). The observational data set of \citet{Yoachim:2006} represent light ratios, 
while the one of \citet{Comeron:2014} are mass ratios computed from photometry using different $M/L$ for the thin and thick discs.}
 \label{figure_thick2thin}
\end{figure}

Notwithstanding the multiple pitfalls when comparing dynamical to photometric mass ratios, 
$M/L$ are very sensitive to the IMF, stellar evolution and star formation rate (hereafter SFR) histories \citep[e.g][]{Bell:2001,Conroy:2009}, 
Figure~\ref{figure_thick2thin} compares the dynamical thick-to-thin values with the observations of \citet{Yoachim:2006} and \citet{Comeron:2014}.
\citet{Comeron:2018} have revised their analysis of the S$^{\rm 4}$G data from \citet{Comeron:2014} and expanded the galaxy sample, 
but their photometrically derived thick/thin disc mass ratios did  not change on average. 
One of the conclusions of \citet{Comeron:2018} is that $82\pm6\%$ of the galaxies in their sample of 141 objects are double thin/thick discs, 
which is higher than the $56\%$ frequency out of our 25 simulated galaxies. 

The take away message from Figure~\ref{figure_thick2thin} is that the thick/thin dynamical mass ratios from simulations are in broad agreement with the photometric ones from observations. 
However, the observations of both \citet{Yoachim:2006} and \citet{Comeron:2014} suggest that more massive systems should have relatively less massive 
thick (photometric) discs, while the mass ratios from our kinematic selection shows no such anticorrelation with $v_{\rm c}$. 
To have a good understanding of the reason for this discrepancy with observations we have to make the step from kinematics to photometry for our simulated galaxies, 
which is the aim of a future study.

\subsection{Velocity dispersion vs rotational velocity}
\label{sigma_vs_vrot}

\begin{figure}
\includegraphics[width=0.47\textwidth]{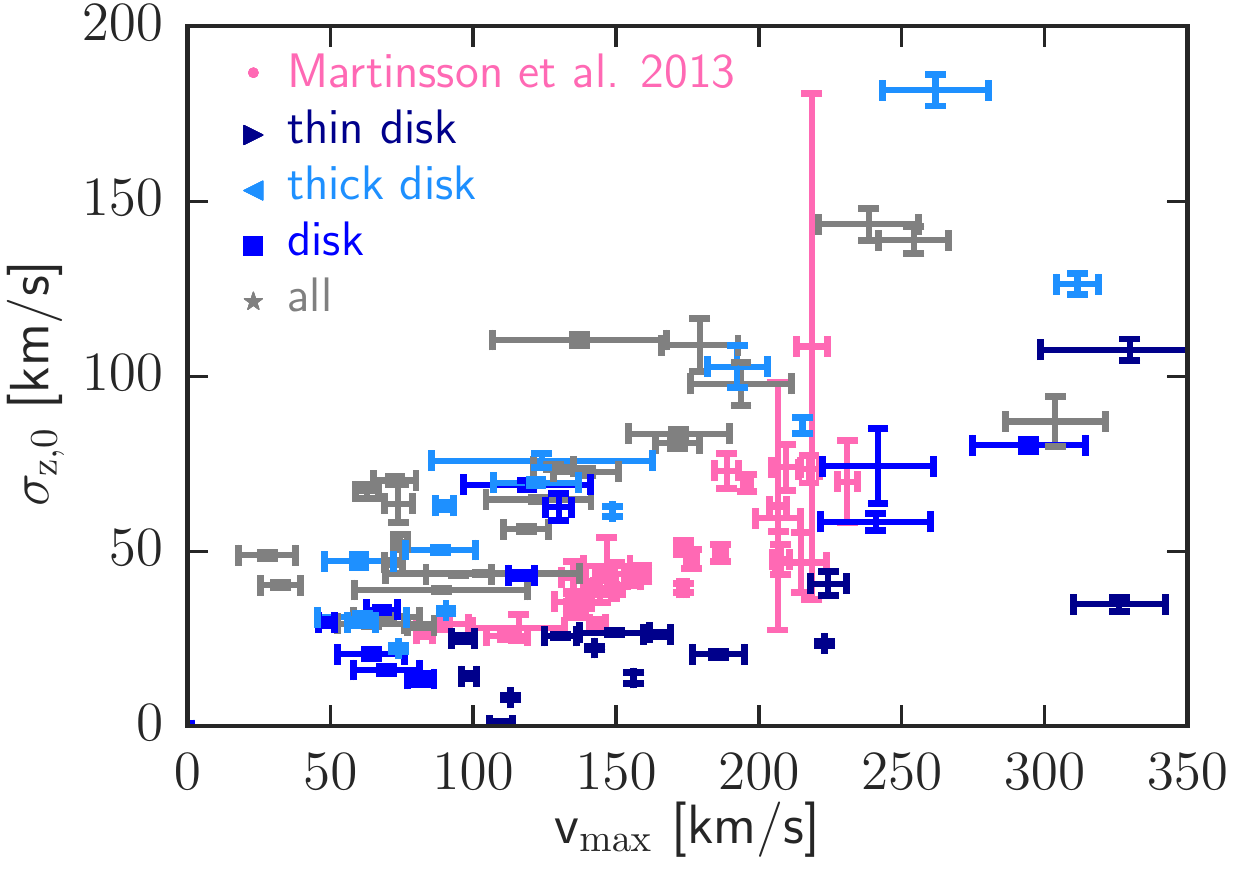}
\caption{The central vertical velocity dispersion $\sigma_{\rm z,0}$ as a function of maximum edge-on line-of-sight velocity $v_{\rm max}$
for the relevant kinematic components of the simulated galaxies. The pink symbols give the observational values from the DiskMass survey \citep{Martinsson:2013}.}
\label{figure_vsig_1}
\end{figure}

To test if the stellar kinematics of the simulated galaxy sample are within observational ranges, we compare the complete stellar distributions and the 
kinematically defined discs with the DiskMass Survey galaxies \citep{Bershady:2010}. In particular, \citet{Martinsson:2013} analysed a subsample of the 
DiskMass galaxies and placed them in a rotation -- velocity dispersion plane. 
The galaxy sample of \citet{Martinsson:2013} was chosen explicitly to contain only the most dynamically cold systems in the DiskMass Survey, 
while no such criteria was applied to the simulated sample.
\citet{Martinsson:2013} use exponential fits to the line-of-sight velocity dispersion data 
corrected for inclination to derive the so-called central vertical velocity dispersions $\sigma_{\rm z,0}$ of the stars. To quantify the stellar rotation, 
they extracted the line-of-sight velocity profiles along the semi-major axes of each galaxy image and used a hyperbolic tangent fit 
$\rm v_{\rm los}=V_{\rm arot}tanh(R/R_{\rm s})$ to obtain the stellar asymptotic rotational velocities $V_{\rm arot}$. These authors caution that 
the hyperbolic tangent is not the optimal choice for all types of galaxies (e.g. galaxies with rotational velocities that decline with radius). 

The edge-on line-of-sight velocities along the semi-major axes $v_{\rm los}(R)$ are shown for the complete stellar distributions, and for the various kinematic disc components 
in Figure~\ref{figure_appendix_3a} of the Appendix. The $v_{\rm los}(R)$ for the complete stellar distributions are well fitted by the
hyperbolic tangent function only in approximately $60\%$ of the simulated galaxies. For the disk components this particular function provides a good approximation 
in a majority of cases: 10 of 14 thin discs, 8 of 14 thick discs, and 6 of 11 discs. By comparison, the hyperbolic tangent is adequate for a larger fraction of observed galaxies
(29 out of 31 DiskMass galaxies). For this reason, we use as an estimation of rotation the maximum velocity instead of the asymptotic values from the fits.
For the simulated galaxies that are not well approximated by the fitting function $v_{\rm max}/V_{\rm arot}\sim1.3$, while for the DiskMass galaxies $v_{\rm max}/V_{\rm arot}\simeq1$. 

The vertical velocity dispersion profiles $\sigma_{\rm z}(R)$ of the simulated galaxies and of the kinematic disc components are given in 
the Figure~\ref{figure_appendix_2a}. 
As it can be appreciated from this figure, the $\sigma_{\rm z}(R)$ can only be approximated by an exponential function for the most massive galaxies. 
Also, at a closer look, the observational data of \citet{Martinsson:2013} is quite noisy and not obviously following an exponential outside of the 
bulge regions. For these reasons we decided to estimate the central vertical velocity dispersion by using the extrapolation to $R=0$ of a second order polynomial instead, 
which provides a very good description of the simulations outside the bulge region for all 25 galaxies, as well as for their disc components. 
The corresponding values of $\sigma_{\rm z,0}$ computed in this manner are given in Figure~\ref{figure_appendix_2a}. 

With these caveats in mind, Figure~\ref{figure_vsig_1} gives the comparison between the simulations (grey and blue symbols) and the observations (pink symbols) 
in the $v_{\rm max}$-$\sigma_{\rm z,0}$ plane. At a fixed maximum rotational velocity, the central vertical velocity dispersion of the complete stellar distributions in the 
simulations (grey stars) is about a factor of two larger than the observational values (pink dots). 
However, the majority of the kinematic large scale single discs (blue squares) and the thin discs (dark blue triangles) are in better agreement with the observational data, 
actually having smaller $\sigma_{\rm z,0}$ than the observations for fixed $v_{\rm max}$.
This behaviour can be naturally explained by the fact that stellar kinematics in observational studies are biased towards younger stellar populations, 
which are more likely to be part of the dynamically cold structures.  
For an even more apple to apple comparison between simulations and observations, 
the former should be analyzed by weighting each stellar particle's contribution to the velocity dispersion and rotational velocity in accordance with 
the amplitudes of its corresponding metal absorption lines given its age and chemical composition. 

We should also note that resolution could play a role in the magnitude of the vertical velocity dispersion 
\citep[e.g.][]{Velazquez:2005}. 
However, the differences we find between the central velocity dispersions 
of the complete stellar populations and those of the kinematic discs are not particular to the NIHAO simulations \citep{Obreja:2016}; 
the more recent higher resolution AURIGA simulations show the same behaviour \citep{Grand:2017}. Given also that stellar feedback 
is known to heavily influence the disc structure \citep[e.g.][]{Roskar:2014}, the numerical effects in the vertical velocity 
dispersion of our galaxies are most likely subdominant.

\begin{figure}
\includegraphics[width=0.48\textwidth]{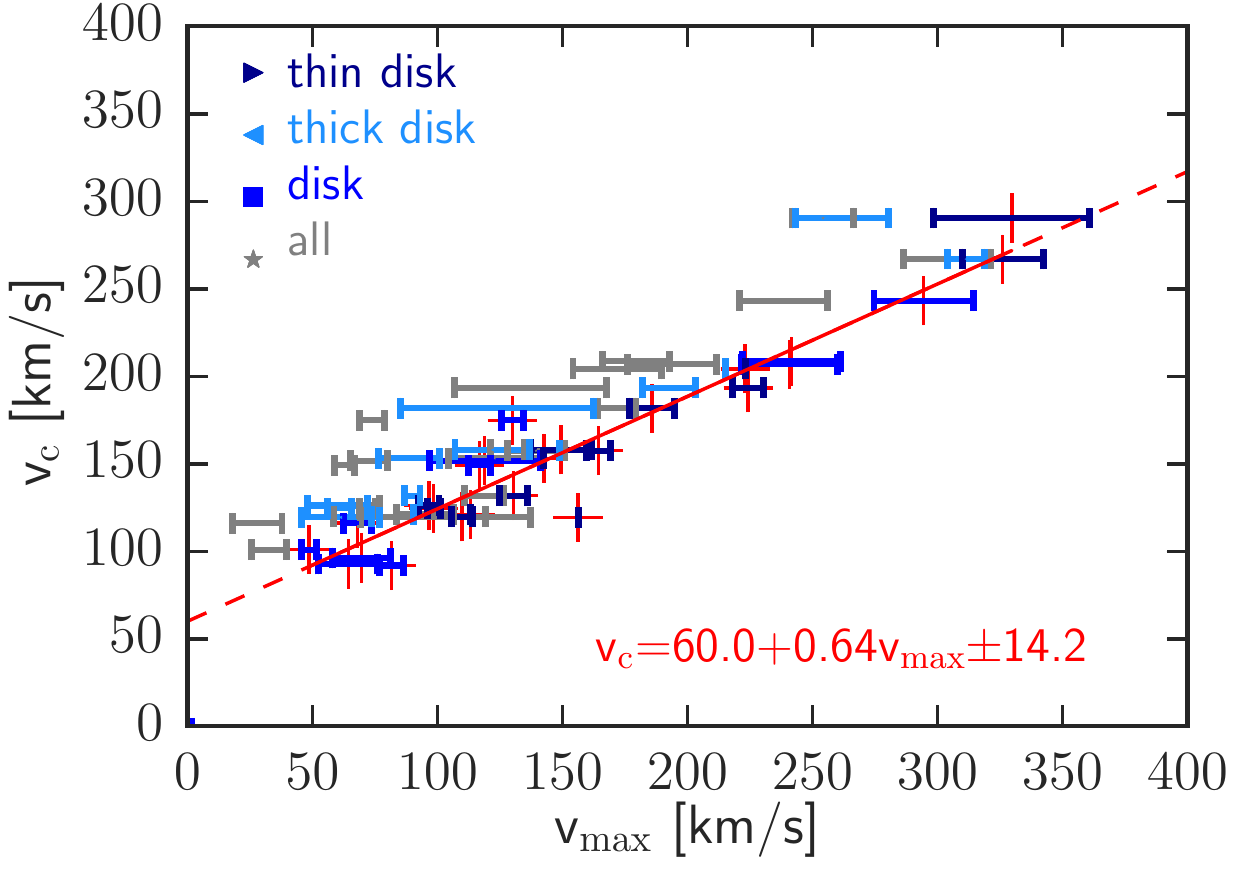}\\
\includegraphics[width=0.48\textwidth]{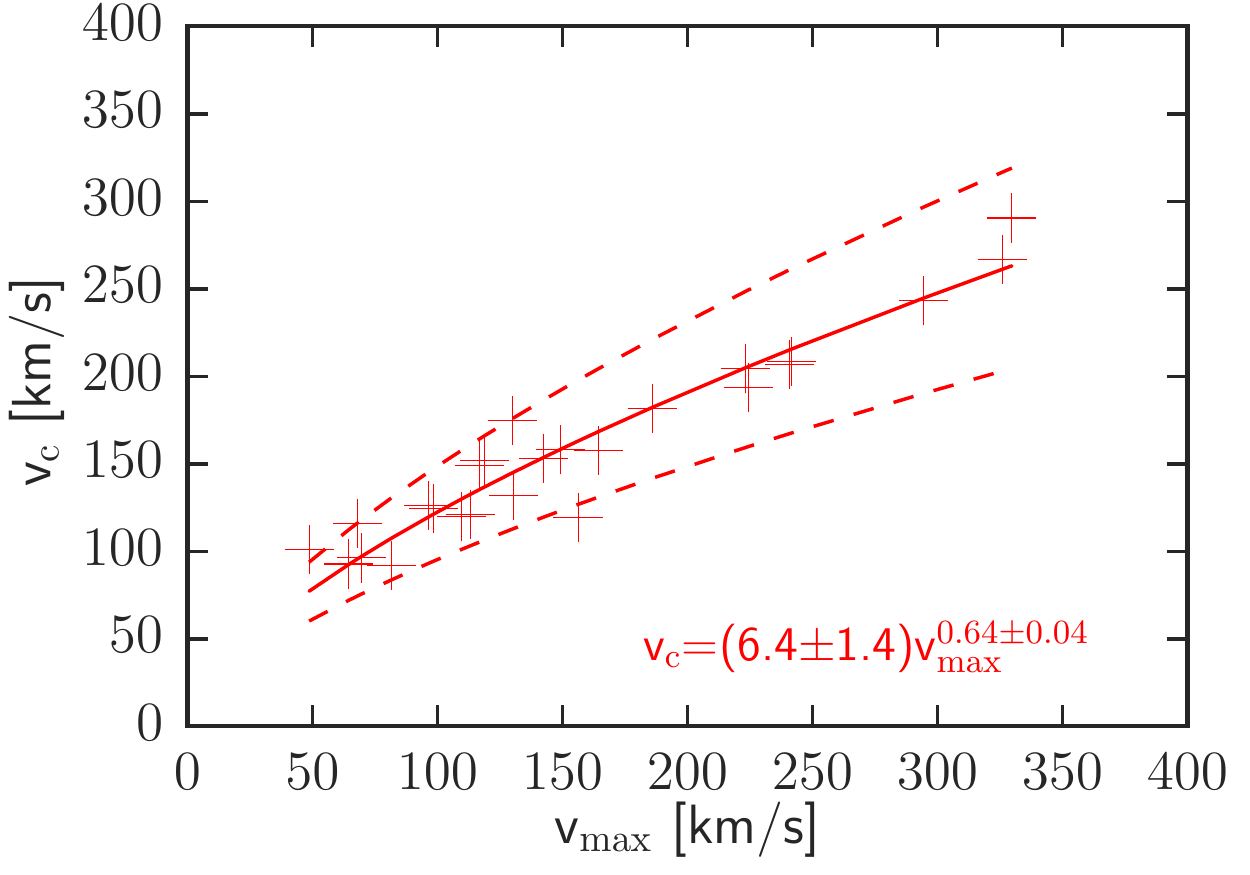}
\caption{The circular velocity $v_{\rm c}$ at $0.15r_{\rm vir}$ as a function of maximum edge-on line-of-sight velocity $v_{\rm max}$.
The top panel shows the values for the complete stellar distributions (grey) and for the various kinematic discs (blue, dark blue and light blue). 
The linear regression in the top panel has a Pearson correlation coefficient $r_{\rm P}=0.96$, 
and  has been made only with the data of the large scale single discs (blue) and thin discs (dark blue), marked by the red cross symbols.
The bottom panel shows an alternative power-law fit to the same (thin and large scale single discs) data as in the top panel.}
\label{figure_vsig_2}
\end{figure}

\subsection{Circular velocity vs rotational velocity}
\label{vc_vs_vrot}

In Figure~\ref{figure_vsig_2} the intrinsic property, circular velocity $v_{\rm c}$ is plotted as a function of the observational accessible parameter
maximum edge-on line-of-sight rotational velocity $v_{\rm max}$. 
As very well exemplified by the left panel of Figure~6 in Paper I, the thick discs always have rotational velocities smaller than their thin counterparts. 
The rotational velocity profiles along the semimajor axes for all the kinematic disc components are given in bottom part of Figure~\ref{figure_appendix_3a} in the Appendix.  

Figure~\ref{figure_vsig_2} clearly shows that the circular velocity correlates tightly with the maximum rotational velocity of the stellar (thin) disc. 
The linear regression in the top panel given by the red line is computed using only the large scale single discs and the thin discs, which are marked by the red crosses:

\begin{equation}
 \rm v_{\rm c} = 60.0 + 0.64 v_{\rm max} \rm\pm 14.2 \; [km/s],
\label{predict_vc_from_vrot}
 \end{equation}
with a Pearson correlation coefficient r$_{\rm P}$=0.96. 
The dashed red lines shows the extrapolation of this correlation at both lower and higher $v_{\rm max}$.
The bottom panel of the same figures shows the same data (large scale single discs and thin discs), now fitted with a power-law:
\begin{equation}
 \rm v_{\rm c} = (6.4 \rm\pm 1.4)v_{\rm max}^{0.64 \rm\pm 0.04} \; [km/s],
\label{predict_vc_from_vrot1}
 \end{equation}
The solid red curve gives the power-law fit, while the two dashed curves show the corresponding $\pm1\sigma$.  
 
Therefore, if the edge-on line-of-sight rotation profile for the stellar kinematic (thin) disc can be measured,
the total mass contained within r$_{\rm c}$ and, implicitly, the inner dark matter mass can be reliably estimated.
This relation, therefore, provides a new independent measure of the dark matter mass, that can be used together with 
other stellar and gas kinematic tracers to properly test the current $\rm\Lambda$CDM model \citep[e.g.][]{Bradford:2016,Maccio:2016,Brooks:2017,Teklu:2018}.

Though powerful, Equations~\ref{predict_vc_from_vrot} or \ref{predict_vc_from_vrot1} are not easy to apply to observational data because in extragalactic observations 
it is still hard to disentangle the kinematics of the thin stellar discs from those of the thick counterparts.
As discussed earlier, the particular stellar kinematic tracers used in observational studies are metal absorption lines that trace better younger stellar population than old. 
As a consequence, we expect observational values of $v_{\rm max}$ to fall somewhere in between our kinematically defined thin and thick discs, but closer to the former 
given that they tend to contain a higher fraction of young stellar populations (e.g. Paper I).

From the point of view of observed double discs galaxies, stellar thick discs show a variety of behaviours \citep[e.g.][]{Yoachim:2008b,Kasparova:2016}.   
\citet{Yoachim:2008b}, for example, studied a sample of nine almost edge-on galaxies with gas and stellar kinematic data, 
and found that their lower mass galaxies have thick discs lagging behind the thin ones. 
In one case the thick disc rotates in the opposite direction than the thin one. 
They conclude that this variety of thick disc kinematics can not likely be explained by secular evolution (thin disc heating), but are rather due to mergers. 
Thus, most of the stars of the thick discs are either formed or accreted during the fast merger epoch at high redshifts, 
a scenario supported by simulations like those of \citet{Brook:2004} or \citet{Dominguez:2015,Dominguez:2017}, as well as NIHAO suite.

We conclude that the stellar kinematic structures of the simulated galaxies are in general good agreement with both photometric and kinematic observational data.
As such, we can use these kinematically defined stellar structures to constrain the possible formation patterns of galactic components observed in the nearby Universe.

\section{Intrinsic properties}
\label{int_prop}

The eight types of stellar components are expected to be differentiated among themselves at least in: 
sizes, shapes, angular momenta, rotational support and characteristic time scales of mass assembly. 
As in Paper I, the positions and velocities of all the progenitor particles $\{i\}$ of each component $(k)$ at any given time $t$ 
are first transformed to the corresponding (time dependent) reference frame of $(k)$.
At each $t$, the simulation box is rotated so that the $z$-axis is the $z=0$ symmetry axis of the galaxy. 
Keeping the $z$-axis constant in time makes it easier to interpret the evolution in quantities such as rotation support. 
To compute all of the above mentioned quantities, and their evolutions, we use the mass of the particles at $z=0$, $m_{\rm i}(t) = m_{\rm i(*)}(z=0)$.
This is an important point, since it allows to draw conclusions on the evolution in properties such as size, shape, rotational support or angular momentum of 
the (conserved) Lagrangian mass of each $z=0$ stellar structure.   

\begin{figure}
\begin{center}
\includegraphics[width=0.48\textwidth]{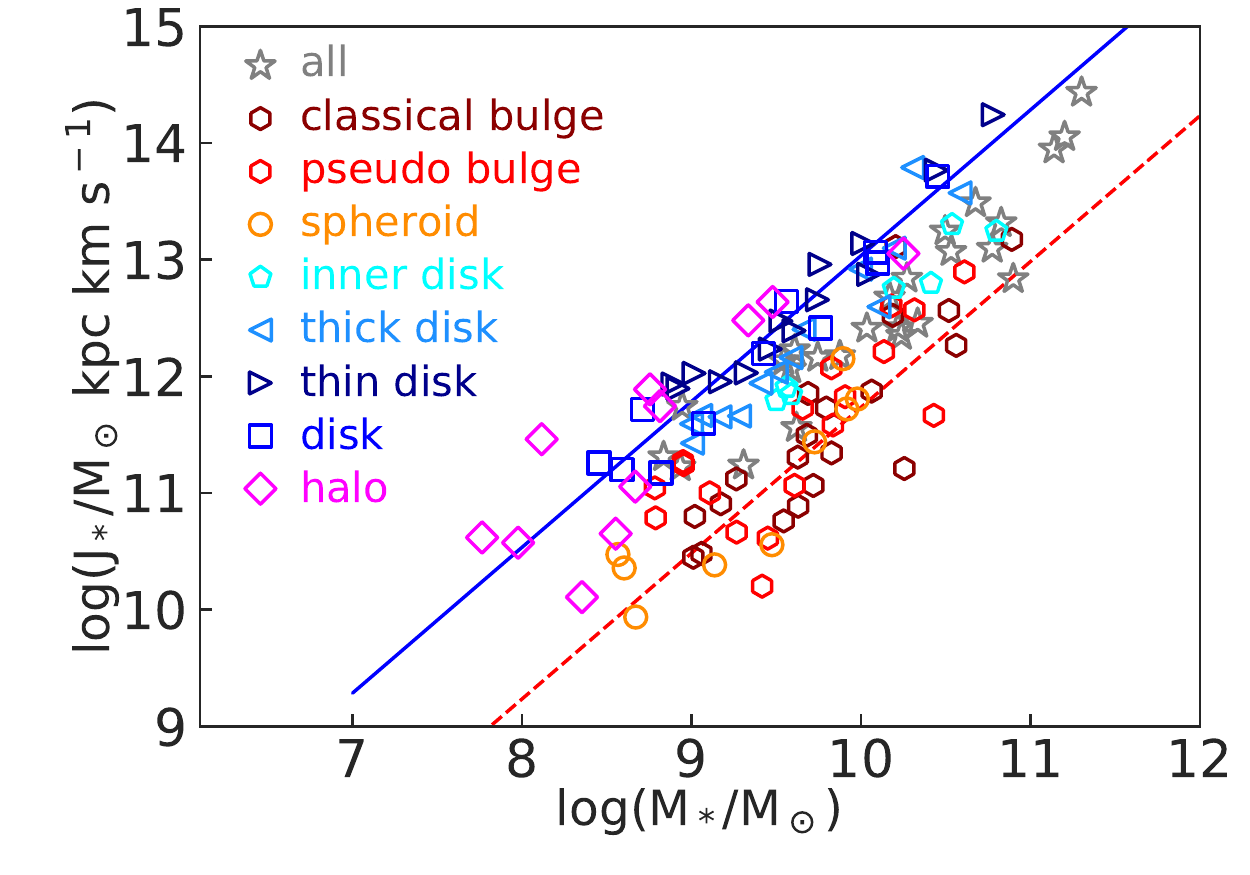}\\
\includegraphics[width=0.48\textwidth]{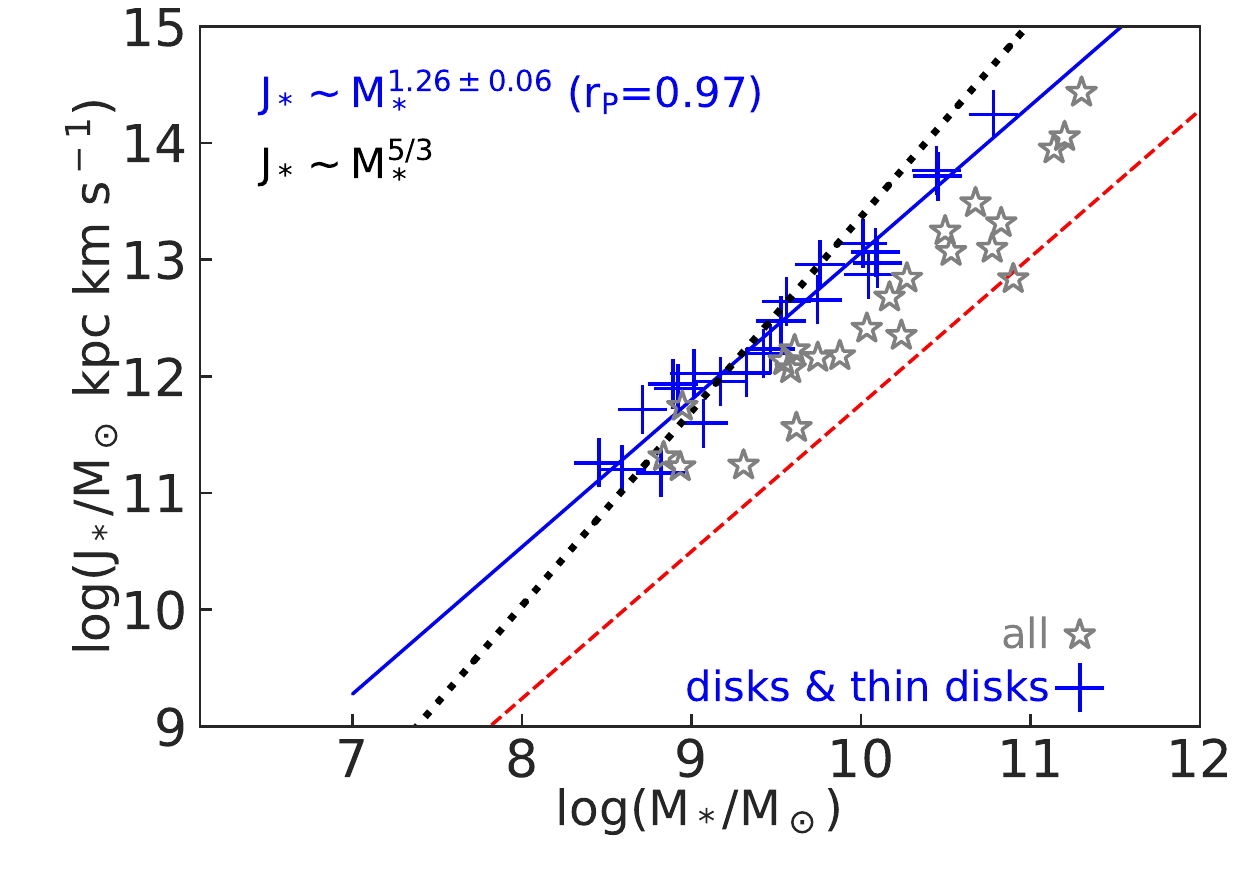}
\caption{The angular momentum $J_{\rm *}$ as a function of the stellar mass for all components of the simulated galaxy sample.
The top panel shows all the components, while the bottom one shows only the data corresponding to all the stars in the galaxy (grey stars), 
the thin discs and the single large scale discs (both with blue crosses). 
The linear regression through the data set containing only the thin discs and large scale discs (marked by blue crosses in the bottom panel) 
is given by the solid blue line. The dashed red line is the regression shifted downwards by 1.3~dex, such that is passes through the 
simulation points corresponding to the dispersion dominated components. 
The dotted black line in the bottom panel shows the tidal torque theory prediction $J\sim M^{\rm 5/3}$ \citep{Shaya:1984}.}
\label{figure_dyn}
\end{center}
\end{figure}

\begin{figure*}
\begin{center}

\includegraphics[width=0.33\textwidth]{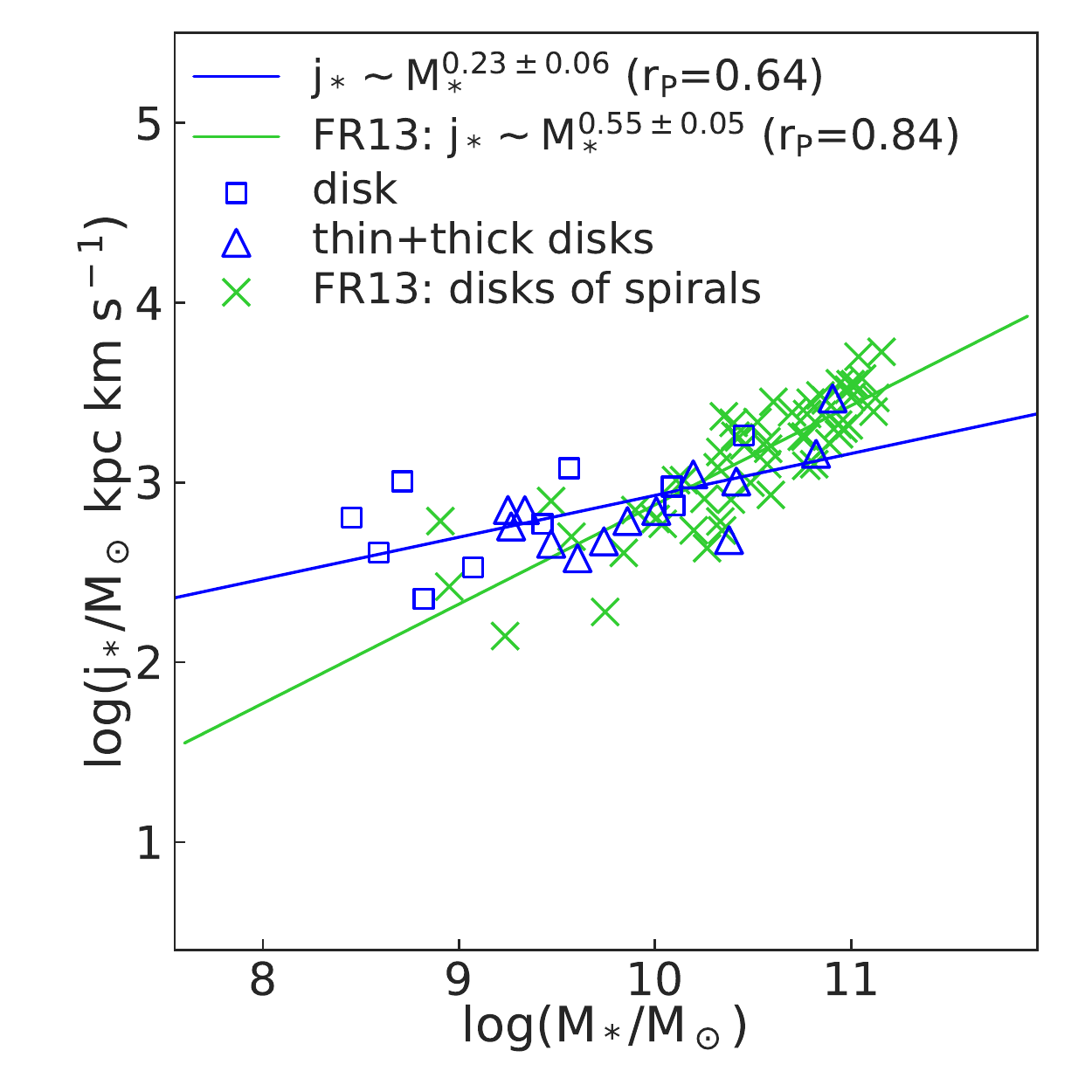}
\includegraphics[width=0.33\textwidth]{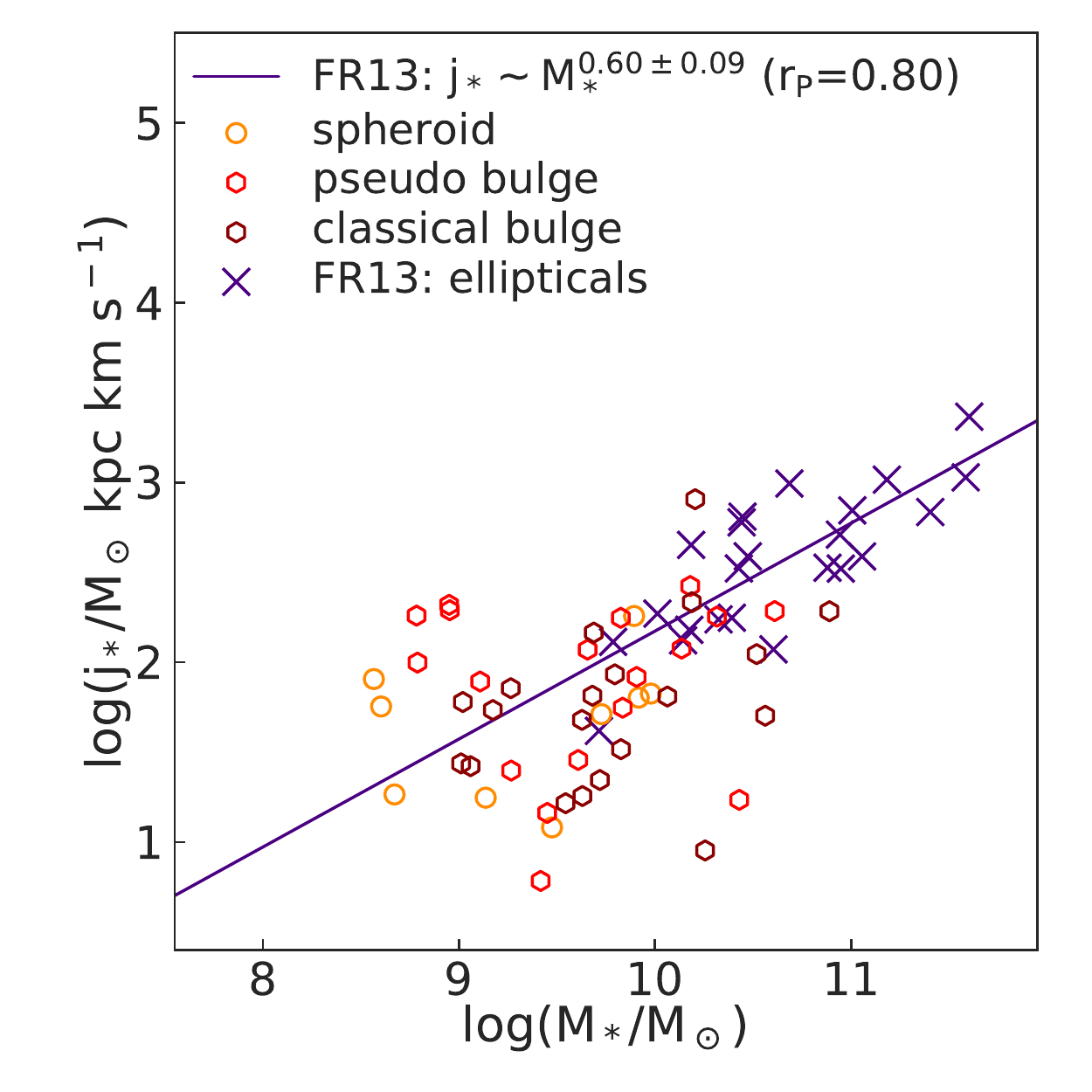}\\
\includegraphics[width=0.33\textwidth]{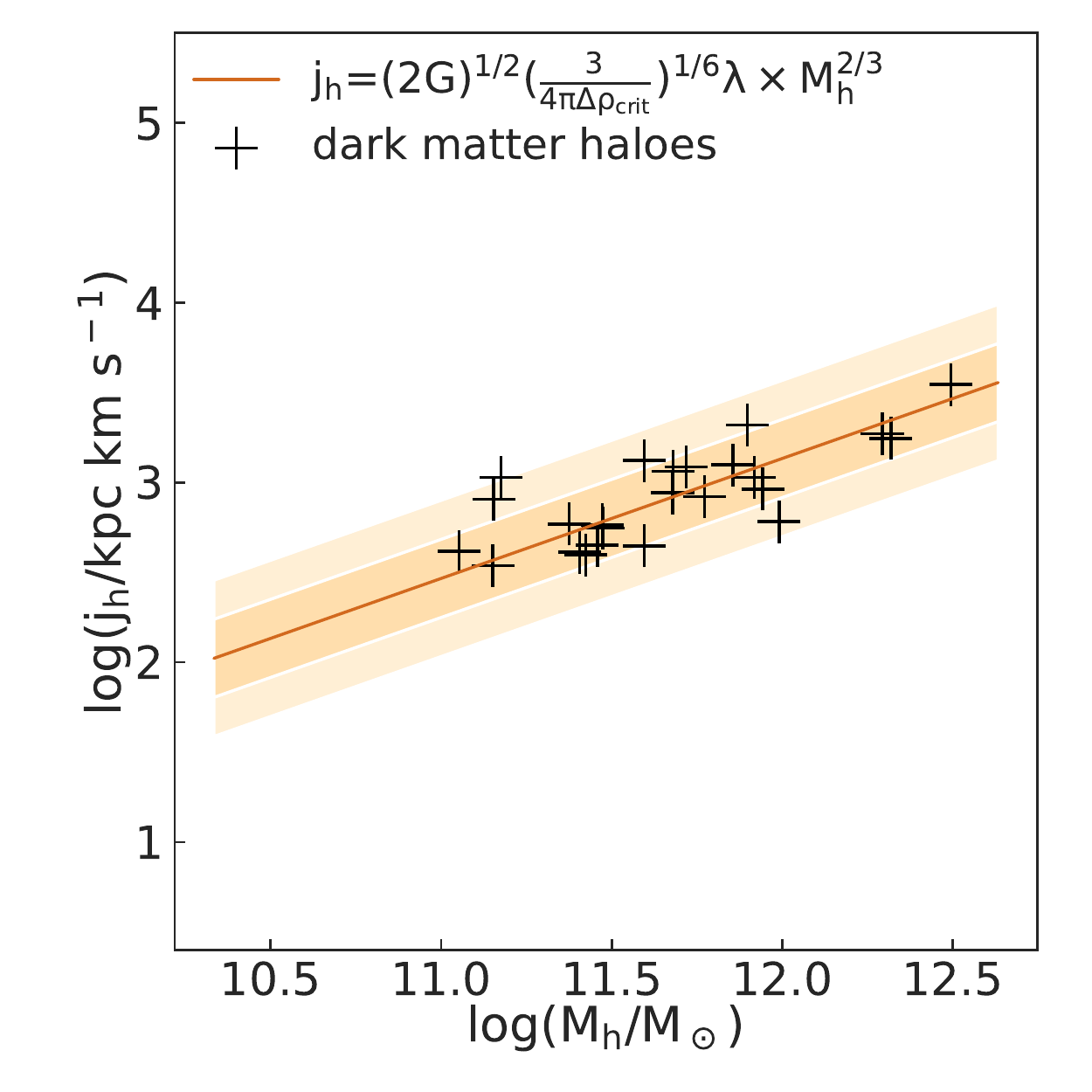}
\includegraphics[width=0.33\textwidth]{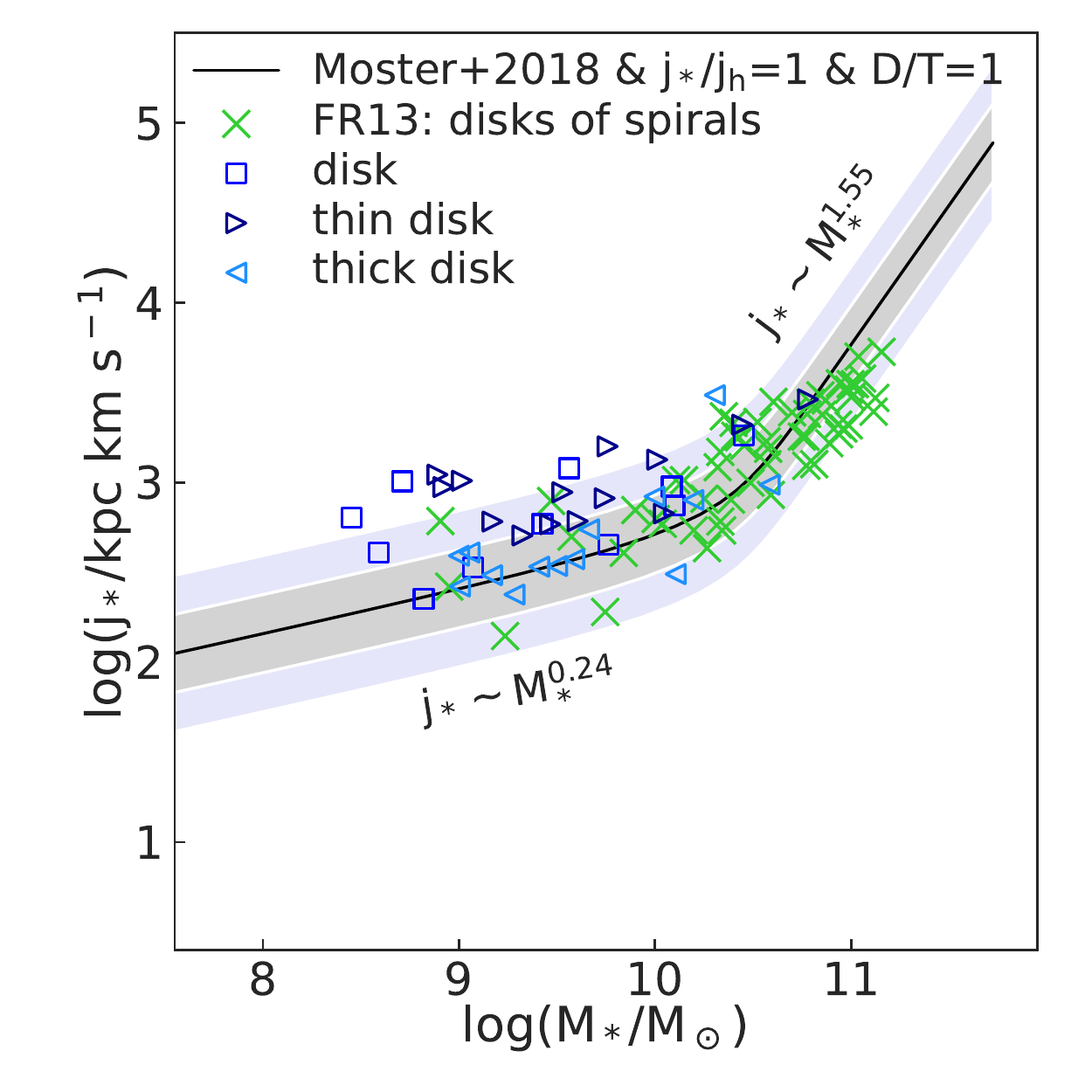}
\includegraphics[width=0.33\textwidth]{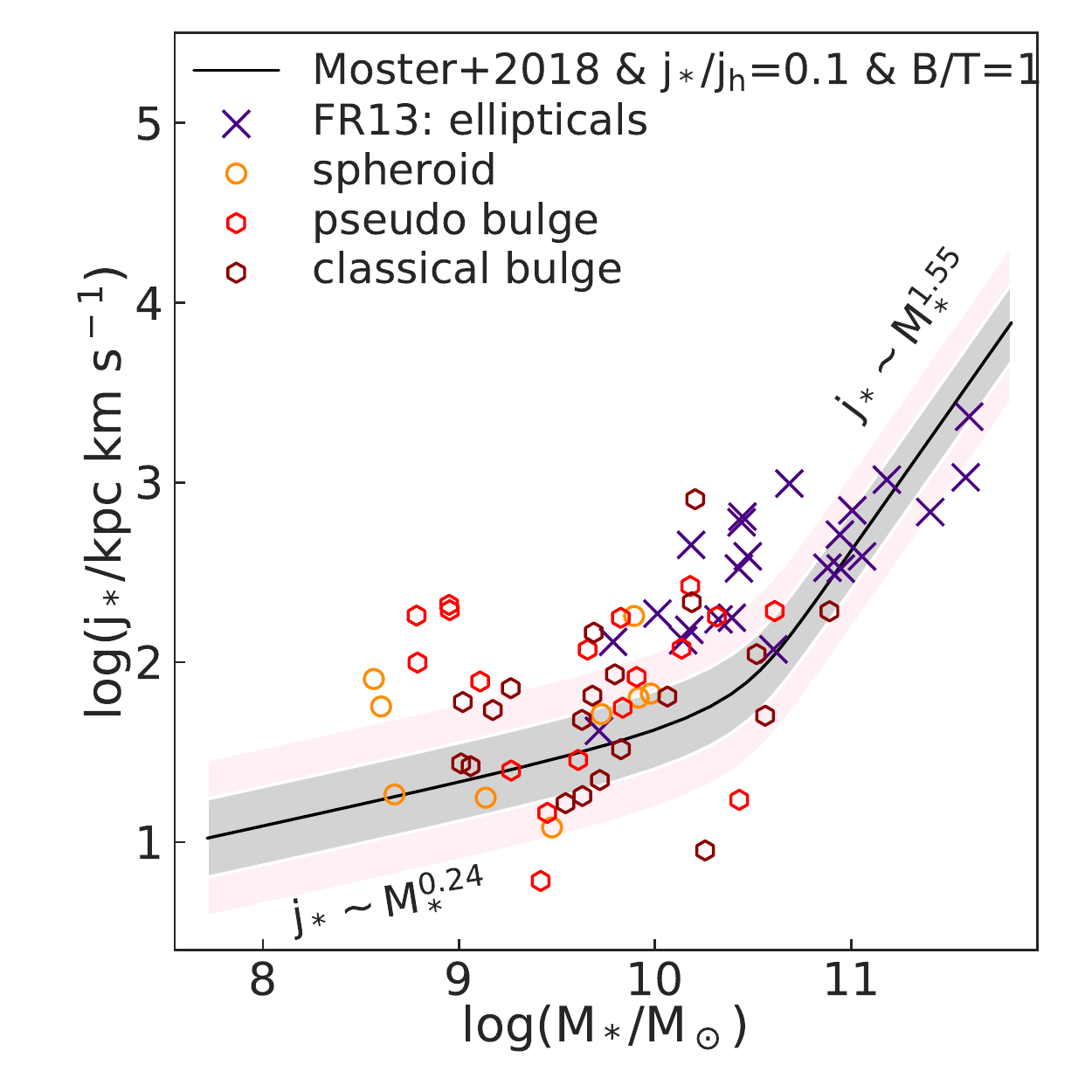}\\
\caption{The specific angular momentum $j_{\rm *}$ as a function of the stellar mass $M_{\rm *}$ 
for the rotation and velocity dispersion dominated components of the simulated galaxy sample, 
together with the observational samples of spiral discs (green symbols) and elliptical galaxies 
(indigo symbols) of FR13. The two top panels show the power-law fits through the corresponding simulated 
and observed discs and spheroids, while the bottom ones show the predictions of the analytical model described in the text, 
where the black curves correspond to $\lambda=\overline{\lambda}=0.035$ \citep{Bullock:2001}, the shaded grey areas to 
$\lambda=\overline{\lambda}\pm1\sigma$, and the shaded blue and pink areas to $\lambda=\overline{\lambda}\pm2\sigma$. 
The bottom left panel gives the specific angular momentum - mass diagram for the simulated dark matter haloes (black symbols), 
together with the $j\sim\lambda M^{\rm 2/3}$ prediction based on the definitions for the virial quantities (brown line) and its 
one and two sigma deviations based on the distribution of dark matter spin parameters $\lambda$ (orange and light orange shaded areas).}
\label{figure_dyn1}
\end{center}
\end{figure*}

\subsection{The angular momentum -- mass relation}
\label{jm_sec}

Apart from mass, one of the most important properties of galaxies is the angular momentum. 
In simulations, the specific angular momentum $j$ is computed as: 
\begin{equation}
j\equiv\frac{|\overrightarrow{J}|}{M}\equiv\frac{\Sigma_i m_i\overrightarrow{r_i}\times \overrightarrow{v_i}}{\Sigma_i m_i} ,
\label{am_definition_sim}
\end{equation}
where $m_i$, $\overrightarrow{r_i}$ and $\overrightarrow{v_i}$ are the mass, position and velocity of particle $i$, and the sum is over all the particles $i$ belonging to a given dynamical component. $\overrightarrow{J}$ and $M$ are the total angular
momentum and mass of a given component. Therefore, when using Equation~\ref{am_definition_sim}, $j$ and $M$ are correlated by definition, 
while $|\overrightarrow{J}|$ and $M$ are not. 

Galaxies are thought to aquire their rotation mainly at high redshifts, through the torques induced 
by misaligments between the large scale tidal field and the inertia tensor of the collapsing patch of the universe representing 
a galaxy's progenitor material 
\citep[tidal torque theory,][]{Hoyle:1951,Peebles:1969,Doroshkevich:1970,Fall:1980,White:1984}. The angular momentum of a dark matter 
halo can thus be computed using the spherical shell approximation in the linear regime of perturbation growth \citep[e.g.][]{Shaya:1984},
and should follow a tight correlation with the mass $J\sim M^{\rm 5/3}$, or alternatively $j\sim M^{\rm 2/3}$.
According to this theory, most of the galactic angular momentum is acquired in the linear regime of perturbation growth, and therefore 
the maximum $J$ of a collapsing patch of matter should be reached approximately at the time of maximum expansion in the 
cosmological evolution of an idealized spherical overdensity. In the non-linear stages of perturbations growth, galactic angular 
momentum can be acquired, lost and/or redistributed due to mergers \citep[e.g.][]{Barnes:1987}. The tidal torque theory assumes that 
baryons are well mixed with the dark matter, at least in the early stages of the universe, and therefore their specific angular 
momenta should be roughly equal at high $z$, $j_{\rm baryon}\simeq j_{\rm dark}$. Moreover, if one further assumes that angular 
momentum is conserved after the time of maximum expansion, as it is often done in analytical studies, the relation between sizes 
and rotation velocities for observed disc galaxies arises naturally \citep[e.g.][]{Dalcanton:1997,Mo:1998}.   

The top panel of Figure~\ref{figure_dyn} shows the positions of the eight kinematic components 
as well as the positions of the 
host galaxies in the $J_{\rm *}$-$M_{\rm *}$ plane, given that these are the two independent quantities from simulations, 
according to Equation~\ref{am_definition_sim}. Since angular momentum conservation is typically expected for discs and not 
for spheroids, which are though to originate mainly through mergers, we fitted with a power-law (blue line in both panels) 
only the large scale disc and thin disc components of our simulated galaxies (blue crosses), as it is shown in the bottom panel 
of the same figure. The result is that our 25 discs have $J_{\rm *}$ very tightly correlated with the mass, with a value for the 
Pearson coefficient of $r_{\rm P}=0.97$, and a root-mean square of 0.04~dex. 
However, the coefficient of the power-law fit:
\begin{equation}
 \rm J_{\rm *} = 3.4 \times M_{\rm *}^{\rm 1.26 \pm 0.06}
\label{predict_Jz}
 \end{equation}
is significantly lower that the $5/3$ value of \citet{Shaya:1984}.
The dotted black line in the bottom panel of Figure~\ref{figure_dyn} shows the relation $J\sim M^{\rm 5/3}$.
The dashed red lines in Figure~\ref{figure_dyn} show the correlation of Equation~\ref{predict_Jz} offsett down by 1.3~dex 
such that it passes through the data points corresponding to the dynamically hot components. 
Therefore, we find that kinematic stellar discs have $\sim20$ times larger $J_{\rm *}$ than the dynamically hot components 
at the same stellar mass. The dynamically hot components (classical/pseudo bulges and spheroids) show a large scatter around the displaced relation (dashed red line) for the `cold' discs. The thick and inner discs also seem to follow a parallel sequence to 
the dynamically coldest components with a slightly smaller normalization.

To explore why we obtain such a different slope for the $J_{\rm *}$-$M_{\rm *}$ relation, we turn to more recent 
observational studies, and in particular to the works of \citet[][hereafter RF12]{Romanowsky:2012} and 
\citet[][hereafter FR13]{Fall:2013} who revised the original relation $j_{\rm *}$-$M_{\rm *}$ of \citet{Fall:1983}.
Our interest in these particular studies resides in the fact that RF12 and FR13 decompose photometrically their galaxies, 
and use the bulge-to-total ($B/T$) ratio to split a galaxy's stellar specific angular momentum and mass into a disc and a 
bulge component. The difference between RF12 and FR13 is that the former adopted a mass-to-light $M/L_{\rm K}$ ratio of one for 
both the disc and the bulge components, while the later used different $M/L$ for the two. 
In this manner, these authors find that discs and spheroids populate distinct regions of the $j_{\rm *}-M_{\rm *}$, 
both groups following power-laws, $j_{\rm *}\sim M_{\rm *}^{\rm\alpha}$, with similar exponent $\alpha\simeq0.6\pm0.1$ (FR13). 
Observationally, at the same stellar mass, discs have $\sim6$ times larger specific angular momenta than spheroids.

It is important to keep in mind that in observations, specific assumptions 
for the mass and velocity profiles are made, such that angular momentum can be estimated by: 
\begin{equation}
j_* = \frac{J_*}{M_*} = A~v_s(R_{s1})~R_{s2},  
\label{am_definition_obs}
\end{equation}
(e.g. RF12) where $A$ is a factor depending on the mass profile and galaxy inclination, 
$v_s$ the rotation velocity at a specific radius $R_{\rm s1}$, and $R_{\rm s2}$ is the characteristic radius of the mass profile. 
RF12 assume that spiral galaxies are well described by infinitely thin discs and exponential surface mass density profiles 
that have flat circular rotation profiles, and therefore $A=2$, $v_s$ is the inclination corrected, asymptotic rotational 
velocity of the ionized gas (thought to estimate well the stellar disc rotation), and $R_{s2}$ is the exponential disc 
scalelength from the K-band photometry. For ellipticals and spheroids of late type galaxies, RF12 assume S\'{ersic} mass density profiles, and get for the $n=4$ case that $A=3C$ (where $C$ is a deprojection factor depending on the galaxy inclination), 
$R_{s2}$ is the effective radius of the S\'{ersic} profile, and $R_{s1}$ is twice the effective radius. For spheroids, 
RF12 use velocites from stellar spectroscopy, planetary nebulae and globular clusters. In Equation~\ref{am_definition_obs} 
the independent quantities are $j_{\rm *}$ and $M_{\rm *}$, and not $J_{\rm *}$ and $M_{\rm *}$ as it is the case for 
Equation~\ref{am_definition_sim}.

Therefore, in most simulations studies of angular momentum in galaxies, the prefered diagram 
is $j_{\rm *}-M_{\rm *}$ \citep[e.g.][]{Teklu:2015,Genel:2015,Pedrosa:2015} to facilitate the comparison with observations. 
\citet{Teklu:2015} have used the Magneticum Pathfinder simulations to study the evolution of
angular momentum in galaxies, and found that the specific stellar angular momentum as a function of mass populates different 
parts of the $j_{\rm *}-M_{\rm *}$ in accordance with the mean stellar circularity $\langle j_{\rm z}/j_{\rm c}\rangle$, 
Their large circularity galaxies ($\langle j_{\rm z}/j_{\rm c}\rangle\sim1$) clusters close to the analytical model of 
RF12 $j_{\rm *}\sim M_{\rm *}^{2/3}$ for discs of spirals, while small circularities ($\langle j_{\rm z}/j_{\rm c}\rangle\sim0$)
are in the region of the elliptical galaxies of RF12. The analytical model of RF12 and FR13 places galaxies along parallel 
$j_{\rm *}\sim M_{\rm *}^{2/3}$ with a normalization depending on the bulge-to-total ratio. 
\citet{Genel:2015} use the Illustris simulation at $z=0$ to populate the $j_{\rm *}-M_{\rm *}$ diagram, 
and study the galaxies by splitting them in two groups according to concentration, specific star formation rate or flatness. 
In this manner, they find two parallel sequences in log($j_{\rm *}$)-log($M_{\rm *}$), where flat, low concentration or 
high specific star formation rate galaxies are close to the $j_{\rm *}\sim M_{\rm *}^{2/3}$ of observed spiral discs, while 
round, concentrated or quiescent objects roughly follow the observational sequence of elliptical galaxies. 
\citet{Pedrosa:2015} go a step further and separate their simulated galaxies into a spheroid and a disc component using a 
dynamical criteria, namely that all particles with $j_{\rm z}/j_{\rm c}>0.5$ belong to the disc. Fitting power-law relations 
for the discs and spheroids separately, they find slopes $\alpha\sim0.9$ for the former and $\alpha\sim1.1$ for the latter, 
and little evolution with redshift for both types of components.

Figure~\ref{figure_dyn1} shows our simulated galaxies together with the observational 
data for discs of spirals (green symbols) and elliptical galaxies (indigo symbols) of FR13 in the $j_{\rm *}-M_{\rm *}$ plane.
To ease the comparison with observations and previous simulations studies, we combine 
the dynamical thin and thick discs together, such that any of the 25 simulated galaxies have only one large scale stellar disc (empty blue 
squares and triangles). The top left panel shows the simulated dynamical discs together with the disc of spirals of FR13, and their
corresponding linear regressions. The observational and simulation disc samples overlap to a great extent, but the linear 
regression through the observations results in $\alpha=0.55\pm0.05$ and a high correlation coefficient $r_{\rm P}=0.84$, while the 
simulated discs have $j_{\rm *}$ less correlated with $M_{\rm *}$ ($r_{\rm P}=0.64$) and result in a lower slope $\alpha=0.23\pm0.06$.
We make no attempt to fit the dynamically hot components in the top right panel, but we can appreciate that they overlap with the 
elliptical sample of FR13, which is well correlated ($r_{\rm P}=0.80$) along the power-law $\alpha=0.60\pm0.09$.  

To understand where the $\alpha=2/3$ relation comes from, we show in the bottom left panel of 
Figure~\ref{figure_dyn1} the $j-M$ relation for the dark matter haloes of our 25 simulated galaxies. The brown power-law in this 
panel \emph{is not a fit} to the data, but the relation that arises naturally taking into account the definitions for halo virial 
mass and velocity, $M_{\rm h}\equiv\Delta\rho_{\rm crit}\frac{4\pi}{3}r_{\rm vir}$ and 
$v_{\rm vir}\equiv\sqrt{G~M_{\rm h}/r_{\rm vir}}$, and spin parameter $\lambda\equiv\frac{j_{\rm h}}{\sqrt{2}r_{\rm vir}v_{\rm vir}}$
\citep{Bullock:2001}, where $\rho_{\rm crit}$ is the critical density of the universe, and $\Delta$ is the overdensity factor, 
that can be either a fixed number (e.g. $\Delta$=200 as it is used in the NIHAO papers) or a function of the cosmology and redshift 
\citep[e.g.][]{Bryan:1998}. \citet{Bullock:2001} find that spin parameters of haloes in N-body simulations follow a log-normal 
distribution with a mean $\overline{\lambda}=0.035$ and a dispersion in ln($\lambda$) of 0.50. Given that \citet{Bullock:2001} use the definition of 
overdensity of \citet{Bryan:1998}, we converted the halo masses back to $M_{\rm h}=M_{\rm 200}$. From the bottom left panel of 
Figure~\ref{figure_dyn1}, it is clear that \emph{the dark matter haloes of our simulated galaxies follow a power-law with $\alpha=2/3$}:
\begin{equation}
 j_h = \sqrt{2G}(\frac{3}{4\pi\Delta\rho_{crit}})^{1/6}\lambda M_h^{2/3}
\end{equation}

To transform the above Equation into a relation for stars we need to convert $j_{\rm h}$ into $j_{\rm *}$, 
and $M_{\rm h}$ into $M_{\rm *}$. In analytical studies of disc galaxies a wide used assumption is $j_{\rm *}\simeq j_{\rm h}$ to 
first approximation \citep[e.g][]{Mo:1998}. For elliptical galaxies and spheroids, various studies (e.g. RF12, FR13, 
\citealt{Genel:2015}) concluded that $j_{\rm *}$ should be a few times smaller than $j_{\rm h}$, so we will assume that for our 
dispersion dominated components a good choice is $j_{\rm *}\simeq 0.1j_{\rm h}$. Finally, we use the parametrizations of the 
\textsc{emerge} model \citep{Moster:2018} to convert the halo mass into a stellar mass. All recent models for the dependence of 
stellar mass on halo mass \citep[e.g.][]{Moster:2013,Kravtsov:2014,Behroozi:2013,Behroozi:2018}, including \textsc{emerge}, 
result in a double power law relation between the integrated star formation efficiency $M_{\rm *}/f_{\rm bar}M_{\rm h}$ and the dark 
matter halo mass $M_{\rm h}$:
\begin{equation}
 \frac{M_*}{f_{bar}M_h} = 2 \epsilon_N \left[ \left(\frac{M}{M_1}\right)^{-\beta} + \left(\frac{M}{M_1}\right)^{\gamma} \right]^{-1},
 \label{abund_match}
\end{equation}
where $f_{\rm bar}$ is the cosmological baryon fraction, and $\beta$ and $\gamma$ give the slope at small and large halo masses. All parameters
of Equation~\ref{abund_match}, namely $\beta$, $\gamma$, and the normalizations $\epsilon_{\rm N}$ and $M_{\rm 1}$ are functions of redshift. 

We use Equation~\ref{abund_match} with the lowest redshift parameters available in \citet{Moster:2018} for 
star forming and quiescent galaxies to construct an analytic prediction for the dynamical discs and dispersion supported components, respectively. 
These analytic predictions are shown by the black curves in the bottom centre and right panels of Figure~\ref{figure_dyn1}. The one and two sigma 
shaded regions take into account only the dispersion in the spin parameter $\lambda$ \citep{Bullock:2001}. 
As expected from the conversion $M_{\rm h}$ to $M_{\rm *}$, the $j_{\rm *}-M_{\rm *}$ also follows a double power-law, 
with a slope at small masses of $\alpha\simeq2/3(1+\beta)=0.24$ in very good agreement with our 
value $\alpha=0.23\pm0.06$, and a significantly steeper slope at high masses $\alpha\simeq2/3(1-\gamma)=1.55$. The 
large offset between the black curves for discs and spheroids comes from the difference between $j_{\rm *}\simeq j_{\rm h}$ 
and $j_{\rm *}\simeq 0.1j_{\rm h}$, and one can see that both the simulations and observations are scattered around these relations.
In this framework, the main reason why the observational data of FR12 and RF13 are well described by a single power law is the relatively 
small range of stellar masses probed, corresponding to a small region around the halo mass where the integrated star formation efficiency peaks:
$M_{\rm max}=M_{\rm 1}(\beta/\gamma)^{1/(\beta+\gamma)}\simeq\rm 10^{\rm 12}M_{\rm \odot}$ for star forming galaxies. 
On the other hand, our simulated galaxies populate the halo mass range roughly up to $M_{\rm max}$, and as such only probe the left
hand side of the $M_{\rm h}$ -- $M_{\rm *}/f_{\rm bar}M_{\rm h}$ relation.

Last, but not least, the analytical predictions based on the \textsc{emerge} model do not take into account: 
i) the $B/T$ ratio dependence on mass, needed to pass from total stellar masses to disc (spheroid) masses \citep[e.g.][]{Dutton:2011},
ii) the scatter in stellar at fixed halo mass in Equation~\ref{abund_match}, and the true distributions of $j_{\rm *}/j_{\rm h}$. 
All these factors can modify the analytical predictions, although they would not erase the double power-law nature of the model. 
In this perspective, we conclude that the apparent disagreement between our data and the observations in terms of slope $\alpha$, and 
in a broader sense among the various results of simulations already published, boils down to sample completeness (e.g. mass range)
and the assumptions used to estimate $j_{\rm *}$ from observations.

\begin{figure}
\begin{center}
\includegraphics[width=0.48\textwidth]{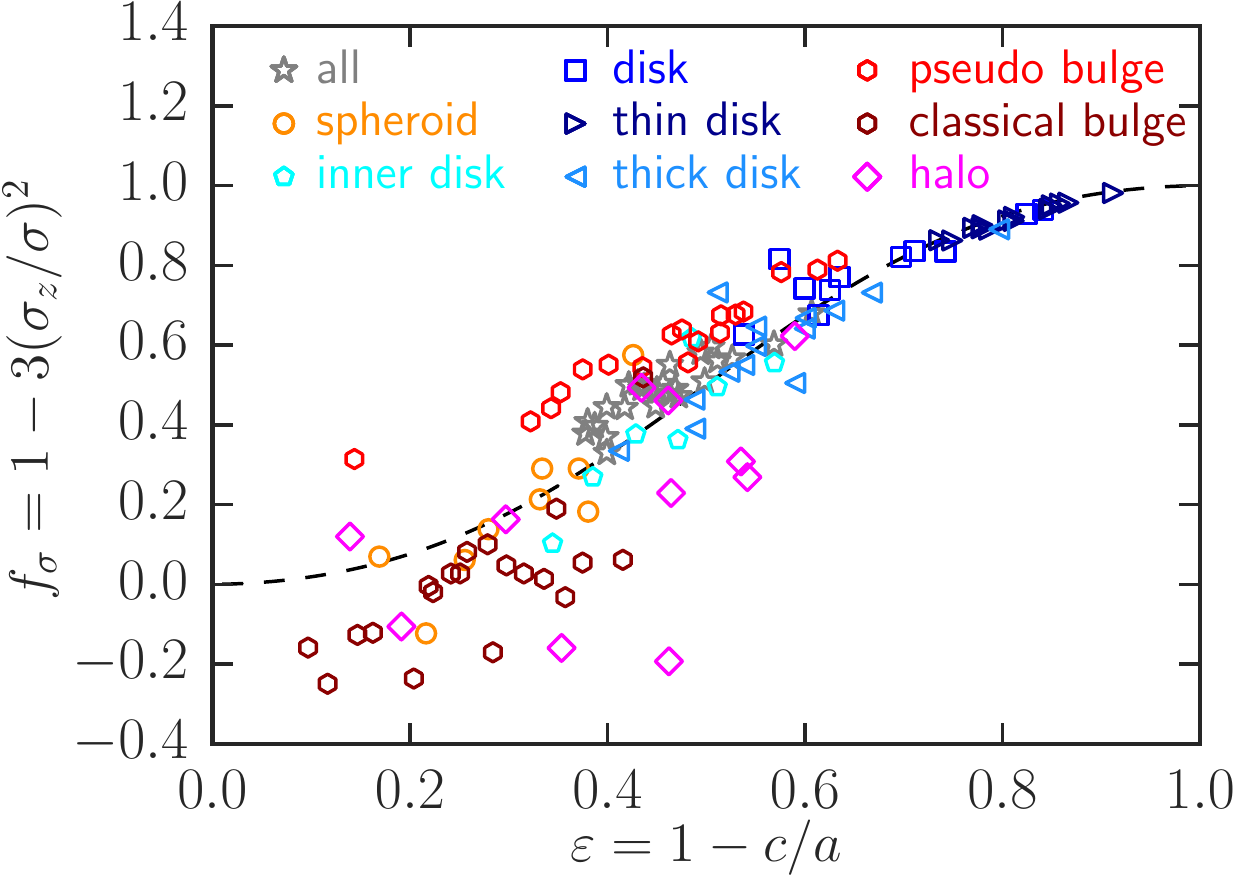}
\caption{The velocity dispersion fractions, $f_{\rm\sigma}=1-3(\sigma_{\rm z}/\sigma)^{\rm 2}$, as a function of intrinsic ellipticities, $\varepsilon$, 
for all the kinematic components in the galaxy sample.
The dashed black curve represents the analytical function $f_{\rm\sigma}(\varepsilon;\beta)=[1+(\varepsilon/(1-\varepsilon))^{\rm -\beta}]^{\rm -1}$ with $\beta=1.8$.}
\label{figure_rotsupp}
\end{center}
\end{figure}

\begin{figure*}
\includegraphics[width=0.48\textwidth]{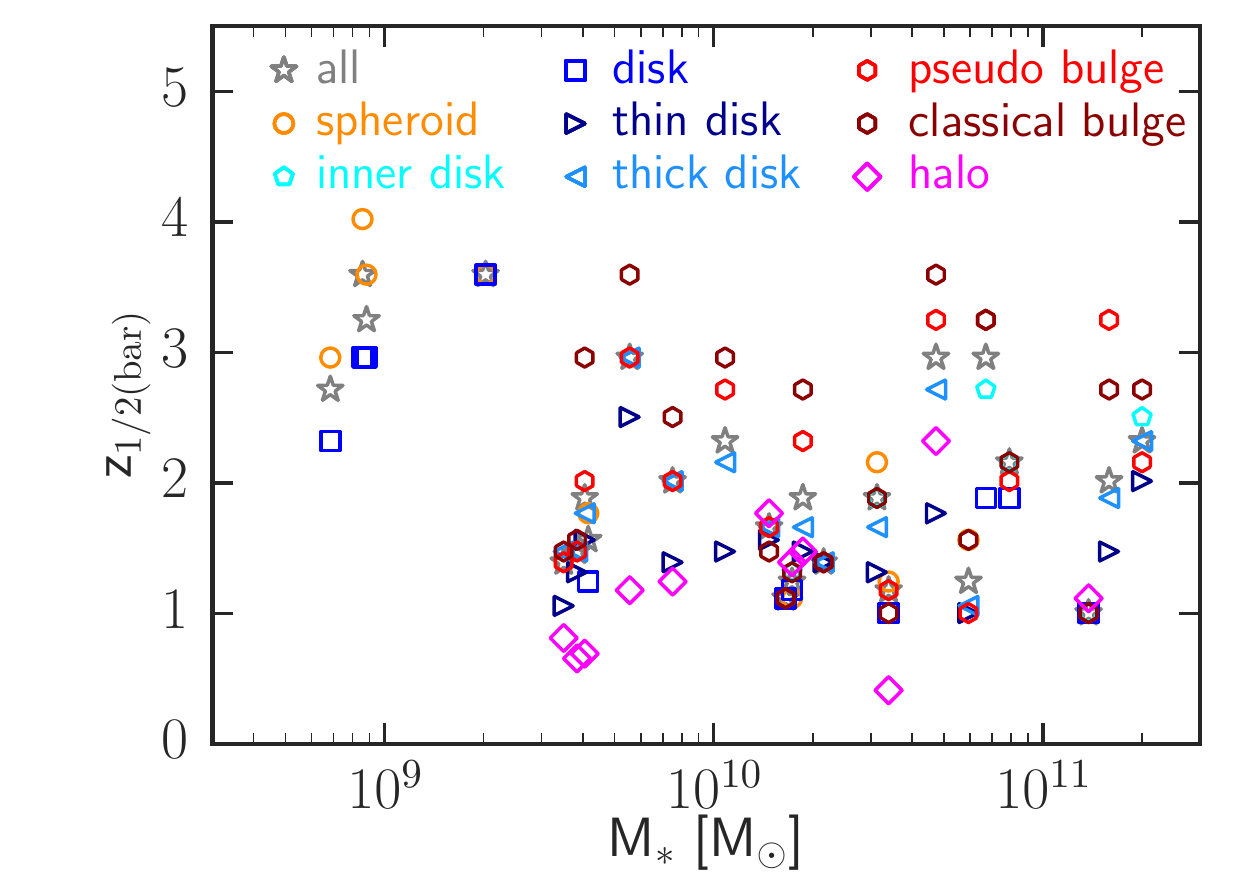}
\includegraphics[width=0.48\textwidth]{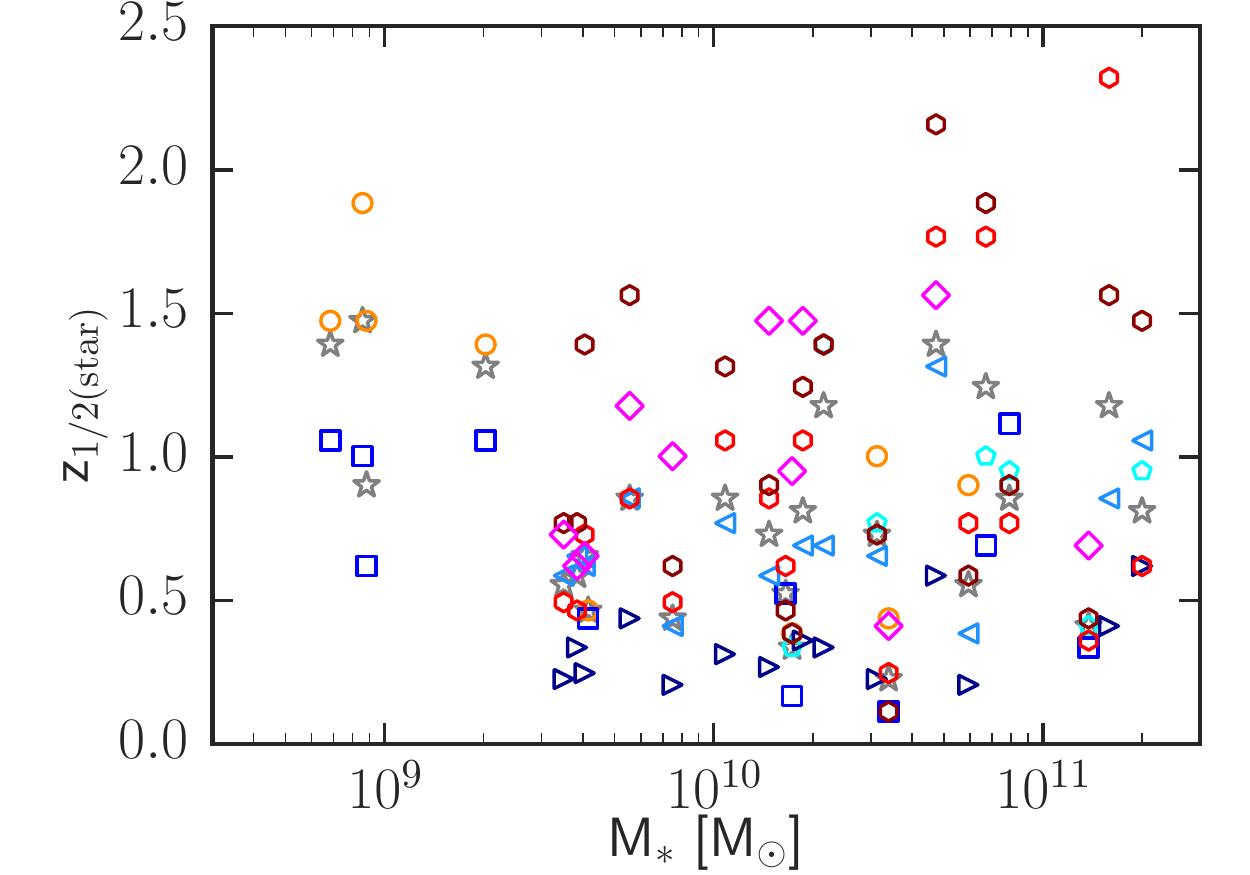}\\
\caption{\textbf{Left:} The redshifts at which half of the \textit{progenitor baryonic mass} of $z=0$ stellar kinematic structures are inside the main progenitor halo as a function 
of total galaxy stellar mass at $z=0$. 
\textbf{Right:} The redshifts at which half of the \textit{progenitor stellar mass} of $z=0$ stellar kinematic structures are inside the main progenitor halo as a function 
of total galaxy stellar mass.}
\label{figure_t}
\end{figure*}

\subsection{Rotational support vs shape}
\label{ell_rotsup}

One often used diagnostic of rotational support is the diagram which relates the isophotal shape $\varepsilon(R_{\rm e})$ 
to the rotation-to-dispersion ratio  $v/\sigma(R_{\rm e})$ \citep{Davies:1983x}, 
or its more modern version $\varepsilon(R_{\rm e})$ - $\lambda(R_{\rm e})$ \citep{Emsellem:2011}.
Both these diagrams use galaxy properties measured in the very inner regions, $R<R_{\rm e}$, and as such do not provide an estimation of a whole galaxy's dynamical state.
It is expected that bulges and discs occupy distinct regions in any plane equivalent to the spin-shape one. 

Since we are interested not only in the spins and shapes of the various stellar kinematic components at $z=0$, but also in the evolution of their Lagrangian masses, 
we use the ellipticity $\varepsilon$ of \citet{GonzalezGarcia:2005} defined by:
\begin{equation}
 \varepsilon^{(k)}=1-c^{(k)}/a^{(k)},
\end{equation}
where $c^{\rm(k)}$ and $a^{\rm(k)}$ are respectively the smallest and largest semiaxes derived from the eigenvalues of the inertia tensor of structure $(k)$:
\begin{equation}
I_{jl}^{(k)} = \sum_{i\in(k)} m_i(\delta_{jl}r_i^2-x_jx_l),
\end{equation}
with $j$ and $l$ looping over the Cartesian coordinates. 

To estimate the rotational support we used the velocity dispersion fraction $f_{\rm\sigma}$:
\begin{equation}
f_{\sigma}^{(k)}=1-3(\sigma_z^{(k)}/\sigma^{(k)})^2, 
\end{equation}
where $\sigma_{\rm z}^{\rm(k)}$ is the vertical velocity dispersion of structure $(k)$, and $\sigma^{\rm(k)}$ is the corresponding modulus of the 3D velocity dispersion. 
The advantage of using $\varepsilon$ and $f_{\rm\sigma}$ is that they can be self consistently computed even at high redshifts 
when the baryonic mass progenitor of a given $z=0$ kinematic component has still a filamentary spatial distribution (see Paper I). 
With this definition for rotational support, razor thin discs are characterized by $f_{\rm\sigma}=1$, and systems with only random motions have $f_{\rm\sigma}=0$.

Figure~\ref{figure_rotsupp} shows all the kinematic components (coloured symbols) as well as the whole galaxies (gray stars) in the  $\varepsilon$-$f_{\rm\sigma}$ plane.
All data points are at redshift $z=0$. 
The interesting aspect of this figure is that the velocity dispersion fraction $f_{\rm\sigma}$ of the various kinematic disc types appear to be very tightly correlated with the shape parameter $\varepsilon$.
The dashed black line represents the function 
$f_{\rm\sigma}(\varepsilon;\beta)=[1+(\varepsilon/(1-\varepsilon))^{\rm -\beta}]^{\rm -1}$ with $\beta=1.8$, 
which was plotted as a reference and was not derived by fitting the data points. 
This test function seems to depict very well the correlation $f_{\rm\sigma}$-$\varepsilon$ for the kinematic disc data, but not so well for the dispersion dominated components. 
Moreover, the various disc types are placed along the $f_{\rm\sigma}(\varepsilon;\beta)$ in a predictable sequence, with thin discs occupying the upper right end, followed by discs, thick discs
and inner discs as $\varepsilon$ and $f_{\rm\sigma}$ decrease, with a variable degree of overlap between neighboring types. 
The thin discs in particular are clustered in a small region with $\varepsilon\in[0.7,0.9]$ and $f_{\rm\sigma}\in[0.8,1.0]$. 
To compute the rotational support $f_{\rm\sigma}$ in this figure, the galaxies have been oriented with the $z$-axis in the direction of the total angular momentum.
Therefore, $z$ is expected to coincide with the symmetry axis of the discs, but not necessarily with that, if it exists, of the dispersion dominated structures. 
Recomputing the symmetry axis for each component separately has only a small effect on the $f_{\rm\sigma}$ of a few of the classical bulges and stellar haloes, 
but does not change their spread around $f_{\rm\sigma}(\varepsilon;\beta)$.

\subsection{Timescales of mass assembly}

The assembly history of the various structures is quantified using the same definitions as in Paper I.
For any given stellar component, we trace back in time the positions of all its progenitor particles, and compute what fraction of its (invariant redshift $z=0$) 
mass $M_{\rm *k}$ is inside the virial radius $r_{\rm vir}(t)$ of the host dark matter halo along the main branch of the merger tree, at each time step $t$. 
Therefore:

\begin{equation}
  \begin{aligned}
 \frac{M_{\rm bar}(<r_{\rm vir}(t))}{M_{\rm *k}}  &=  \frac{\sum_{i\in (k)}m_i(r_i<r_{\rm vir}(t))}{ M_{\rm *k} }\\
 \frac{M_{\rm star}(<r_{\rm vir}(t))}{M_{\rm *k}} &=  \frac{\sum_{i\in(k)}m_i(r_i<r_{\rm vir}(t)|i=*)}{ M_{*k} }\\
  \end{aligned}
  \label{z50eq}
\end{equation}

Figure~\ref{figure_t} gives the redshifts at which half of the \textit{progenitor baryonic mass} (left) and half of the \textit{progenitor stellar mass} (right) of each kinematic structure 
is within the main progenitor dark matter halo as functions of the total galaxy stellar mass at redshift zero $M_{\rm *}$.
In other words, the redshifts at which the fraction in Equations~\ref{z50eq} are equal to $0.5$. 
Using the total stellar mass instead of the mass of each component facilitates the comparison among the various timescales within one galaxy. 

The first impression from the two panels of this figure is that there is no obvious correlation between the half mass redshifts and the total stellar mass. 
However, it is clear that thin discs have lower half mass baryonic and stellar redshifts than thick discs, 
and that all disc types have lower half mass redshifts than the dispersion dominated kinematic components. 
This is in agreement with the fact that extragalactic thin discs have been found to be younger than their thick counterparts \citep{Yoachim:2008a}. 
One visible tendency it that the more massive a galaxy is, the earlier the progenitor baryons of the thin disc were accreted and the later it formed its stars. 

Regarding the two bulge types, a very fuzzy picture emerges, with some pseudo bulges being older structures than their classical bulge counterparts.
Older stellar ages of pseudo bulges as compared to those of their corresponding companion classical bulges are at odds with the secular evolution formation scenario of \citet{Kormendy:2004}.
However, we note that it is more difficult for {\tt gsf} to disentangle the dispersion dominated stellar components in the inner galactic regions
than doing the same for the rotation dominated structures.

The other interesting finding is that stellar haloes turn out to be late assembled structures throughout all this NIHAO galaxy subsample.
These stellar haloes have among the lowest values for the baryonic half mass redshift, when compared to all other types of kinematic structures.

Overall, Figure~\ref{figure_t} proves that even though the various kinematic structures have distinct spatial and dynamical properties, 
they do not show such clearly distinct temporal properties in terms of assembly histories and stellar ages.

\section{Formation of galactic structures}
\label{glabal_evolution}

After showing that the redshift $z=0$ properties of the kinematic groups are indeed consistent with the observed stellar structures of galaxies, we analyze their evolutionary patterns. 
We use the stellar mass function of \citet{Moustakas:2013} for $z\sim0.01$, $\Phi(M_{\rm *})$, to compute the average evolutionary pattern of all these properties,
separately for each kinematic stellar structure.
For any property $X$, its mean and variance at any time $t$, are calculated as:

\begin{align}
 \langle X(t)\rangle &=  \frac{\Sigma_i w_i X_i(t)}{\Sigma_i w_i}\nonumber\\
 \Delta^2_X(t)         &=  \frac{\Sigma_i w_i ( X_i(t)- \langle X(t)\rangle )^2}{\Sigma_i w_i}
 \label{eq_avdisp}
\end{align}
where the weight is given by the value of $\Phi(M_{\rm *})$ at the total $z=0$ stellar mass, $M_{\rm *}$, of the galaxy $i$ containing the specific kinematic component, 
$w_{\rm i}=\Phi(M_{\rm *}^{\rm (i)})$. 
This weighting procedure is used to take into account the fact that the masses of this simulated sample do not populate uniformly the observed mass function of observed galaxies.  

\begin{figure}
\includegraphics[width=0.48\textwidth]{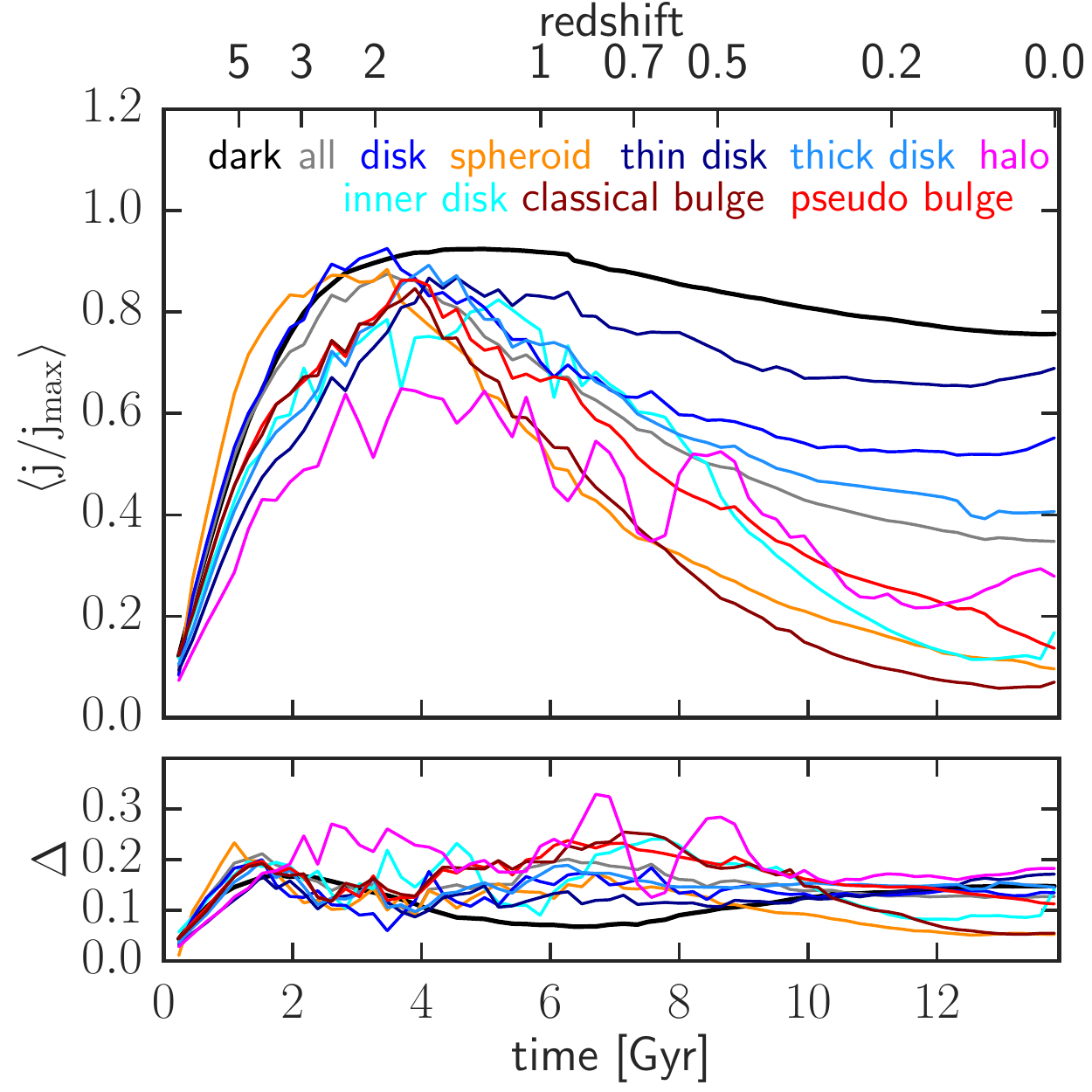}
\caption{The mass function weighted global evolutions of the normalized spins for the Lagrangian masses comprising
all stars in the galaxies at $z=0$ (gray curves), the stars in the various kinematic components at $z=0$ (colored curves), and the dark matter haloes at $z=0$ (black curves). 
The bottom small panel give the corresponding weighted standard deviations, $\Delta$.}
\label{figure9_spin}
\end{figure}

\begin{figure}
\begin{center}
\includegraphics[width=0.47\textwidth]{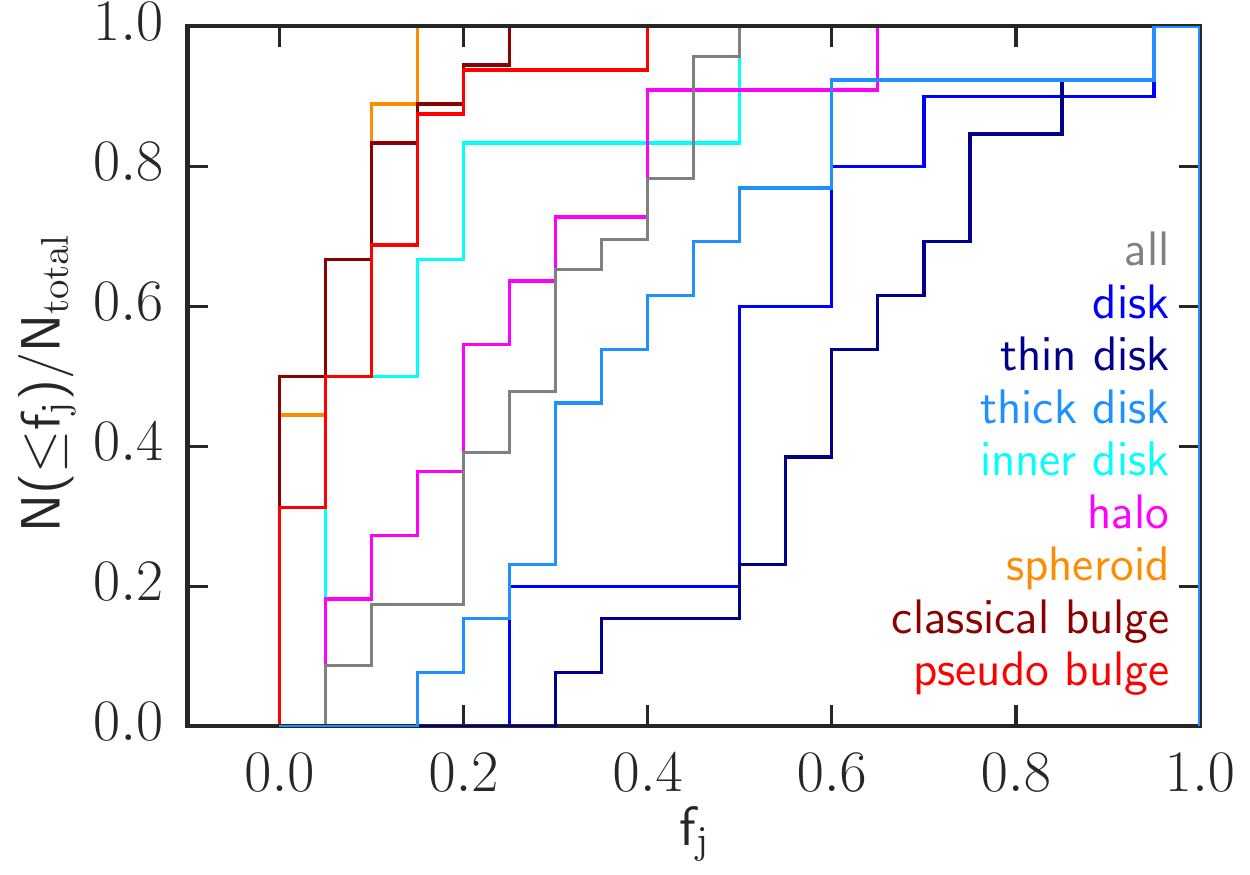}
\caption{The normalized cumulative histograms of the fraction of retained specific angular momentum $f_j$ 
for the various stellar kinematic components.}
\label{fig_fj}
\end{center}
\end{figure}

\subsection{The spin evolution}
\label{spin_conservation}

Whether part of the accreting baryonic material ends up forming an extended rotation dominated component depends greatly 
on how much of its maximum angular momentum is conserved up to $z=0$. 
The \textit{fraction of retained angular momentum} $f_j$ is one way to quantify this loss. 
In this work, this parameter is computed as:
\begin{equation}
 f_j = \frac{j}{j_{\rm max}}
 \label{fjdef}
\end{equation}
separately for each component of each galaxy, where $j$ is the component's specific angular momentum at $z=0$ and $j_{\rm max}$ is its maximum along all the universe's lifetime. 
However, the more widely used definition introduced by \citet{Mo:1998}: 
\begin{equation}
 \eta_j = \frac{j}{j_{\rm dark}}
 \label{fjdefRF12}
\end{equation}
normalizes the specific angular momentum of a particular component to the dark matter halo value at the same time instance. 
The definition in Equation~\ref{fjdefRF12} is based on the work of \citet{Fall:1980} who showed that the large scale tidal torques in the early universe  
should endow both dark matter and baryons with nearly the same specific angular momentum. 
If this prediction of the tidal torque theory is correct, 
the (unobserved) sizes and spins of the virialized dark matter haloes can be related to the (observed) sizes of the 
centrifugally supported baryonic discs \citep[e.g.][]{Mo:1998}.

There are various reasons why we use Equation~\ref{fjdef} instead of \ref{fjdefRF12}. 
The most important reason is that we trace back in time the (constant mass) Lagrangian patches of each component individually, 
be it a stellar dynamical component or its hosting dark matter halo at $z=0$. At high-$z$ the material in any such patch is partially 
inside haloes and partially in the more diffuse filamentary cosmic structure (see Figure 11 in Paper I), 
therefore at these epochs we can not apply Equation~\ref{fjdefRF12} simply because the material we trace 
is not associated with a single halo. Equation~\ref{fjdef} is best suited for our purposes because it can be applied at any redshift and 
irrespective if the particular Lagrangian mass is all inside a single halo.
Another reason is that we do not assume that the large scale tidal torques in the high redshift universe endow both the 
dark matter and baryonic Lagrangian patches that collapse to form the galactic components of a present day galaxy with the same angular momentum. 
Instead we want to test this assumption of the tidal torque theory. Last but not least, there are various physical processes (e.g. 
violent relaxation, angular momentum transfer between baryons and dark matter) that could also alter the angular momentum of the 
collapsing dark matter Lagrangian patches which form the $z=0$ haloes. Therefore, we also want to compute what is the evolution, if any, 
in the dark matter angular momentum between decoupling from the Hubble flow and $z=0$. In any case, for an easier connection with previously 
published studies we provide a comparison between $\eta_{\rm j}$ and $f_{\rm j}$ at $z=0$ in Section~\ref{summary}.

Figure~\ref{figure9_spin} shows the averaged evolutionary patterns for the normalized spins of all eight kinematic components (colored curves), 
as well as of the galaxies as a whole (gray curves), and of the $z=0$ dark matter haloes hosting them (black curves). 
The normalization is the corresponding angular momentum maximum $j_{\rm max}$ of each particular component throughout the universe's history. 
The small bottom panel gives the corresponding standard deviations $\Delta$.  

The averaged evolution of normalized angular momentum is very similar to that of the test galaxy g8.26e11 from Paper I.  
The generic thin and thick discs, and the classical and pseudo bulges have an initial growth phase, 
coeval with the spin acquisition of the dark matter progenitor patch, as predicted by tidal torque theory.
This theory states that at high redshifts, both dark matter and baryons acquire angular momentum via the 
torques produced by the collapse of neighboring regions of the universe.
The reason that the $j/j_{\rm max}$ curves in Figure~\ref{figure9_spin} 
never reach unity is because they represent averages over the same components detected in several galaxies. 
In other words, for each component of each galaxy, the redshift at which $j=j_{\rm max}$ has a different value, thus leading to a 
damping of the peak in the average curves for each type of component shown in Figure~\ref{figure9_spin}.

After they reach their maximum angular momentum at $z\sim1.7$, as can be seen from Figure~\ref{figure9_spin}, 
the thin/thick discs and the classical/pseudo bulges lose part of it.
The biggest loss is suffered by the classical bulge, which ends up with $f_{\rm j}=0.07\pm0.05$, and the smallest by the thin disc, $f_{\rm j}=0.69\pm0.17$. 
The dark matter, on the other hand, retains most of its angular momentum, $f_{\rm j}=0.75\pm0.15$. 
The single large scale discs have a normalized specific angular momentum evolution very similar to that of thin discs, ending up with only a little lower $f_{\rm j}$, 
while the spheroids look like the classical bulges, but with the maximum shifted to higher redshifts, $z\sim3.1$ as compared to $z\sim2.0$. 
The redshifts at which each type of component reaches its maximum $j$ also varies: the discs have significantly lower $z(j_{\rm max})=2.11\pm0.98$
than spheroids for which $z(j_{\rm max})=3.14\pm1.31$, while the classical bulges, pseudo bulges, thick discs and thin discs form a sequence of decreasing 
$z(j_{\rm max})$ from $1.96\pm0.83$ for classical bulges to $1.37\pm0.46$ for thin discs. The stellar haloes reach their maximum $j$ at lower redshift $z(j_{\rm max})=1.29\pm0.67$. For comparison $z(j_{\rm max})=1.44\pm0.65$ for the dark matter.
The large scale disc components have standard deviations that vary very little with time, $0.10\lesssim\Delta\lesssim0.15$ supporting the idea that stellar 
kinematic discs in general, irrespective of their mass, have a well constrained self-similar angular momentum evolution. 
However, this is not true for the two bulge types, who display a peak in the normalized angular momentum standard deviation, 
$\Delta\simeq0.25$, at the approximate epoch of halo virialization, $z\simeq0.7$.
This feature is a direct consequence of the large variety in the formation patterns of bulges.  
In this respect, stellar haloes have even more diverse assembly histories, 
showing a rapidly changing average spin evolution with a large dispersion $\Delta$ fluctuating around $0.2$. 
Since bulge and spheroid stars are expected to form mainly at high redshift, the most probable cause for their angular momentum loss is its transfer to 
the dark matter component during the halo collapse. However, further study is necessary in order to estimate how strong this process is.

Therefore, in Figure~\ref{fig_fj} we show the normalized cumulative histograms of $f_j$ for 
all the kinematic components present in the simulation sample (colored lines), as well as for the whole galaxies' stars (grey line). 
This figure demonstrates that the gravitationally tightly bound dispersion dominated stellar components lost most of their angular momentum, acquired at high redshifts through tidal torques.
Within this group, spheroids lose the most, $f_j<0.16$,  while pseudo bulges the least, $f_j<0.41$. 
Globally, $90\%$ of the velocity dispersion dominated components in this galaxy sample retained less than $10\%$ of their initial angular momentum. 
The stellar haloes display a different behavior, spanning a much broader range of $0.07<f_j<0.69$, 
which is to be expected given the variety in their formation histories.

For the baryons that end up forming large scale rotation dominated structures the angular momentum is much better conserved. 
Of thin discs $85\%$ retain more than half of their initial spin, while $80\%$ of single large scale discs have $f_{\rm j}>0.50$. 
The thick discs, on the other hand, are more prone to angular momentum loss, $30\%$ having $f_j>0.50$ with one of our thick discs reaching as low as $f_j=0.16$.  
The inner discs span the range $0.08\leq f_j\leq0.51$. 

\begin{figure}
\includegraphics[width=0.48\textwidth]{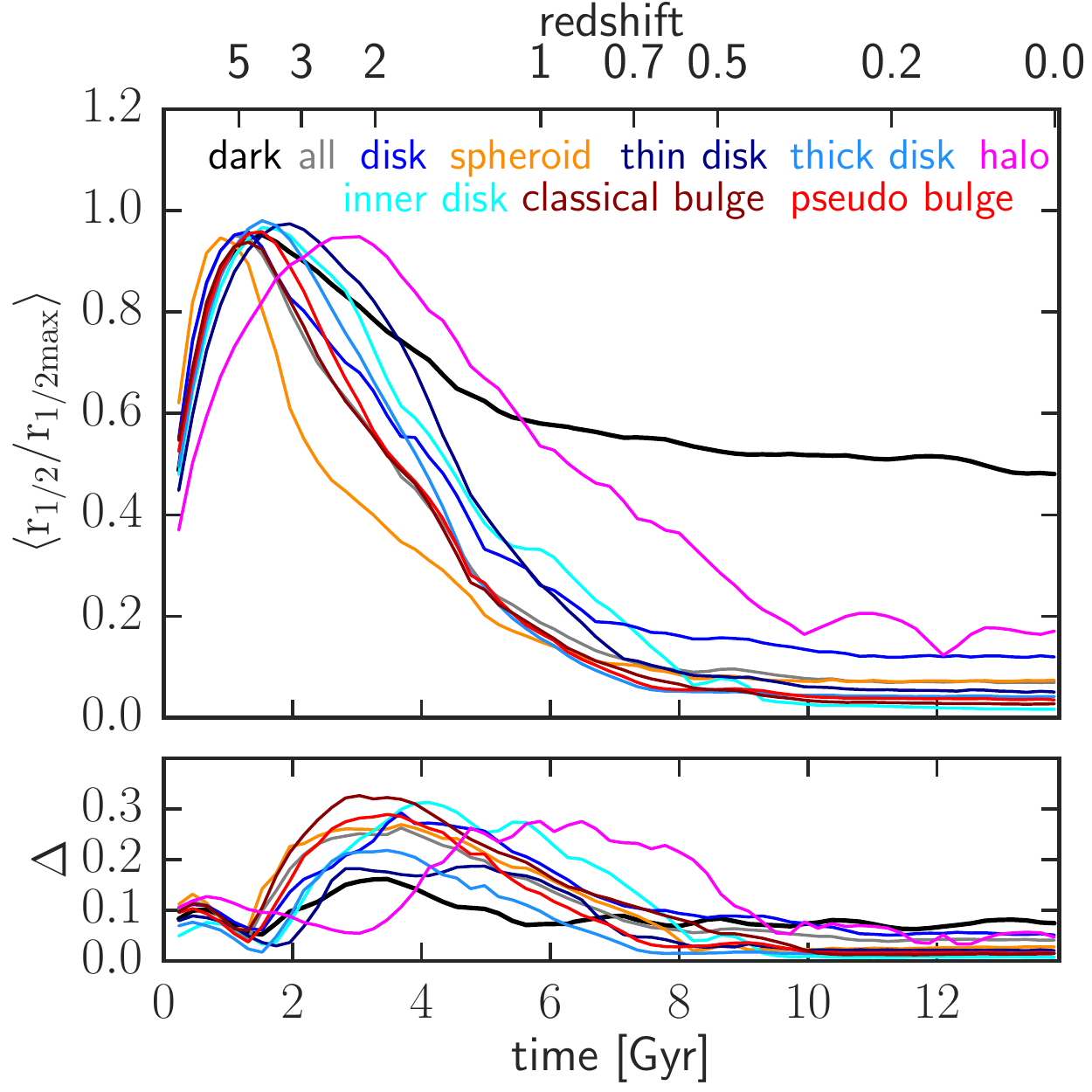}
\caption{The mass function weighted global evolutions of the normalized sizes (3D half mass radii) for the Lagrangian masses comprising
all stars in the galaxies at $z=0$ (gray curves), the stars in the various kinematic components at $z=0$ (colored curves), and the dark matter halo at $z=0$ (black curves). 
The small bottom panel gives the corresponding weighted standard deviations $\rm\Delta$.}
\label{figure9_size}
\end{figure}

\begin{figure}
\includegraphics[width=0.48\textwidth]{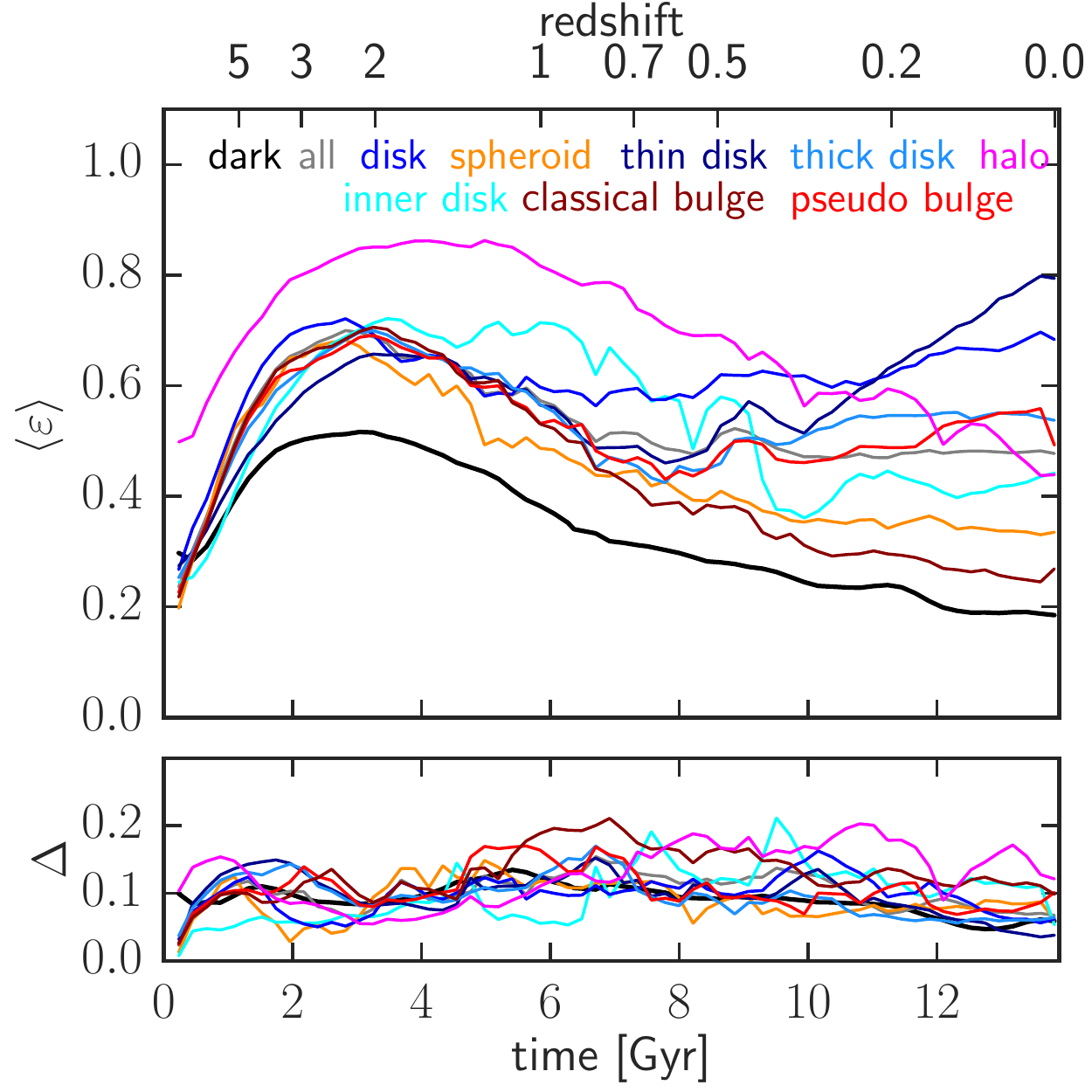}
\caption{The mass function weighted global evolutions of the normalized shapes for the Lagrangian masses comprising
all stars in the galaxies at $z=0$ (gray curves), the stars in the various kinematic components at $z=0$ (colored curves), and the dark matter halo at $z=0$ (black curves). 
The small bottom panel gives the corresponding weighted standard deviations $\rm\Delta$.}
\label{figure9_shape}
\end{figure}

\subsection{The size evolution}

The intrinsic size evolution is quantified by the 3D half mass radius $r_{\rm 1/2}$.
The average evolution of the normalized half mass radii and their corresponding dispersions is plotted in Figure~\ref{figure9_size}. 
The spheroids, classical and pseudo bulges collapse at earlier than the rotation dominated structures,  
the redshift at which $r_{\rm 1/2}$ reaches its maximum varying from $z(r_{\rm 1/2max})=5.70\pm1.92$ for spheroids 
to $3.99\pm1.02$ for pseudo bulges.
Among the latter, thin discs have a delayed collapse with respect to the thick ones, $z(r_{\rm 1/2max})=3.17\pm0.61$ vs $3.74\pm0.74$.
Overall the stellar haloes are the ones collapsing the latest ($z(r_{\rm 1/2max})=2.14\pm0.60$). 
However, the standard deviations of all stellar components but the stellar halo show a broad peak between $z\sim3.0$ and $z\sim1.0$. 
The peak of the stellar halo is shifted towards later times, $2.0<z<0.7$.

The large variations in the evolutionary patterns of the various components' sizes do not allow us to draw any definite conclusions.
On the other hand, the dark matter material shows a small and relatively constant $\Delta\sim0.10$, its averaged normalized size reaching its maximum at about the same time as the baryonic components $z(r_{\rm 1/2max})=3.70\pm1.13$. 
Due to its collisionless nature, its collapse stops much earlier than that of the baryons, a little before $z\sim1.0$. 
Therefore, $z\simeq1.0$ can be taken as the approximate end of the virialization epoch for all the dark matter haloes in the current simulation sample. 

\begin{figure}
\includegraphics[width=0.48\textwidth]{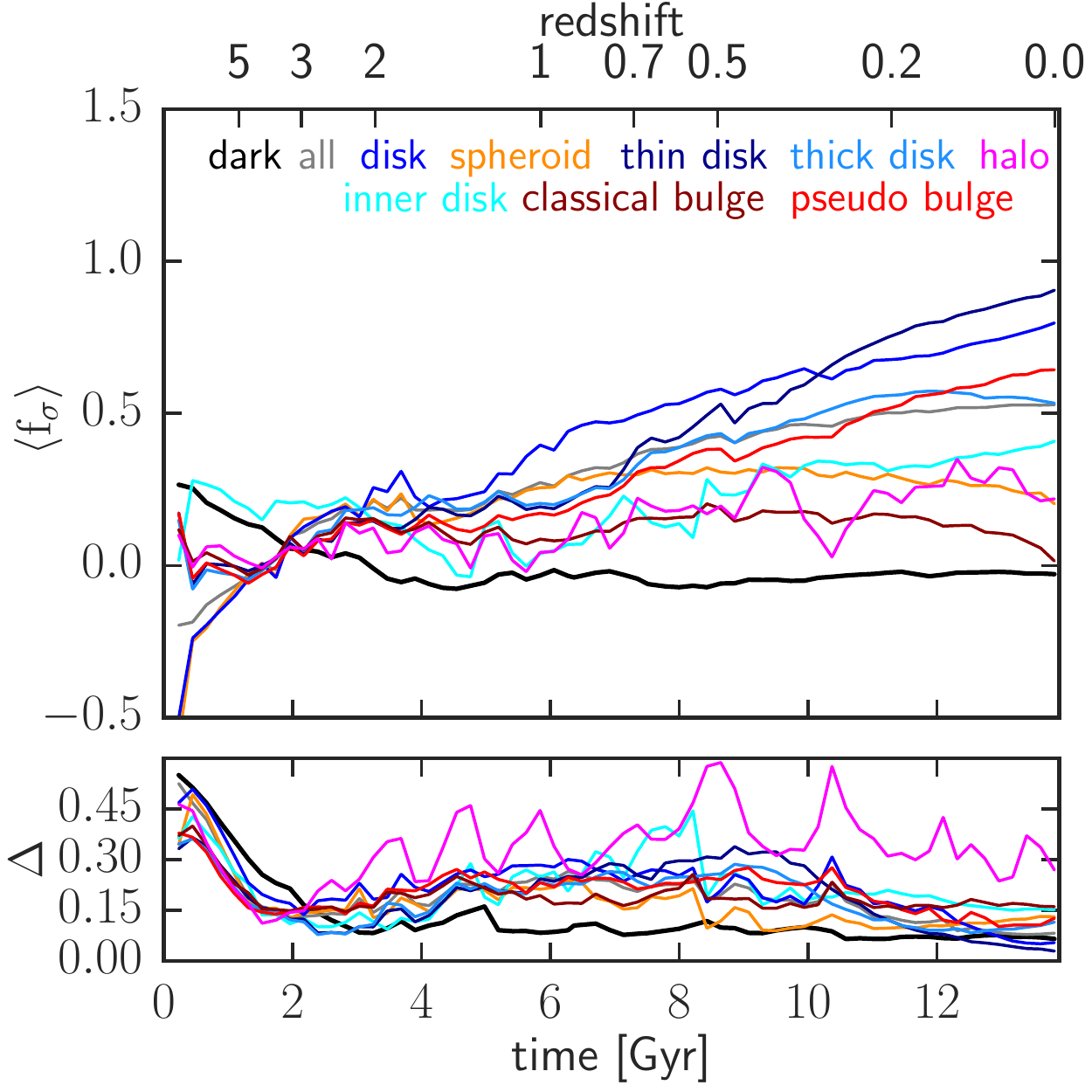}
\caption{The mass function weighted global evolutions of the vertical velocity dispersion fractions for the Lagrangian masses comprising
all stars in the galaxies at $z=0$ (gray curves), the stars in the various kinematic components at $z=0$ (colored curves), and the dark matter halo at $z=0$ (black curves). 
The small bottom panel gives the corresponding weighted standard deviations $\rm\Delta$.}
\label{figure9_rotsup}
\end{figure}

\subsection{The evolution of shapes and rotational support}

As discussed in Section~\ref{ell_rotsup}, we quantify the shape of the Lagrangian mass of each kinematic structure using the 
ellipticity $\langle\varepsilon\rangle$ derived from the eigenvalues of the inertia tensor. 
The evolution of the average ellipticities shown in Figure~\ref{figure9_shape} display high redshift peaks ($3.0<z<2.0$) for all baryonic components, 
but for the stellar halo whose maximum is broader and displaced towards later times, $z\sim1.5$. 
The dark matter shows also a peak at $z\sim2.5$ and a monotonic decrease afterwards. 
The standard deviations of all components are relatively constant, fluctuating around $\Delta\sim0.1$. 

The dark matter, classical bulges and spheroids decrease their ellipticities $\langle\varepsilon\rangle$ all 
the way to values close to spherical symmetry, $\sim0.2$, $\sim0.25$ and $\sim0.35$, respectively, at redshift zero. 
The large scale disc components and the pseudo bulges, on the other hand, 
reach minima at $z\sim0.7$, after which their asymmetry increases. 
The late time evolution of $\langle\varepsilon\rangle$ for thin discs is particularly steep. 

The relatively large value of the high-$z$ ellipticity baryonic peaks ($\sim0.65$ for the thin disc up to $\sim0.85$ for the stellar halo) reflect 
a highly non-isotropic early assembly, in agreement with the material of all $z=0$ structures being part of a collapsing filamentary structure. 
Comparatively, the dark matter shows a more isotropic collapse, the maximum $\langle\varepsilon\rangle$ being approximately $0.5$. 

Figure~\ref{figure9_rotsup} gives the evolution of the rotational support, as quantified by $\langle f_{\rm\sigma}\rangle$. 
The Lagrangian mass of the $z=0$ dark matter halo has large values at $z\geq2.0$, after which it stabilizes at $\langle f_{\rm\sigma}\rangle\sim0$ 
with a constant and small standard deviation $\Delta\sim0.1$. 
Among the baryonic components, the classical bulges have the smallest velocity dispersion asymmetry throughout the Universe's lifespan. 
At $z=0$ classical bulges are consistent with having a perfectly isotropic velocity dispersion, same as the dark matter.
The large scale discs, on the other hand, monotonically increase their $\langle f_{\rm\sigma}\rangle$, the thin discs reaching final values close to one, 
as expected of systems with little to no vertical motions. 

Interestingly, the thick discs and the pseudo bulges show a very similar evolution of their corresponding rotational support. 
The two also have similar shape evolutions, as quantified by the ellipticity $\langle\varepsilon\rangle$. 
Based on these properties pseudo bulges look like the extension of (thick) discs to the very inner galactic regions.

\begin{figure}
\includegraphics[width=0.48\textwidth]{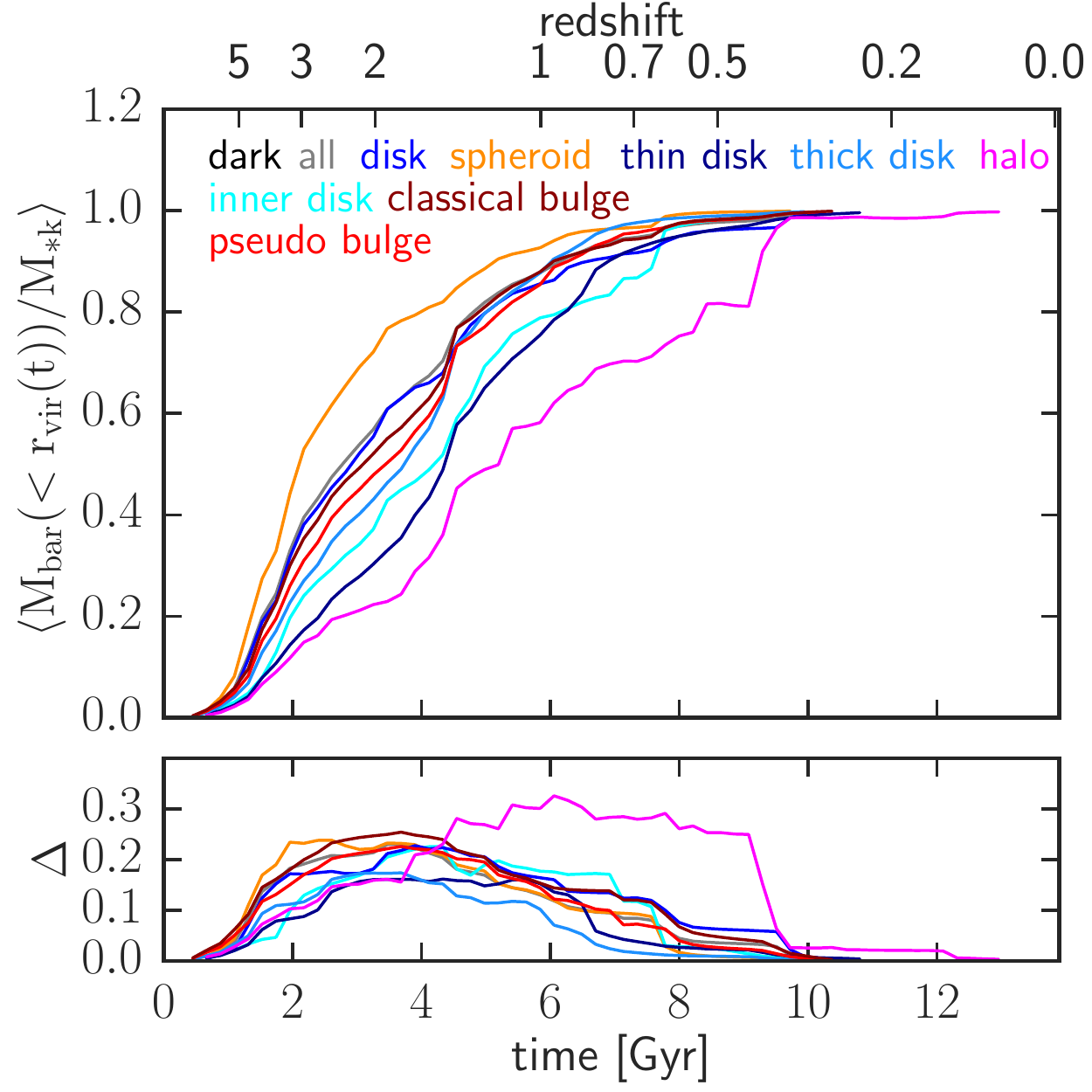}
\caption{The mass function weighted global evolutions of the progenitor baryon assembly along the main branch of the merger tree for the various $z=0$ kinematic components. 
The small bottom panel gives the corresponding weighted standard deviations $\Delta$.}
\label{figure10_bar}
\end{figure}

\begin{figure}
\includegraphics[width=0.48\textwidth]{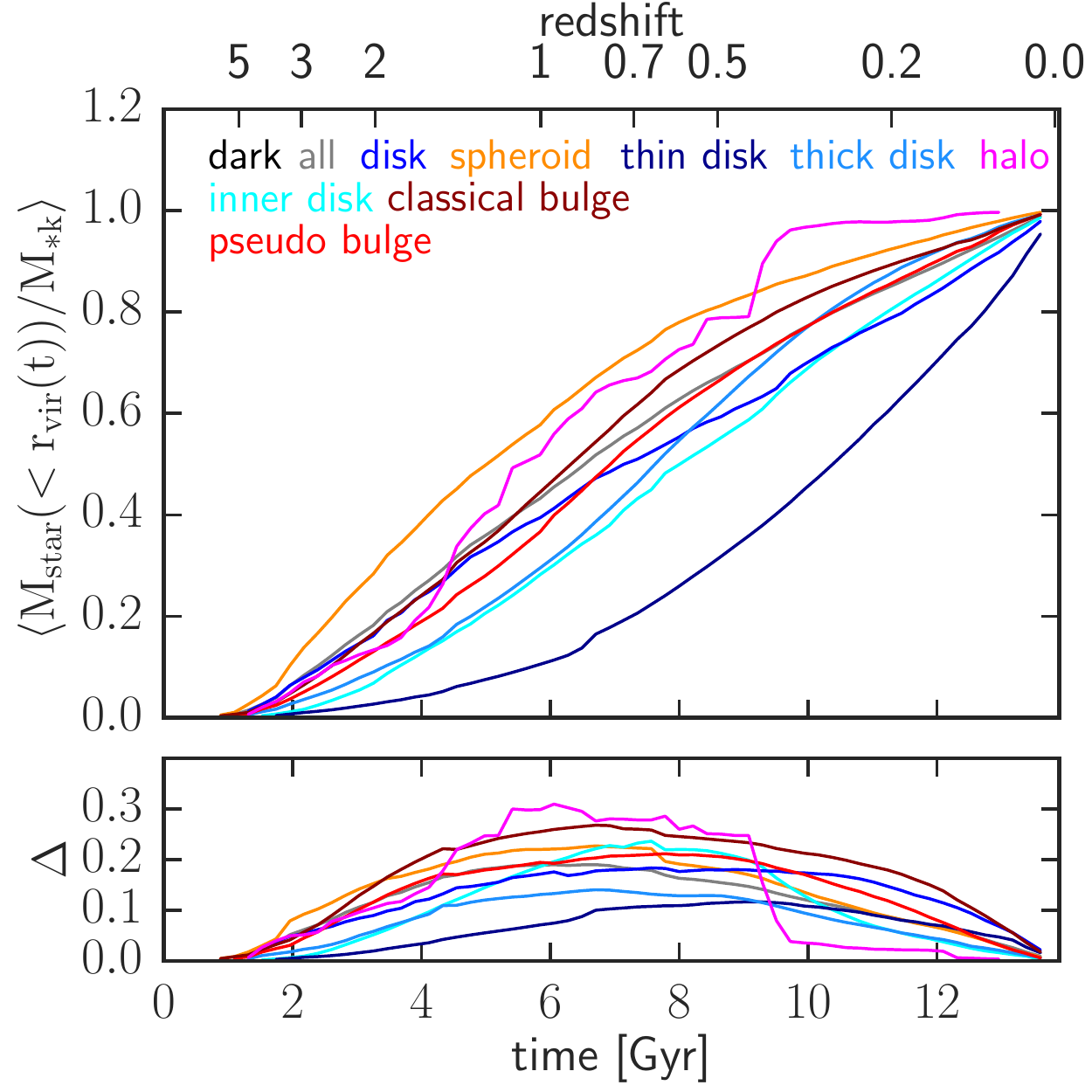}
\caption{The mass function weighted global evolutions of the progenitor stars assembly along the main branch of the merger tree for the various $z=0$ kinematic components. 
The small bottom panel gives the corresponding weighted standard deviations $\Delta$.}
\label{figure10_star}
\end{figure}

\subsection{The mass accretion histories}

The averaged evolutions of the baryonic and stellar mass along the main branch of the merger tree (Equations~\ref{z50eq}) 
are shown in Figures~\ref{figure10_bar} and \ref{figure10_star}, respectively.
In Figure~\ref{figure10_bar}, the classical and pseudo bulges, as well as the large scale single discs and the thick discs follow 
the evolution of the baryonic progenitors of the full $z=0$ stellar populations (grey curves). 
With respect to them, the progenitor material of spheroids is accreted faster and at earlier times, while that of thin and inner discs is incorporated later. 
The stellar haloes clearly stand out as being accreted at later times and during a longer timescale.

The clear outliers in Figure~\ref{figure10_star} are the thin discs which form their stars much later on than any other kinematic component. 
At the other extreme are the spheroids which form their stars the earliest. However, all components show large standard deviations that vary strongly with time. 
Among them, the thin discs have the smallest $\Delta$. 

One interesting aspect of Figure~\ref{figure10_bar} is that thick discs on the average accrete their progenitor baryons at the same time as the bulges, especially the pseudo ones, 
but form their stars slightly later than pseudo bulges (see Figure~\ref{figure10_star}). 
This result is in line with the conclusions drawn by \citep{Comeron:2014} from observations. 
However, as shown in other studies before \citep[e.g.][]{Oklopcic:2017,Buck:2017} our simulations disfavor the clumpy disc scenario \citep{Bournaud:2007}
for the formation of thick discs and bulges \citep{Bournaud:2009,Elmegreen:2008,Inoue:2012}, used by \citet{Comeron:2014} to interpret their data.   
Instead, our results suggest that pseudo bulges are likely to be the seed around which thick discs or single large scale discs form later on. 

\begin{table*}
\scriptsize
\centering
\begin{tabular}{cccccccccccccc}
\hline
Component & disc & thin disc & thick disc & inner disc & spheroid & classical bulge & pseudo bulge & halo & all stars\\
\hline
frequency (\%) & 44 (30) & 56 (70) & 56 (70) & 28 (35) & 36 (20) & 80 (100) & 72 (90) & 44 (55) & 100 \\
$\langle f_{\rm mass}\rangle$ & 0.40$\rm\pm$0.13 & 0.24$\rm\pm$0.03 & 0.27$\rm\pm$0.03 & 0.22$\rm\pm$0.09 & 0.51$\rm\pm$0.15 & 0.27$\rm\pm$0.06 & 0.18$\rm\pm$0.04 & 0.04$\rm\pm$0.02 & 1.00$\rm\pm$0.00\\
$\langle f_{\rm j}\rangle$ & 0.55$\rm\pm$0.13 & 0.69$\rm\pm$0.17 & 0.41$\rm\pm$0.15 & 0.17$\rm\pm$0.14 & 0.10$\rm\pm$0.05 & 0.07$\rm\pm$0.05 & 0.14$\rm\pm$0.11 & 0.28$\rm\pm$0.18 & 0.34$\rm\pm$0.13\\
$\langle\varepsilon\rangle$ & 0.68$\rm\pm$0.06 & 0.79$\rm\pm$0.04 & 0.54$\rm\pm$0.06 & 0.44$\rm\pm$0.05 & 0.33$\rm\pm$0.06 & 0.27$\rm\pm$0.10 & 0.49$\rm\pm$0.10 & 0.44$\rm\pm$0.12 & 0.48$\rm\pm$0.07\\
$\langle f_{\rm\sigma}\rangle$ & 0.80$\rm\pm$0.05 & 0.90$\rm\pm$0.03 & 0.53$\rm\pm$0.11 & 0.41$\rm\pm$0.15 & 0.20$\rm\pm$0.13 & 0.01$\rm\pm$0.16 & 0.64$\rm\pm$0.13 & 0.22$\rm\pm$0.27 & 0.53$\rm\pm$0.08\\
$\langle z_{\rm 1/2}\rangle_{\rm bar}$ & 2.44$\rm\pm$0.74 & 1.51$\rm\pm$0.40 & 1.89$\rm\pm$0.51 & 1.83$\rm\pm$0.53 & 2.98$\rm\pm$0.91 & 2.25$\rm\pm$0.84 & 2.05$\rm\pm$0.69 & 1.13$\rm\pm$0.49 & 2.47$\rm\pm$0.81\\
$\langle z_{\rm 1/2}\rangle_{\rm star}$ & 0.81$\rm\pm$0.31 & 0.31$\rm\pm$0.10 & 0.68$\rm\pm$0.20 & 0.67$\rm\pm$0.27 & 1.32$\rm\pm$0.47 & 1.00$\rm\pm$0.50 & 0.80$\rm\pm$0.41 & 0.97$\rm\pm$0.36 & 0.95$\rm\pm$0.39\\
$\langle\tau\rangle_{\rm bar}$ [Gyr] & 4.55$\rm\pm$1.96 & 4.85$\rm\pm$0.67 & 4.04$\rm\pm$0.84 & 4.83$\rm\pm$1.17 & 3.37$\rm\pm$1.11 & 3.90$\rm\pm$1.36 & 4.13$\rm\pm$1.08 & 6.19$\rm\pm$1.59 & 4.18$\rm\pm$1.48\\
$\langle\tau\rangle_{\rm star}$ [Gyr] & 9.21$\rm\pm$1.58 & 7.21$\rm\pm$1.32 & 7.52$\rm\pm$1.24 & 7.80$\rm\pm$1.45 & 6.91$\rm\pm$1.50 & 6.25$\rm\pm$1.71 & 7.48$\rm\pm$1.89 & 4.91$\rm\pm$1.50 & 8.63$\rm\pm$0.97\\
$\langle\kappa\rangle_{\rm bar}$ & 0.68$\rm\pm$0.46 & 1.22$\rm\pm$0.52 & 1.21$\rm\pm$0.74 & 0.80$\rm\pm$0.32 & 0.55$\rm\pm$0.34 & 1.30$\rm\pm$2.14 & 1.54$\rm\pm$3.29 & 1.57$\rm\pm$1.05 & 0.72$\rm\pm$0.51\\
$\langle\kappa\rangle_{\rm star}$ & 0.94$\rm\pm$0.71 & 1.51$\rm\pm$0.43 & 0.98$\rm\pm$0.28 & 0.81$\rm\pm$0.32 & 0.65$\rm\pm$0.35 & 0.89$\rm\pm$0.43 & 1.01$\rm\pm$0.38 & 3.96$\rm\pm$4.99 & 0.71$\rm\pm$0.35\\
$\langle f_{\rm in-situ}\rangle$ & 0.89$\rm\pm$0.06 & 0.97$\rm\pm$0.05 & 0.96$\rm\pm$0.03 & 0.96$\rm\pm$0.03 & 0.95$\rm\pm$0.05 & 0.96$\rm\pm$0.04 & 0.96$\rm\pm$0.05 & 0.35$\rm\pm$0.14 & 0.93$\rm\pm$0.04\\
$\langle \eta_{\rm j}\rangle$ & 0.67$\rm\pm$0.32 & 1.30$\rm\pm$0.76 & 0.61$\rm\pm$0.39 & 0.25$\rm\pm$0.20 & 0.05$\rm\pm$0.02 & 0.10$\rm\pm$0.12 & 0.19$\rm\pm$0.19 & 0.81$\rm\pm$0.43 & 0.38$\rm\pm$0.25\\
$\langle j_{\rm max}^{\rm(k)}/j_{\rm max}^{\rm(dark)} \rangle$ & 1.20$\rm\pm$0.36 & 1.22$\rm\pm$0.33 & 0.99$\rm\pm$0.30 & 1.05$\rm\pm$0.16 & 0.66$\rm\pm$0.35 & 0.83$\rm\pm$0.34 & 0.84$\rm\pm$0.30 & 2.19$\rm\pm$0.096 & 0.93$\rm\pm$0.31\\
$\langle z(j_{\rm max}) \rangle$ & 2.11$\rm\pm$0.98 & 1.37$\rm\pm$0.46 & 1.65$\rm\pm$0.32 & 1.70$\rm\pm$0.72 & 3.14$\rm\pm$1.31 & 1.96$\rm\pm$0.83 & 1.69$\rm\pm$0.68 & 1.29$\rm\pm$0.67 & 1.93$\rm\pm$0.89\\
$\langle z(r_{\rm 1/2max}) \rangle$ & 4.38$\rm\pm$1.17 & 3.17$\rm\pm$0.61 & 3.74$\rm\pm$0.74 & 3.55$\rm\pm$0.86 & 5.70$\rm\pm$1.92 & 4.26$\rm\pm$1.37 & 3.99$\rm\pm$1.02 & 2.14$\rm\pm$0.60 & 4.60$\rm\pm$1.19\\
\hline
\end{tabular}
\centering\caption{The average properties of the eight types of stellar kinematic structures (columns two to nine), as well as of the galaxies' stellar populations as a whole (column ten): 
frequency in the complete 25 galaxy sample (the numbers in parenthesis are the corresponding frequencies in the subsample of the 20 most massive dark matter haloes), 
mass fraction normalized to the total stellar mass $\langle f_{\rm mass}\rangle$, fraction of retained angular momentum $\langle f_{\rm j}\rangle$, 
ellipticity $\langle\varepsilon\rangle$, rotational support $\langle f_{\rm\sigma}\rangle$, 
redshift of half progenitor baryonic mass inside the dark matter halo $\langle z_{\rm 1/2}\rangle_{\rm bar}$, 
redshift of half progenitor stellar mass inside the dark matter halo $\langle z_{\rm 1/2}\rangle_{\rm star}$, 
baryons assembly time scale $\langle\tau\rangle_{\rm bar}$, stars assembly time scale $\langle\tau\rangle_{\rm star}$, 
skewness of baryons assembly time scale $\langle\kappa\rangle_{\rm bar}$, skewness of stars assembly time scale $\langle\kappa\rangle_{\rm star}$, 
global mass fraction of stars born in-situ $\langle f_{\rm in-situ}\rangle$, 
retained fraction of angular momentum according to the definition in Equation~\ref{fjdefRF12} $\langle \eta_{\rm j}\rangle$, 
ratio between the maximum angular momentum of a baryonic component's Lagrangian patch and the maximum of the corresponding dark matter one 
$\langle j_{\rm max}^{\rm(k)}/j_{\rm max}^{\rm(dark)}\rangle$,  
redshift of maximum angular momentum $\langle z(j_{\rm max}) \rangle$, and redshift of maximum 3D half mass radius 
$\langle z(r_{\rm 1/2max}) \rangle$.}
\label{table_z0}
\end{table*}

\begin{figure*}
\begin{center}
\includegraphics[width=0.33\textwidth]{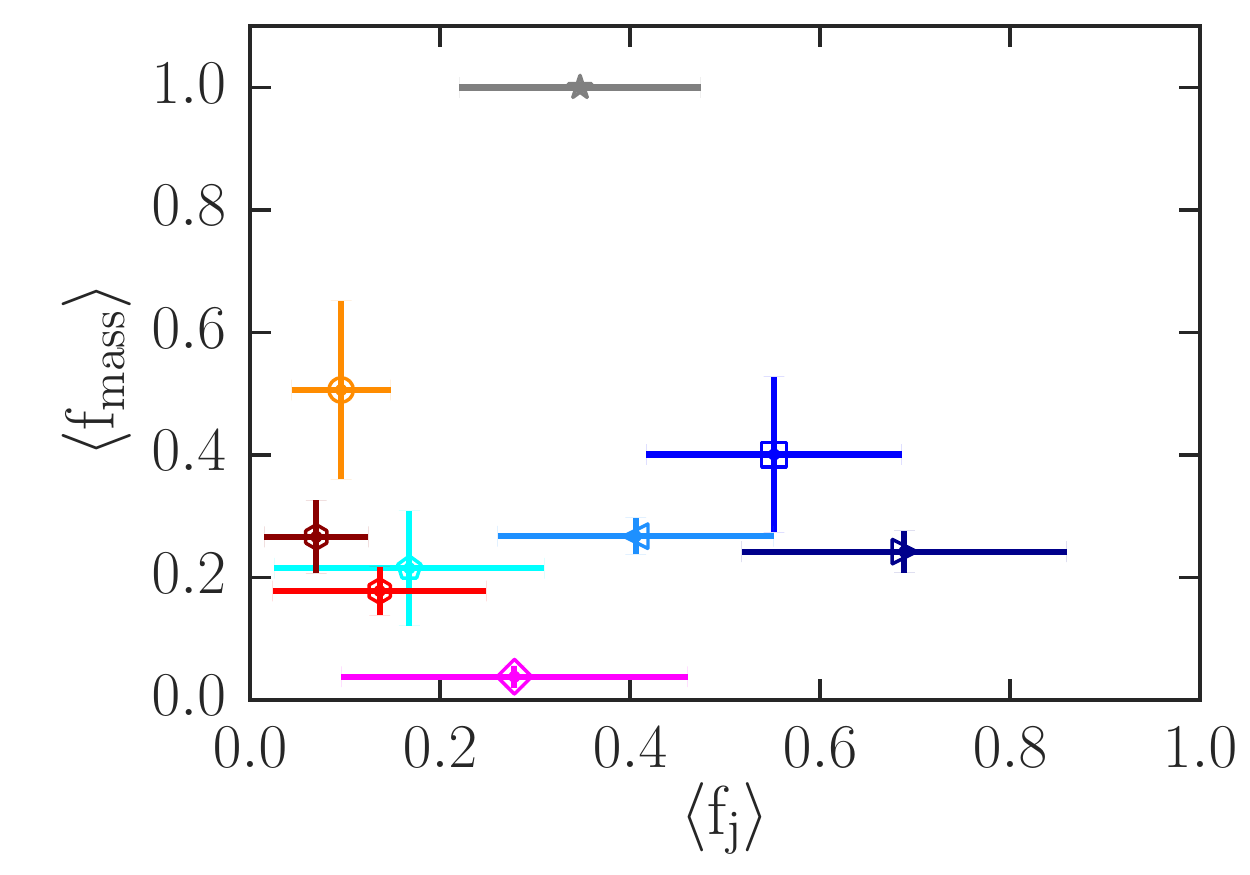}
\includegraphics[width=0.33\textwidth]{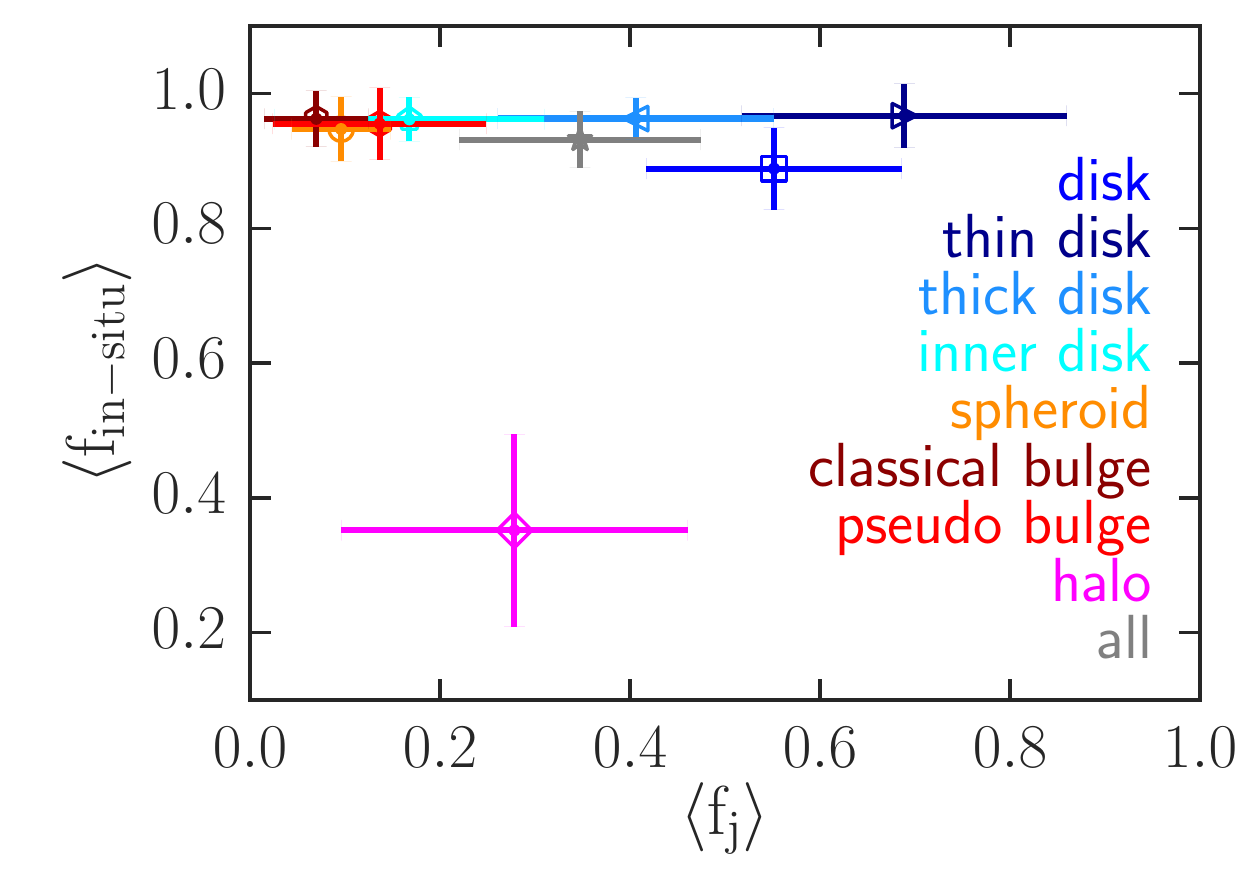}
\includegraphics[width=0.33\textwidth]{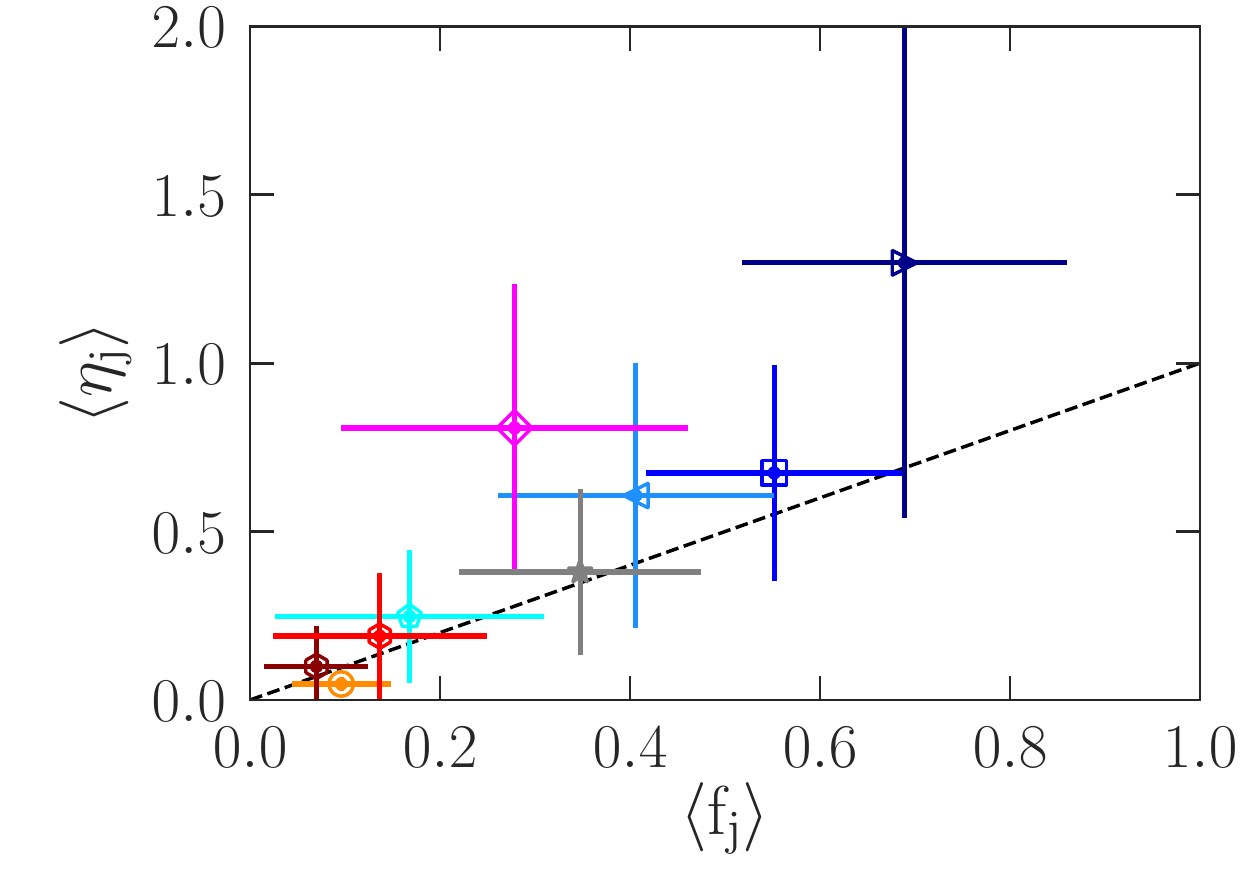}\\
\includegraphics[width=0.33\textwidth]{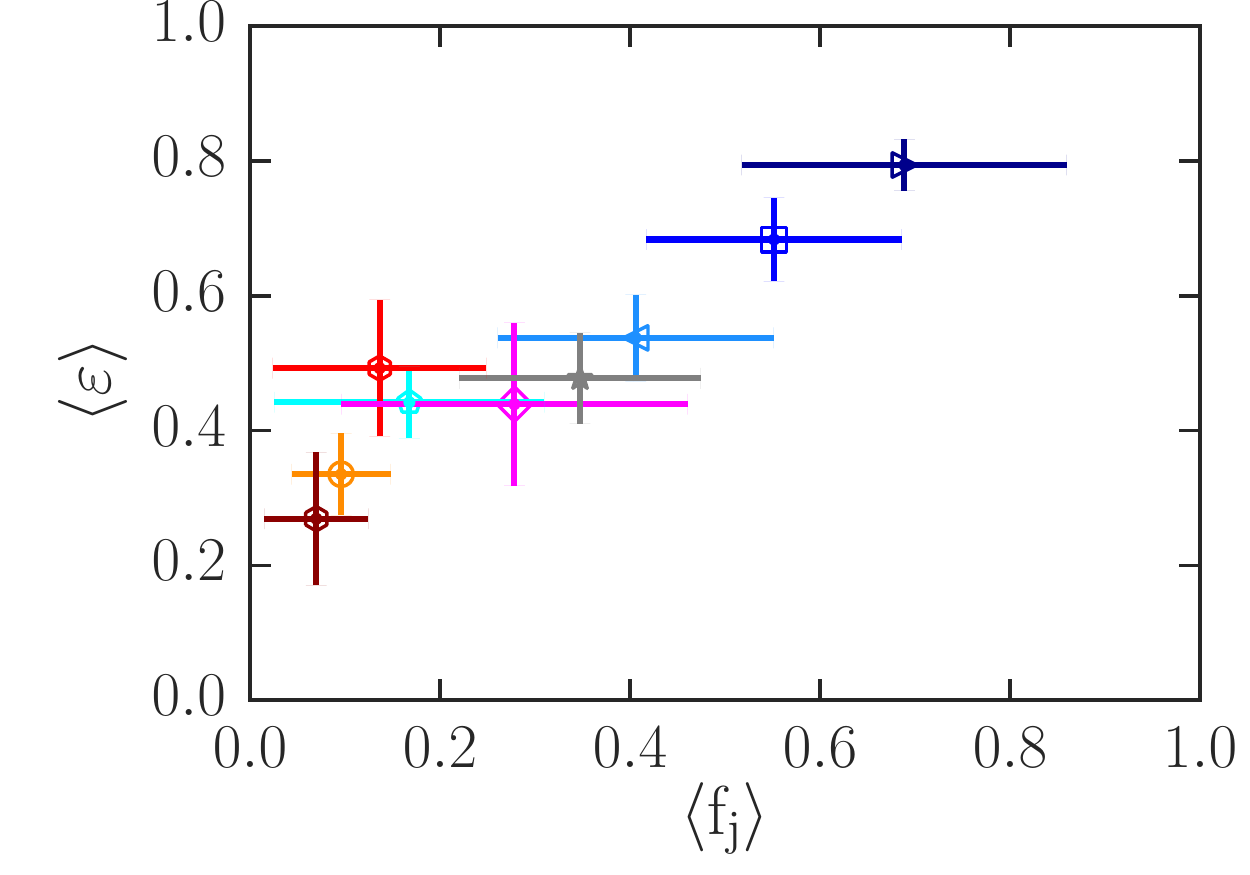}
\includegraphics[width=0.33\textwidth]{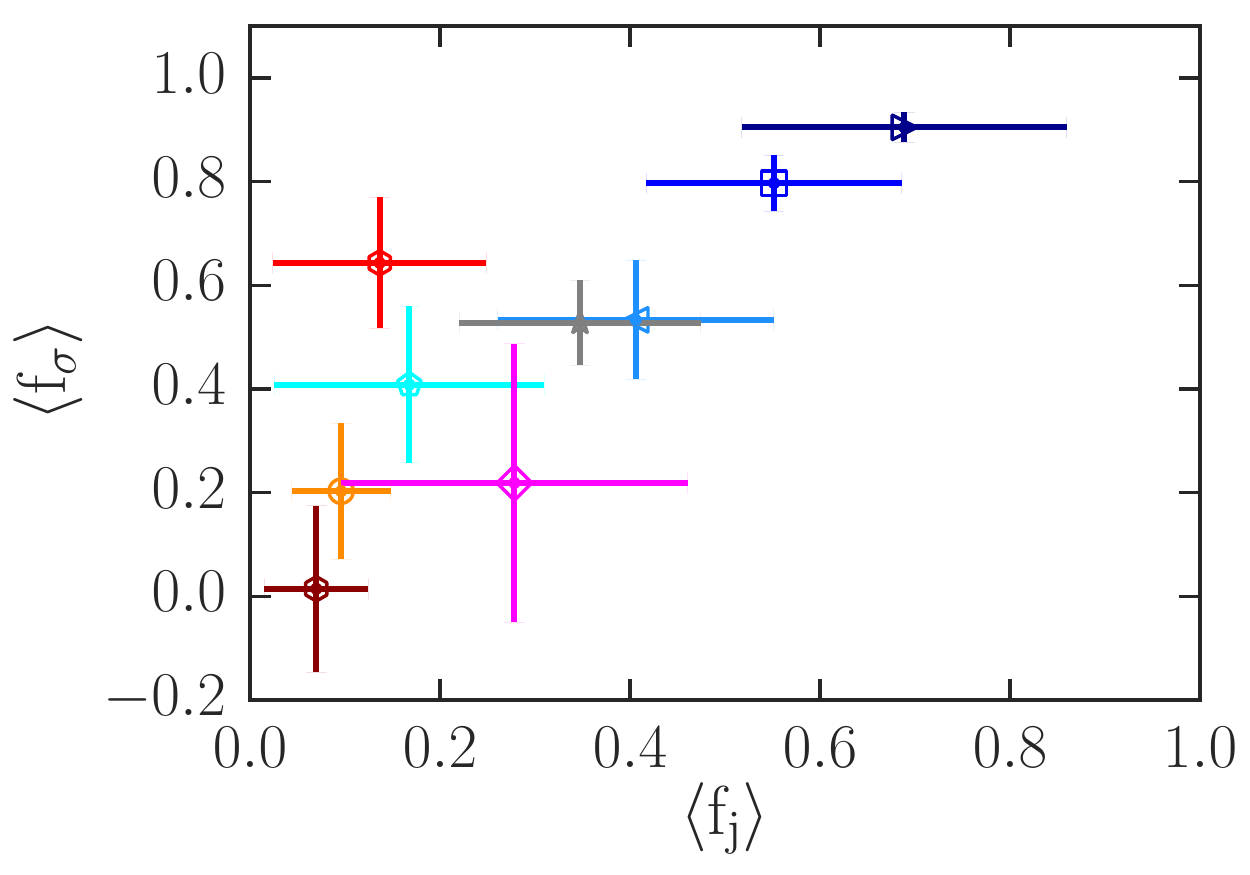}
\includegraphics[width=0.33\textwidth]{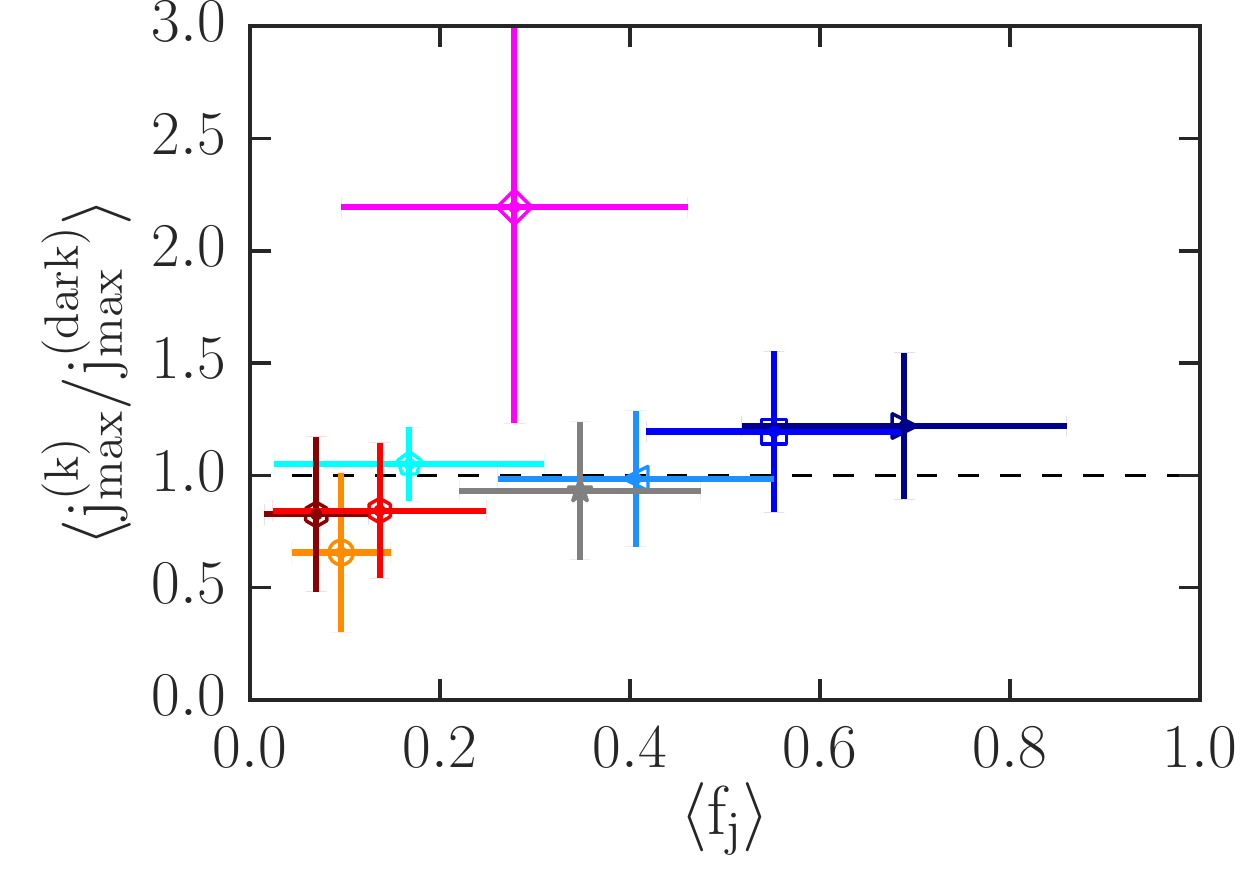}\\
\caption{The average mass fractions normalized to the total stellar mass $\langle f_{\rm mass}\rangle$ (top left),
average global mass fractions of stars born in-situ $\langle f_{\rm in-situ}\rangle$ (top center), 
average fraction of retained angular momentum according to Equation~\ref{fjdefRF12} definition $\langle \eta_{\rm j}\rangle$ (top right),
average ellipticities $\langle\varepsilon\rangle$ (bottom left), average rotational support $\langle f_{\rm\sigma}\rangle$ (bottom center), 
and average ratio of baryonic to dark matter maximum angular momentum $\langle j_{\rm max}^{\rm(k)}/j_{\rm max}^{\rm(dark)}\rangle$ (bottom right) 
for the eight types of kinematic structures as functions of the average retained angular momentum fraction $\langle f_{\rm j}\rangle$.
The dashed lines in the two right panels give the 1:1 correspondence between the two definitions of $f_{\rm j}$ (top) and the theoretical assumptions that 
baryons and dark matter reach the same maximum in specific angular momentum (bottom).}
\label{figure_summary1}
\end{center}
\end{figure*}

\section{Average properties of structures}
\label{summary}

In the light of the $z=0$ properties we discussed in Sections~\ref{int_prop} and \ref{obs_prop}, and of their evolutions in Section~\ref{glabal_evolution}, 
we conclude that the fundamental factor in setting the $z=0$ distinct properties of the eight stellar kinematic components is the retained fraction of angular momentum. 

We use the weighted averaging procedure described in the previous section, where a component's weight is given by the $z\sim0$ volume density of galaxies with the same total stellar mass as its host galaxy, 
to compute the mean and dispersion for fifteen properties. 
These properties are: (i) stellar mass fractions $\langle f_{\rm mass}\rangle$, (ii) retained angular momenta fractions $\langle f_{\rm j}\rangle$ (Equation~\ref{fjdef}),
(iii) ellipticities $\langle\varepsilon\rangle$, (iv) rotational support $\langle f_{\rm\sigma}\rangle$, (v) redshifts of half progenitor baryonic mass inside the dark matter halo 
$\langle z_{\rm 1/2}\rangle_{\rm bar}$, (vi) redshifts of half progenitor stellar mass inside the dark matter halo $\langle z_{\rm 1/2}\rangle_{\rm star}$, 
(vii) baryons assembly time scale $\langle\tau\rangle_{\rm bar}$, (viii) stars assembly time scale $\langle\tau\rangle_{\rm star}$, 
(ix) skewness of baryons assembly time scale $\langle\kappa\rangle_{\rm bar}$, (x) skewness of stars assembly time scale $\langle\kappa\rangle_{\rm star}$, 
(xi) global mass fraction of stars born in-situ $\langle f_{\rm in-situ}\rangle$,
(xii) retained fraction of angular momentum according to the definition in Equation~\ref{fjdefRF12} $\langle \eta_{\rm j}\rangle$, 
(xiii) ratio between the maximum angular momentum of a baryonic component's Lagrangian patch and the maximum of the 
corresponding dark matter one $\langle j_{\rm max}^{\rm(k)}/j_{\rm max}^{\rm(dark)}\rangle$, 
(xiv) redshift where the maximum angular momentum is reached $\langle z(j_{\rm max}) \rangle$, 
and (xv) redshift where the maximum 3D half mass radius is reached $\langle z(r_{\rm 1/2max}) \rangle$.

The assembly timescale $\tau$ is taken to be the time passed between the moments when 10 and 90 per cent of the progenitor mass was inside the dark matter halo:
\begin{equation}
\tau \equiv t_{\rm 90\%} - t_{\rm 10\%} .
\label{tau_eq}
\end{equation}

In a similar way the skewness of the assembly time scale $\kappa$ is the ratio between the early and late assembly timescales: 
\begin{equation}
\kappa \equiv \frac{ t_{\rm 50\%} - t_{\rm 10\%} }{ t_{\rm 90\%} - t_{\rm 50\%} } .  
\label{kappa_eq}
\end{equation}
The global mass fraction of stars born in-situ $f_{\rm in-situ}$ represents the ratio between the mass of a particular component $k$ stars born inside $r_{\rm vir}(t)$ along the main branch of the merger tree 
and the final stellar mass $M_{\rm *k}$. This definition is equivalent to the one used by \citet{Pillepich:2015}.

\begin{figure*}
\begin{center}
\includegraphics[width=0.33\textwidth]{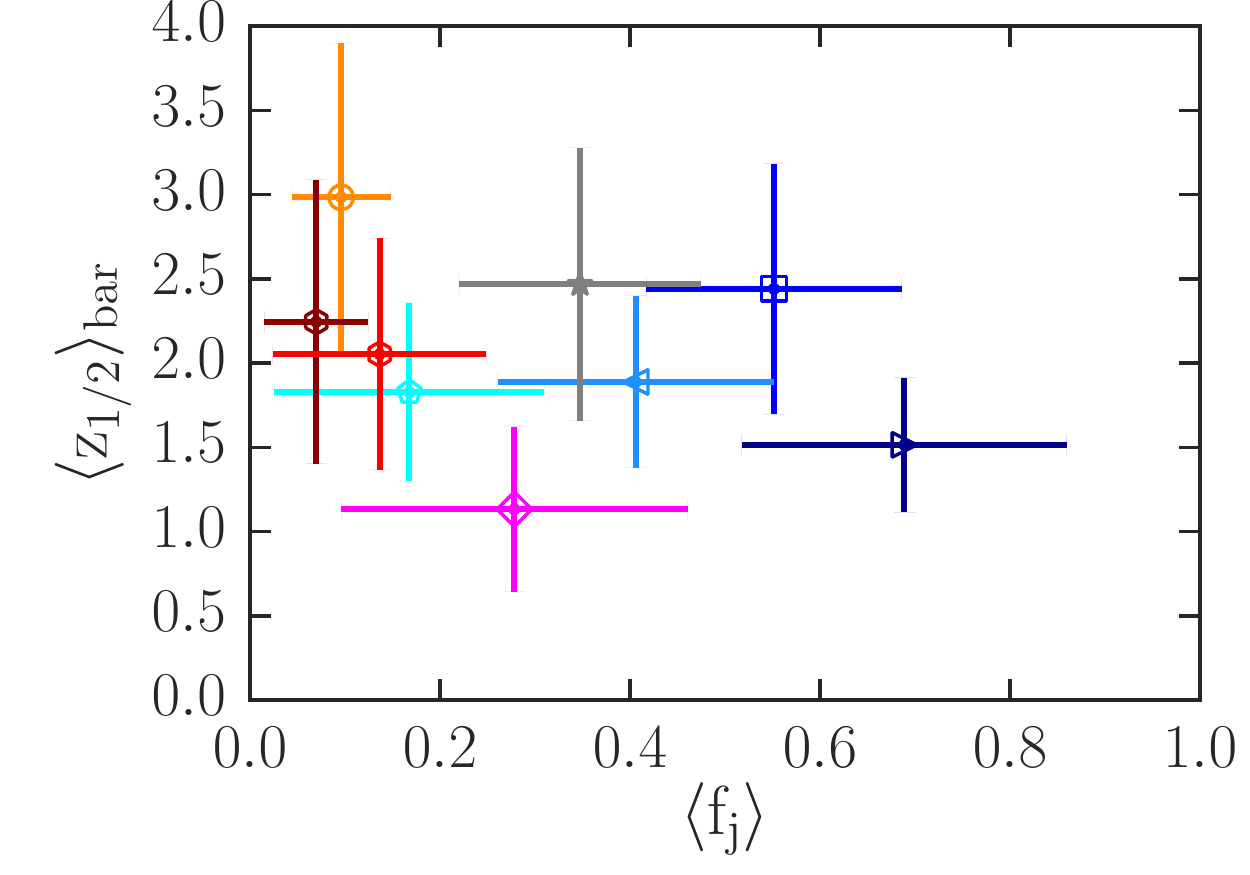}
\includegraphics[width=0.33\textwidth]{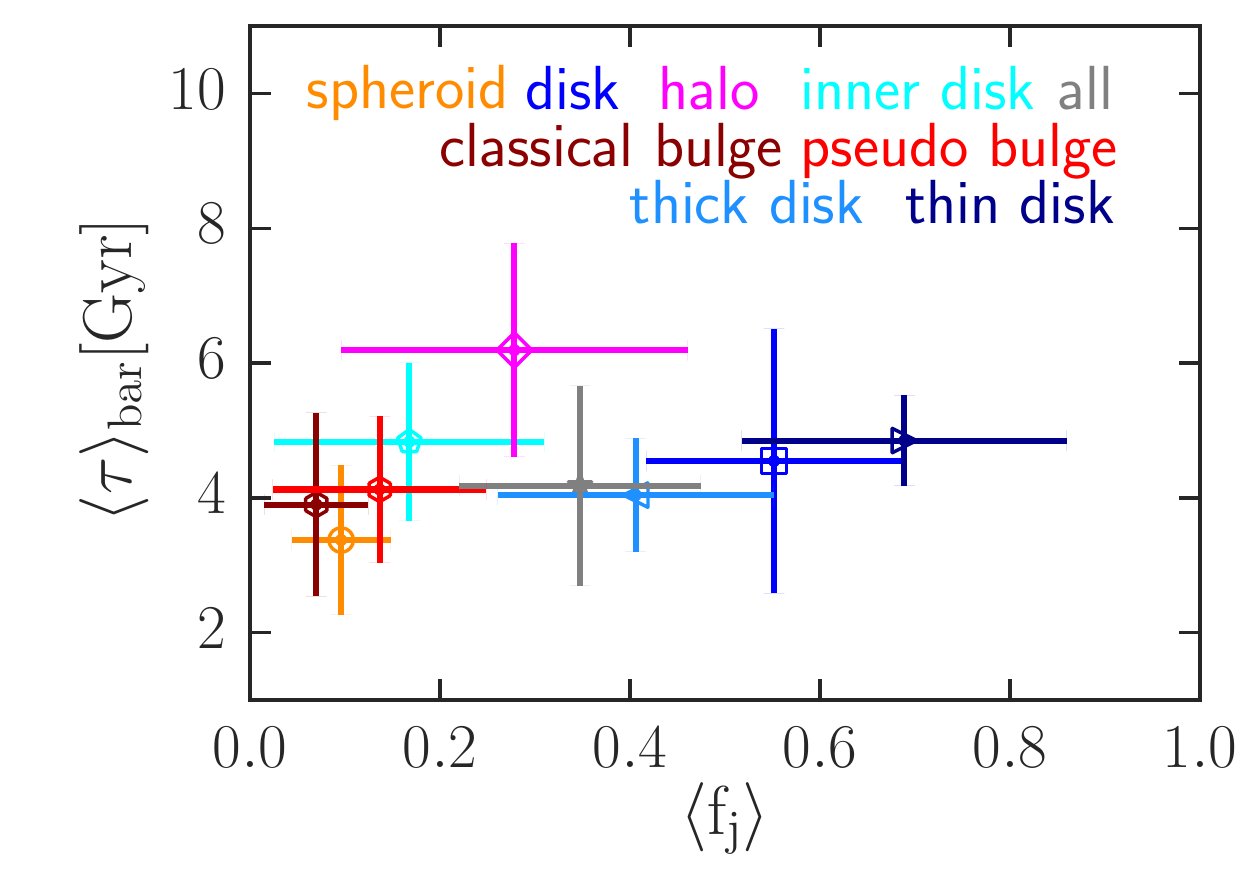}
\includegraphics[width=0.33\textwidth]{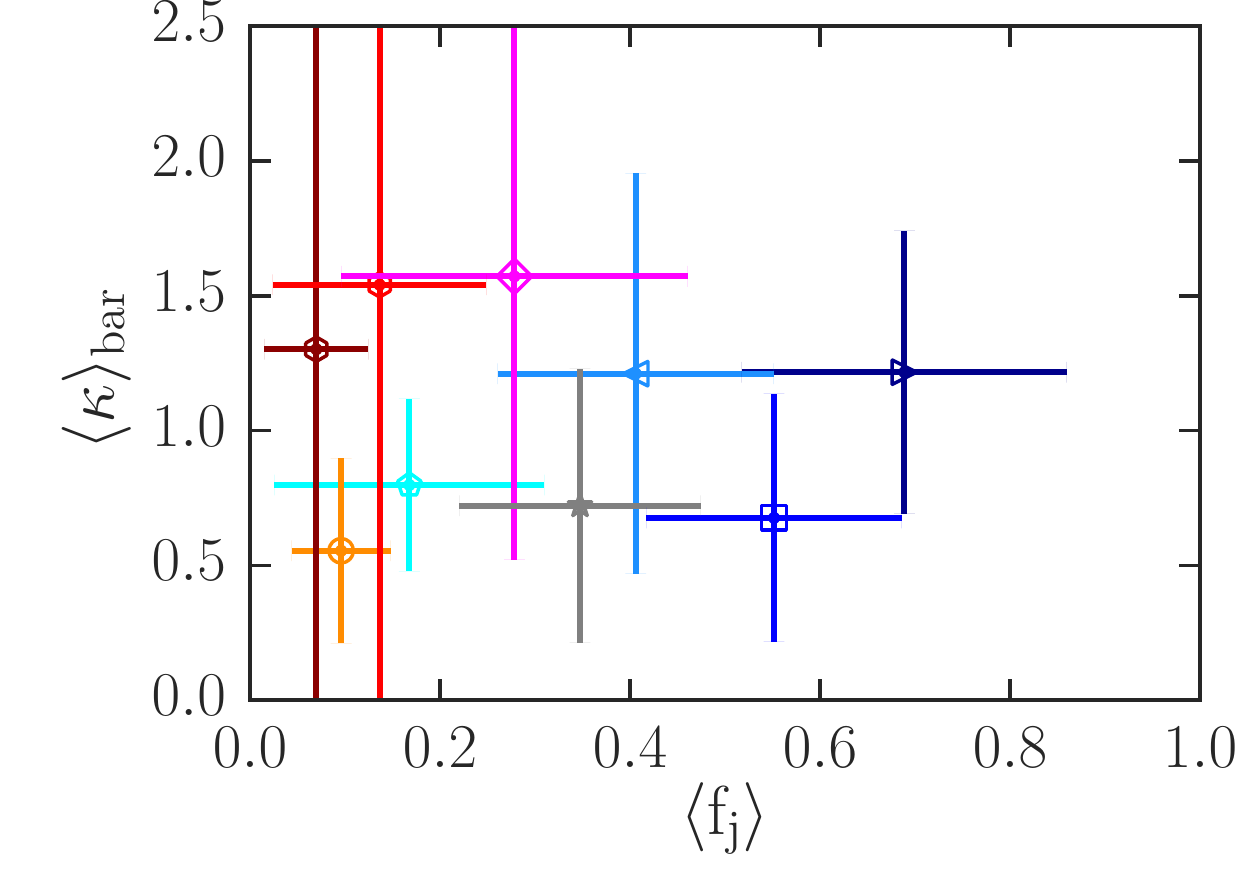}\\
\includegraphics[width=0.33\textwidth]{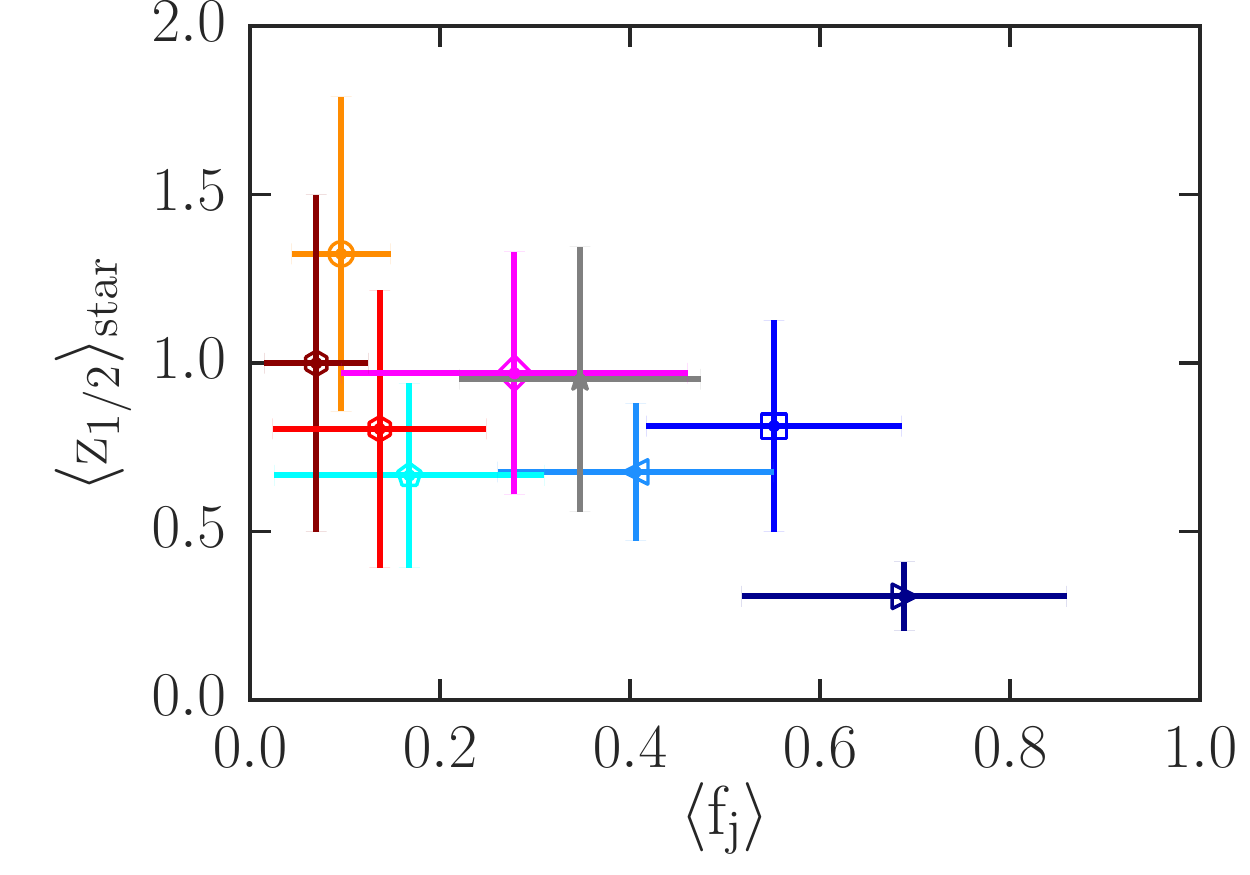}
\includegraphics[width=0.33\textwidth]{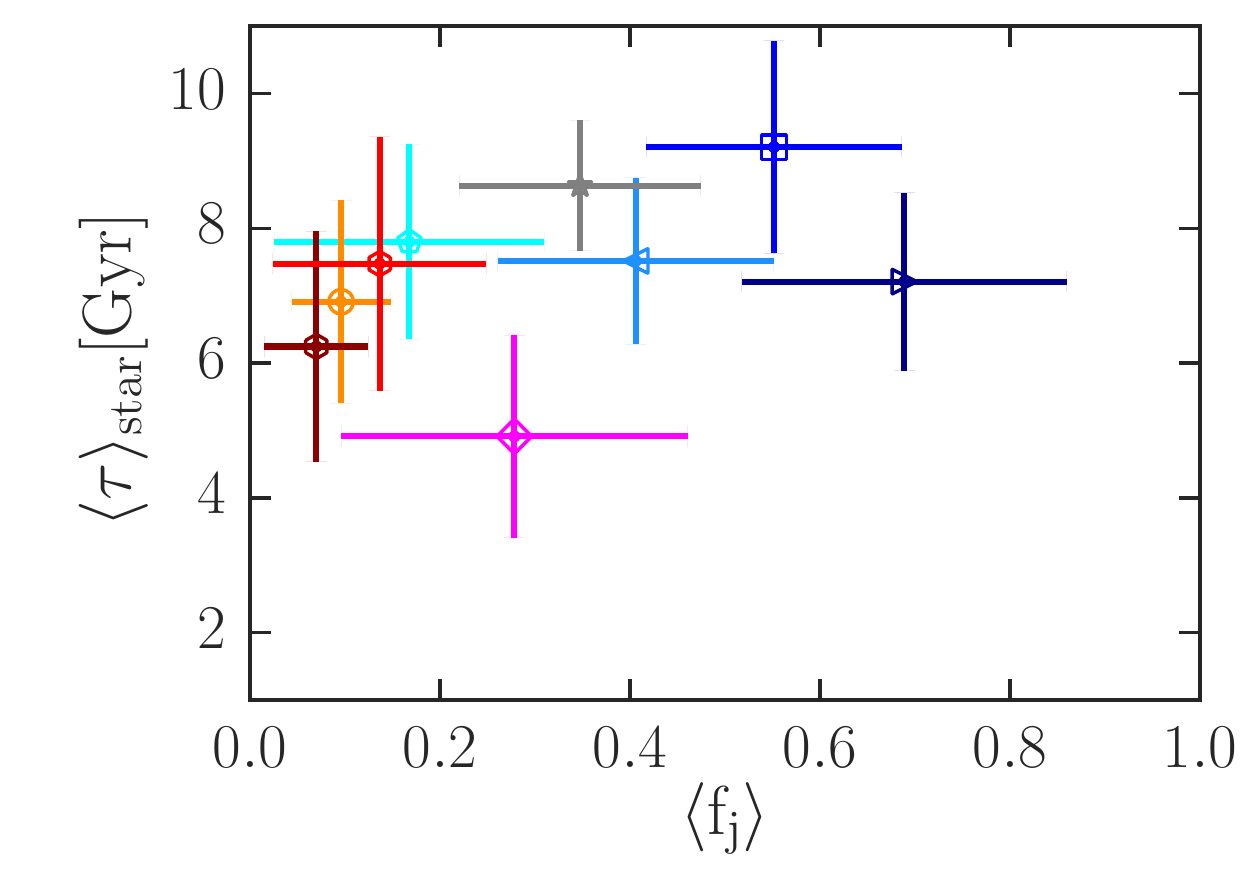}
\includegraphics[width=0.33\textwidth]{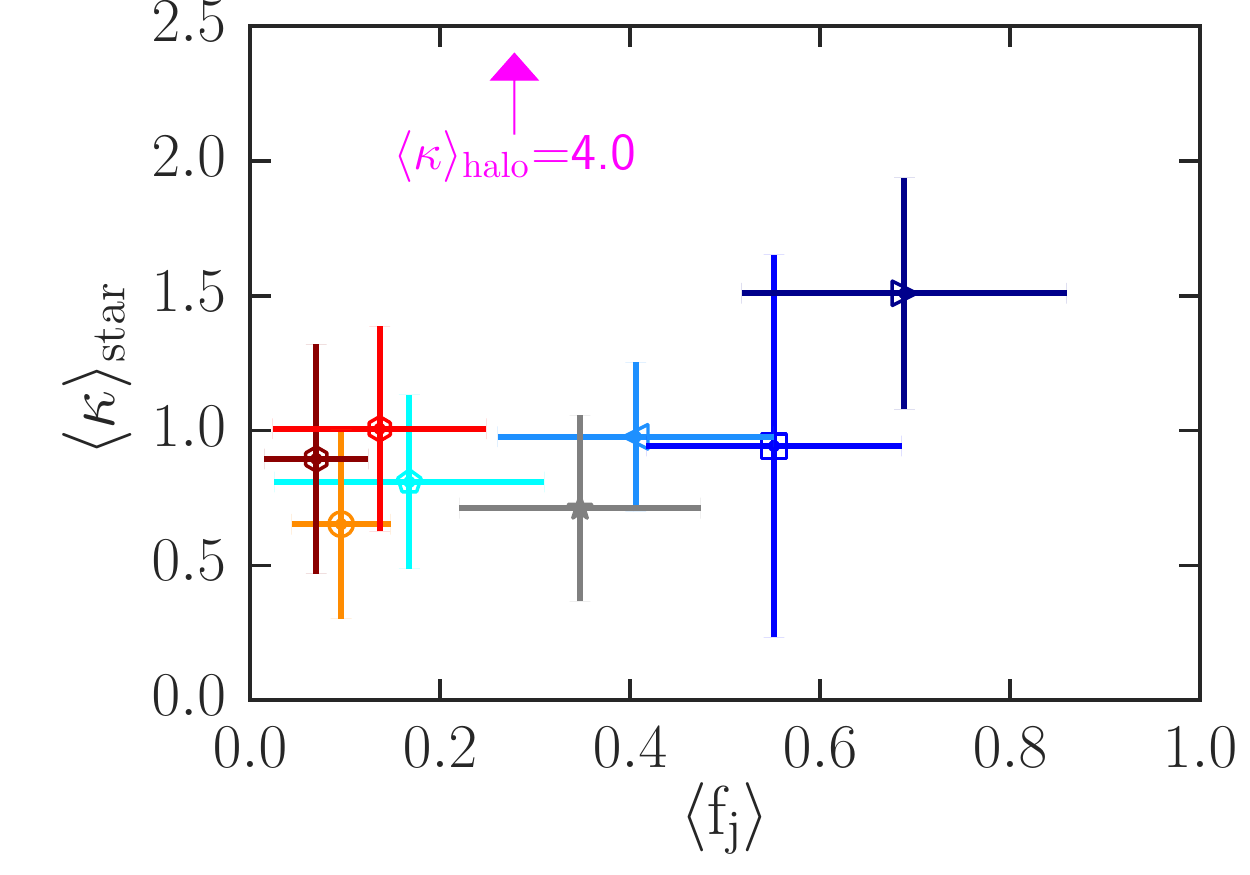}\\
\caption{The average redshift of half mass within the dark matter halo $\langle z_{\rm 1/2}\rangle$ (left), average assembly timescale $\langle\tau\rangle$ (centre), 
and average assembly skewness $\langle\kappa\rangle$ (right) for the baryonic (top) and stellar (bottom) progenitors of the eight types of kinematic structures 
as functions of the average retained angular momentum fraction $\langle f_{\rm j}\rangle$.
The stellar halo in the bottom right panel is represented by an arrow because its value of $\rm\langle\kappa\rangle_{\rm star}$ is much larger than for the rest of the structures.}  
\label{figure_summary2}
\end{center}
\end{figure*}

All these average properties for the eight kinematic components are listed in Table~\ref{table_z0}. 
Two galaxies, g3.61e11 and g5.31e11 have been excluded because they do not have 
enough timesteps at high redshift to constrain well their evolutions. 
Figures~\ref{figure_summary1} and \ref{figure_summary2} show the average properties of the eight types of kinematic components 
as functions of their respective average retained angular momentum fractions. 

The top left panel of Figure~\ref{figure_summary1} shows the average stellar mass fractions enclosed in each type of kinematic component. 
The galaxies hosted by the smallest dark matter haloes have only two distinct kinematic components, a disc and a spheroid. The spheroid makes up about $0.51\pm0.15$ of the mass, 
while the single large scale discs have $\langle f_{\rm mass}\rangle=0.40\pm0.13$. All 20 most massive dark matter haloes (80\% of the sample) host galaxies with an important classical bulge component 
$\langle f_{\rm mass}\rangle=0.27\pm0.06$, while pseudo bulges found in 72\% of the sample have smaller mass fractions $0.18\pm0.04$. 
Both thin and thick discs with  approximately equal mass fractions ($f_{\rm mass}\sim0.25$) are present in 56\% of the sample. 
The stellar haloes, found in 11 out of the 20 most massive dark matter haloes, as expected, encompass only a small part of the total stellar mass 
$\langle f_{\rm mass}\rangle=0.04\pm0.02$. Across the various components, the mass fractions do not correlate with the fractions of retained angular momentum.

The top {center} panel of the same figure gives the mass fraction of the in-situ born stars $\langle f_{\rm in-situ}\rangle$ as a function of $\langle f_{\rm j}\rangle$. 
All kinematic components but the stellar haloes form most of their stars in-situ ($f_{\rm in-situ}\gtrsim0.9$). 
Similarly to the first panel, there is no correlation between $\langle f_{\rm in-situ}\rangle$ and $\langle f_{\rm j}\rangle$. 
The stellar haloes on the other hand have much fewer stars born in-situ $\langle f_{\rm j}\rangle=0.35\pm0.14$ \citep[e.g.][]{Zolotov:2009,Cooper:2015}. 
In this respect, the Milky Way analogue g8.26e11 discussed in Paper I is an outlier with $f_{\rm in-situ}=0.55$.
Globally, $0.93\pm0.04$ of the stellar mass in this 25 galaxy sample is born in-situ.
This $f_{\rm in-situ}$ parameter, however, is bound to be dependent on the resolution, because of the highly stochastic nature of star formation in low mass dark matter haloes.
We intend to test this parameter in a future study using much higher resolution simulations of the same NIHAO galaxies. 

The {left and center} bottom panels of Figure~\ref{figure_summary1} give the ellipticities $\langle\varepsilon\rangle$ and the rotational support $\langle f_{\rm\sigma}\rangle$, 
respectively. Both $\langle\varepsilon\rangle$ and $\langle f_{\rm\sigma}\rangle$ follow a clear predictable sequence with $\langle f_{\rm j}\rangle$, 
from classical bulges that have low ellipticities and little rotational support for small $\langle f_{\rm j}\rangle$, 
to thin discs that have large values of $\langle\varepsilon\rangle$ and $\langle f_{\rm\sigma}\rangle$ for large $\rm\langle f_{\rm j}\rangle$ (top central panel). 
The pseudo bulges stand out from this sequence having considerable rotational support, $\langle f_{\rm\sigma}\rangle=0.64\pm0.13$, even if they lost most of their angular momentum, 
$\langle f_{\rm j}\rangle=0.14\pm0.11$.
The large scale single discs have $\langle f_{\rm j}\rangle$, $\langle\varepsilon\rangle$ and $\langle f_{\rm\sigma}\rangle$ in between the thin and the thick discs, 
while the spheroids are in between classical and pseudo bulges.  

The two panels in the right column of Figure~\ref{figure_summary1} provide the links between our definition of retained fraction of angular momentum (Equation~\ref{fjdef}) 
and that of RF12 (Equation~\ref{fjdefRF12}). The difference between $f_{\rm j}$ and $\eta_{\rm j}$ comes from the choice of normalization. 
While we normalize the present day specific angular momentum of each galaxy component $j_{\rm final}$ to its own maximum over the universe's lifetime $j_{\rm max}^{\rm(k)}$, 
the more widely used definition in Equation~\ref{fjdefRF12} normalizes it to the host dark matter halo specific angular momentum at the same epoch $j_{\rm dark}$.
The two right panels of the figure show that when considering all stars of the galaxies together (grey points) the two definitions for the 
retained fraction of angular momentum lead to the same value on average: $\langle \eta_{\rm j}\rangle\simeq\langle f_{\rm j}\rangle$, 
and both dark matter and baryonic progenitors reach the same maximum in specific angular momentum 
$\langle j_{\rm max}^{\rm(all)}/j_{\rm max}^{\rm(dark)}\rangle\simeq1$. 

When taken individually, most dynamical components deviate from the $\langle j_{\rm max}^{\rm(all)}/j_{\rm max}^{\rm(dark)}\rangle=1$ line. 
These deviations can be understood in terms of the differences between the assembly of the various galaxy components in the early universe. 
The first hint was already provided by Figure~\ref{figure9_size} where one can see that, on average, the Lagrangian patches of $z=0$ 
spheroids start their collapse (maximum at $z\sim6$) before those of the dark matter (maximum at $z\sim5$), which at their turn are only 
later followed by those of the thin discs (maximum at $z\sim3$). This sequence in collapse times together with the fact that the galaxy 
assembly process is more chaotic and violent at higher redshifts explain why the spheroid material reaches lower maximum specific angular 
momentum than the dark matter, while the thin disc reaches higher values, as shown by the 
$\langle j_{\rm max}^{\rm(k)}/j_{\rm max}^{\rm(dark)}\rangle$ vs $\langle f_{\rm j}\rangle$ panel of Figure~\ref{figure_summary1}.
The sequence in $\langle j_{\rm max}^{\rm(k)}/j_{\rm max}^{\rm(dark)}\rangle$ from values $<1$ for spheroids and bulges to values $>1$
for (thin) discs indicates that the dynamical structures of present day galaxies have a non-negligible cosmological origin.
In other words, the progenitor gas of $z=0$ dynamical stellar discs preferentially spans the tail at large values of the total 
baryonic angular momentum distribution at high-$z$, while the spheroid and bulges progenitor gas samples more the peak at low $j$ of the distribution.

Also, if we consider separately each of the eight stellar dynamical components, we find that for most of them    
$\langle \eta_{\rm j}\rangle \neq \langle f_{\rm j}\rangle$. The biggest discrepancies between the two values of retained angular
momentum fractions occur for the thin(thick) discs and for the stellar haloes. Thin discs have $\langle \eta_{\rm j}\rangle=1.30\pm0.76$, 
while their average $f_j$ is only $0.69\pm0.17$. Thick discs, on the other hand, have $\langle \eta_{\rm j}\rangle=0.61\pm0.39$
compared with $\langle f_{\rm j}\rangle=0.41\pm0.15$. Therefore, dynamical thin/thick discs have on average $\rm\sim30\%$ more/less specific 
angular momentum than their dark matter haloes.

In both observations and theoretical studies focusing on stellar angular momentum, galaxies are separated in no more than two
components: a disc and a spheroid. Therefore, in order to compare our results with these studies we need to average over the rotation (or dispersion) 
dynamical components. In this perspective, if we consider together the thin discs, thick discs and discs we get $\langle j_{\rm max}^{\rm(disk)}/j_{\rm max}^{\rm(dark)}\rangle=1.14\pm0.19$, and therefore the assumption of $j_{\rm max}^{\rm(disk)}\simeq j_{\rm max}^{\rm(dark)}$ of \citet{Fall:1980} holds within our errors. Also, the averaged value of $\eta_{\rm j}$ when the three disc types are considered together 
$\langle \eta_{\rm j}^{\rm disk}\rangle=0.86\pm0.30$
is in good agreement with the observational disc value of $\rm$0.8 estimated by FR13.

Figure~\ref{figure_summary2} shows the temporal properties used to quantify the assembly of stellar structures. 
The panels give the average redshift of half mass inside the main halo progenitor $\langle z_{\rm 1/2}\rangle$ (left, from Equation~\ref{z50eq}), 
the average assembly timescale $\langle\tau\rangle$ (centre, Equation~\ref{tau_eq}), and the average skewness of the assembly timescale $\langle\kappa\rangle$ (right, Equation~\ref{kappa_eq}), 
respectively, as functions of $\langle f_{\rm j}\rangle$ for the baryonic (top) and stellar (bottom) progenitors of the eight types of kinematic structures. 
In each panel, the corresponding properties for the full stellar populations are given as grey symbols.   

The progenitor baryons of stellar spheroids are the earliest to be incorporated to the dark matter haloes, $\langle z_{\rm 1/2}\rangle_{\rm bar}\sim3$, 
their assembly time scales are the shortest, $\langle\tau\rangle_{\rm bar}\sim3.4$~Gyr, 
and their assembly is twice as fast before $\langle z_{\rm 1/2}\rangle_{\rm bar}$ than after, $\rm\langle\kappa\rangle_{\rm bar}\sim0.5$. 
At the other extreme, the progenitor baryons of stellar haloes are incorporated the latest, $\langle z_{\rm 1/2}\rangle_{\rm bar}\sim1.1$, 
their assemblies have the longest timescales, $\langle\tau\rangle_{\rm bar}\sim6.2$~Gyr, and proceed faster at late times, $\langle\kappa\rangle_{\rm bar}\sim1.6$. 
The large dispersion of $\langle\kappa\rangle_{\rm bar}$ for the stellar haloes is an indicator of a large variety in assembly patterns. 

The $\langle z_{\rm 1/2}\rangle_{\rm bar}$ for thin discs is smaller than for thick ones. Both are significantly less than $\langle z_{\rm 1/2}\rangle_{\rm bar}$ of single large scale discs. 
On average, thin and thick discs incorporate their baryons in almost equal proportion between late ($z<z_{\rm 1/2}$) and early times ($z>z_{\rm 1/2}$). 
The assembly of large scale single discs is clearly biased towards early times $\langle\kappa\rangle_{\rm bar}=0.68\pm0.46$. 
Therefore, regarding the baryonic assembly of large scale discs' progenitors, our results indicate that larger $z_{\rm 1/2(bar)}$ 
and a bias towards early assembly times (small $\kappa_{\rm bar}$) most likely result in a single large scale stellar dynamical disc rather than and a double thin/thick one. 

On average, the classical and pseudo bulges in this simulated galaxy sample accrete their progenitor baryons at earlier times than the thin and thick discs 
(larger vs smaller $z_{\rm 1/2}^{\rm bar}$) over similar time scales $\tau$. There is only a slight tendency of earlier baryonic assembly for the classical bulges as 
compared to the pseudo ones. The two types of bulges have very large dispersions for $\langle\kappa\rangle_{\rm bar}$ indicating a large variety in the individual 
assembly histories. 

The bottom panels in Figure~\ref{figure_summary2} show the same three quantities for the stellar assembly, 
namely $\langle z_{\rm 1/2}\rangle_{\rm star}$ (left), $\langle\tau\rangle_{\rm star}$ (center), $\langle\kappa\rangle_{\rm star}$ (right), respectively. 
As compared to the equivalent quantities for baryons in the top panels, most of the kinematic components have smaller dispersions. 
For the $\langle z_{\rm 1/2}\rangle_{\rm star}$ vs $\langle f_{\rm j}\rangle$ plane this leads to a clear anti-correlation. 
The component that stands out the most in this case is the thin disc with a significantly smaller half stellar mass assembly redshift, $\langle z_{\rm 1/2}\rangle=0.31\pm0.10$. 
Thick discs have significantly larger half stellar mass assembly redshift $\langle z_{\rm 1/2}\rangle=0.68\pm0.20$ than thin ones.
The two disc types have very similar stellar assembly time scales, $\langle\tau\rangle_{\rm star}\sim7.3$~Gyr.  
In the $\langle\kappa\rangle_{\rm star}$ parameter, all but the thin discs and the stellar haloes are either symmetric or biased towards early times. 
Given the large mass fractions of the in-situ born stars for all components but the stellar haloes, the $\kappa_{\rm star}$ property can be taken as
describing the SFR history. Thus, $\kappa_{\rm star}<1$ means higher levels of SFR before $z_{\rm 1/2}^{\rm star}$ than after, 
while the opposite is true if $\kappa_{\rm star}>1$. The farthest from one this number is, the more biased towards early/late times the SFR. 
The Milky Way analogue g8.26e11 is again an outlier with a thin disc constant SFR ($\kappa_{\rm star}\simeq1.0$).  

Based on the trends among the various stellar structures in their temporal properties as illustrated by Figure~\ref{figure_summary2}, 
an obvious future line of study is the metamorphosis of simulated galaxies' dynamical structures along their main branch of the merger tree. 
By running {\tt gsf} at different times in the evolution of a galaxy, it is possible to quantitatively constrain the migrations of stars from one component to the other, 
thus answering questions like how many of $z=0$ pseudo bulge stars were formed in a high-$z$ dynamical (thin/thick) disc?

\section{Conclusions}
\label{conclusions}

We use a subsample of 25 NIHAO galaxies \citep{Wang:2015} with stellar masses in the range $[7\times10^{\rm 8},2\times10^{\rm 11}]M_{\rm\odot}$ to look for and characterize
kinematic stellar structures. To define the structures we use {\tt galactic structure finder} \citep[{\tt gsf}, ][]{Obreja:2018} which employs the Gaussian Mixture Model
clustering algorithm in the 3D space of stellar normalized angular momentum (azimuthal and in-plane projections) and normalized binding energy.  
In this simulated galaxy sample, {\tt gsf} disentangled eight types of kinematic structures: large scale single discs, thin discs, thick discs, inner discs, 
spheroids, classical bulges, pseudo bulges and stellar haloes. The simulated galaxies range from having two to five distinct stellar structures.  
The masses of the structures are in good agreement with the observational data of: \citet{Yoachim:2006,Comeron:2014,Comeron:2018,DSouza:2014,Harmsen:2017} 
(Figures~\ref{figure_com2} and \ref{figure_thick2thin}), \citet{Martinsson:2013} (Figure~\ref{figure_vsig_1}), and \citet{Romanowsky:2012,Fall:2013} (Figure~\ref{figure_dyn}). We should emphasize that the dynamical stellar components 
obtained with {\tt gsf}, especially the inner discs and pseudo bulges, 
can differ substantially from those identified in observations using various definitions and criteria.    

By tracing back in time the Lagrangian mass enclosing all the progenitors of a given $z=0$ structure, 
we constructed the detail evolutions in a wide range of properties. They include: angular momentum (spin $j$, Figure~\ref{figure9_spin}), 
size (half mass 3D radius $r_{\rm 50}$, Figure~\ref{figure9_size}), shape (ellipticity $\varepsilon$ derived from the inertia tensor, Figure~\ref{figure9_shape}), 
rotational support (anisotropy of the global velocity dispersion $f_{\rm\sigma}$, Figure~\ref{figure9_rotsup}), 
or the baryonic and stellar mass fractions within the virial radius of the main branch dark matter halo progenitor, Figures~\ref{figure10_bar} and \ref{figure10_star}).

\subsection{The angular momentum evolution}

One of the most import results of this study is quantifying the evolution of angular momentum for each of the eight types of redshift $z=0$ stellar structures. 
The averaged evolution of the $j$s for the eight types of structures, the stars of the galaxies as a whole, and the host dark matter haloes are shown in Figure~\ref{figure9_spin}.
The cumulative distributions of the retained fractions of angular momentum $f_{\rm j}$ (Equation~\ref{fjdef}) are illustrated in Figure~\ref{fig_fj}.
Our results can be summarized as follows:
\begin{itemize}
 \item All types of stellar structures acquire their angular momentum before the time of the host dark matter halo turn around, mirroring the $j$-evolution of the dark matter, 
 as predicted by the tidal torque theory \citep{Hoyle:1951,Peebles:1969,Doroshkevich:1970,Fall:1980,White:1984}.
 \item The widely used assumption that the large scale structure torques imprint dark matter and baryons with roughly equal specific angular momentum 
 \citep{Fall:1980} is valid  when considering together all the progenitor baryons of $z=0$ galaxy stars. 
 However, among the individual components of $z=0$ galaxies, 
 only the thick and inner discs have $j_{\rm max}\simeq j_{\rm max}^{\rm(dark)}$. 
 The (thin) discs have a $\sim21\%$ higher value of maximum specific angular momentum than the dark matter, 
 while the bulges and spheroids have a $\sim16\%$ and $\sim34\%$ lower values, respectively. 
 The deviations from $j_{\rm max}=j_{\rm max}^{\rm(dark)}$ for the galaxy subcomponents can be explained 
 in terms of the differences in the collapse/assembly time of the progenitor material. The spheroids and bulges 
 assemble at higher redshifts, with $z(r_{\rm 1/2max})\sim5.7$ and $\sim4.1$ respectively, while the (thin/thick) discs and dark matter haloes collapse later on: $z(r_{\rm 1/2max})\sim3.7$. Given that angular momentum is mostly acquired prior to the 
 turn-around redshift ($z(r_{\rm 1/2max})\sim3.2$ for the thin discs and $\sim3.7$ for the dark matter), 
 the progenitor disc and dark matter halo material are  still gaining angular momentum when the progenitor of the 
 dispersion dominated components have already started their collapse.  
 \item For these 25 NIHAO galaxies the epoch of dark matter halo turn around is at $2\lesssim z_{\rm turn}\lesssim3$, and the virialization ends by $z_{\rm vir}\sim1$. 
 \item Between $z_{\rm turn}$ and $z=0$ all Lagrangian masses of $z=0$ stellar structures lose a fraction of their peak angular momentum. The biggest losses occur before $z_{\rm vir}$.
 \item The structures that lose the highest fraction of their spins are the classical bulges ($\langle f_{\rm j}\rangle=0.07\pm0.05$), 
 while the ones that lose the least are the thin discs ($\langle f_{\rm j}\rangle=0.69\pm0.18$). 
 \item The $\langle f_{\rm j}\rangle$ of thick discs are significantly smaller than those of thin discs ($\langle f_{\rm j}\rangle_{\rm thick}=0.41\pm0.14$). 
 The large scale single discs are in between the thin and thick ones with $\langle f_{\rm j}\rangle=0.55\pm0.13$.
 \item The dark matter follows a similar evolution as the stellar structures' progenitors, but retains a higher fraction of its maximum angular momentum ($\langle f_{\rm j}\rangle=0.75\pm0.15$).
 \item The retained fraction of angular momentum does not correlate with the stellar mass \citep[e.g.][]{Dutton:2012}. 
 \item At $z=0$, the thin discs have higher specific angular momentum than the dark matter haloes 
 ($\langle\eta_{\rm j}\rangle=1.30\pm0.76$). This can be explained as an effect of SNe II feedback, 
 which preferentially removes low angular momentum gas from the inner galaxy regions \citep{Dutton:2009a,Brook:2011}. 
 This gas is subsequently spinned  up in the corona and re-accreted at larger radii to fuel the stellar 
 disc growth \citep{Brook:2012}. 
 \item The average value of the retained fraction of angular momentum for all (thin/thick) discs, 
 estimated using the definition in Equation~\ref{fjdefRF12} is $\langle \eta_{\rm j}\rangle=0.86\pm0.30$, 
 close to the $\approx0.8$ estimated from observations by FR13. For the spheroid and 
 classical/pseudo bulges together $\langle \eta_{\rm j}\rangle=0.11\pm0.08$, close as well to the 
 FR13 value for spheroids $\approx0.1$. 
\end{itemize}

\subsection{Scaling relations at $z=0$}

\begin{itemize}
 \item The circular velocity $v_{\rm c}$ is tightly correlated with the maximum edge-on line-of-sight velocity of the stellar (thin) disc: 
$v_{\rm c} = 60.0 + 0.64 v_{\rm max} \pm 14.2$~[km~s$^{\rm -1}$], or alternatively $v_{\rm c} = (6.4 \pm 1.4)v_{\rm max}^{0.64 \pm 0.04}$~[km~s$^{\rm -1}$].
 \item The angular momenta of the discs and thin discs scale tightly ($r_{\rm P}=0.97$) with the stellar mass according to $J_{\rm *}=3.4M_{\rm *}^{\rm 1.26 \pm 0.06}$. For the halo mass range covered by our simulation sample, the exponent of this relation is in very good agreement with the scaling for 
 dark matter haloes $j_{\rm h} = \sqrt{2G}(\frac{3}{4\pi\Delta\rho_{\rm crit}})^{\rm 1/6}\lambda M_{\rm h}^{\rm 2/3}$ converted to $j_{\rm *}$--$M_{\rm *}$ 
 using the semi-empirical model \textsc{emerge} \citep{Moster:2018}: $j_{\rm *}\sim M_{\rm *}^{\rm 0.24}$.
 \item All disc components at $z=0$ follow the relation $f_{\rm\sigma}=[1+(\varepsilon/(1-\varepsilon))^{\rm -1.8}]^{\rm -1}$ between the intrinsic shape $\varepsilon$
 and the amount of rotational support $f_{\rm\sigma}$. This prediction has recently been confirmed by observations \citep{Zhu:2018c}.
\end{itemize}

\subsection{Generic properties of $z=0$ stellar structures}

The redshift $z=0$ average properties of the eight types of stellar structures, and of the full stellar populations are listed in Table~\ref{table_z0} and illustrated in Figures~\ref{figure_summary1}
and \ref{figure_summary2}. Our most important findings are:
\begin{itemize}
 \item The 25 NIHAO galaxies have the stellar mass split roughly equal between the rotational dominated components and the velocity dispersion dominated ones.
 \item For the galaxies that have both a thin and a thick disc ($56\%$ of the sample), the masses of the two are also roughly equal (Figure~\ref{figure_thick2thin}). 
 \item Stellar haloes make up only a few percent of the total mass ($\langle f_{\rm mass}\rangle=0.04\pm0.02$).
 \item The intrinsic shapes $\varepsilon$ and rotational support $f_{\rm\sigma}$ correlate with the global fraction of angular momentum retained during a galaxy's evolution $f_{\rm j}$ 
 (bottom left and central panels of Figure~\ref{figure_summary1}). 
 The sequence from low to high values of the three properties is: classical bulges, spheroids, pseudo bulges, inner discs, stellar haloes, thick discs, 
 large scale single discs, and thin discs. In these three properties, the complete stellar populations most resemble the thick discs. 
 \item Star formation occurs mostly in-situ for all but the stellar haloes for which $\langle f_{\rm in-situ}\rangle=0.35\pm0.14$. 
 The thin discs have the highest fraction of stars born in-situ $\langle f_{\rm in-situ}\rangle=0.97\pm0.05$. 
 The classical and pseudo bulges have very similar in-situ fractions of $\sim0.95$.
 \item All components but the stellar haloes assemble their baryons in the main dark matter haloes over roughly similar timescales 
 ($\tau_{\rm bar}\sim4\pm2$~Gyr), the fastest ($3.37\pm1.11$~Gyr) being the spheroids, and the slowest ($4.85\pm0.67$~Gyr) the thin discs. 
 \item The redshift of half progenitor baryonic mass within the virial radius varies mildly from classical bulges, to pseudo bulges, 
 to thick discs and finally to thin discs ($z_{\rm 1/2(bar)}$ from $\sim2.3$ to $\sim1.5$).
 Spheroids have the largest $\langle z_{\rm 1/2(bar)}\rangle=2.98\pm0.91$. 
 The large scale single discs have $\langle z_{\rm 1/2(bar)}\rangle$ bigger than both thin and thick discs. 
 \item Similar tendencies are visible for the half stellar mass redshift $\langle z_{\rm 1/2(star)}\rangle$. 
 The thin discs with a $\langle z_{\rm 1/2(star)}\rangle$ significantly smaller than any other component are the outliers.     
 \item The stellar haloes have their progenitor baryons assembled the latest ($\langle z_{\rm 1/2(bar)}\rangle=1.13\pm0.49$) 
 and over the longest timescales ($\langle\tau\rangle_{\rm bar}=6.19\pm1.59$~Gyr). 
 \item Considering the skewness of stellar assembly timescales $\kappa_{\rm star}$ to be also an indicator of the SFR history, 
 the SFR histories of the thin discs in the current galaxy sample range from constant SFRs ($\kappa=1$) to SFRs increasing with time ($\kappa>1$).
 For all other structures but the stellar haloes the SFR histories tend to be biased towards early times.
 \item Most of the dispersions of the temporal properties $z_{\rm 1/2}$, $\tau$ and $\kappa$ are large. This translates into a considerable overlap in stellar ages among the various kinematic components.  
\end{itemize}

\subsection{Future prospects}

Given the tight scaling relations followed by the kinematic large scale discs (single, thin and thick) we are confident that {\tt gsf} is reliable for these kind of structures.  
Regarding the classification of the inner galactic components, we do not claim to have put the final nail. 
As documented extensively by \citet{Kormendy:2004}, the morphological diversity of the inner components of observed galaxies is quite off putting.
For this reason and for the sake of simplicity we either named our inner kinematic components classical or pseudo bulges, or even spheroids. 
This topic surely deserves more detailed studies.

One obvious question is to what extent stellar kinematic structures are stable against an increase in resolution. 
In this regard, \citet{Buck:2018a} have re-simulated at higher resolution some of the most massive NIHAO galaxies, which we will use to study the resolution effect on stellar structures. 
Another import question is whether the generic properties of stellar structures uncovered in the NIHAO sample are common to galaxies simulated with different codes, 
like for example the very high resolution Auriga \citep{Grand:2017} or FIRE \citep{Hopkins:2017} galaxies. 
The frame work described in this study might also prove useful in revealing additional constraints for cosmological simulated Local Group galaxy analogues \citep[e.g.][]{Sawala:2016}.

The kinematic structures of observed galaxies can only be inferred through dynamical modeling. 
In this respect we are looking to bridge the gap between theoretical studies like the current one and observations. 
Thus, we intend to use mock photometry fitting on galaxy simulations to search for quantitative links between morphological tracers of stellar structures and the intrinsic kinematic ones.
The possible imprints of stellar dynamics on the abundance space (e.g. [Fe/H], [$\rm\alpha$/Fe]) are another interesting line of study, 
which can help in interpreting the results of large spectroscopic surveys of Milky Way stars. 

On the observational side, the very recent results of \citet{Zhu:2018a, Zhu:2018b} show that the inner regions ($\leq1R_{\rm e}$) 
of large samples of observed galaxies can be dynamically analyzed in an unified manner by means of Schwarzschild modeling \citep{Schwarzschild:1979}. 
The revised dynamical Hubble diagram proposed by these authors opens the possibility for a much more straightforward comparison of simulations and observations. 
The results of such modeling can be compared more directly with cosmological simulations analysed in a similar manner to the one put forward in this study, 
to build up a coherent and comprehensive picture of galaxy formation and evolution.

\section*{Acknowledgments}

We thank the anonymous referee for helping us to improve the clarity of the manuscript. 
We would also like to thank Rosa Dom\'{\i}nguez Tenreiro, Chris Brook, Fabrizio Arrigoni Battaia and Roberto Decarli for useful discussions. 
This research was carried out on the High Performance Computing resources at New York University Abu Dhabi; 
on the \textsc{theo} cluster of the Max-Planck-Institut f\"{u}r Astronomie and on the \textsc{hydra} clusters at the Rechenzentrum in Garching. 
We greatly appreciate the contributions of these computing allocations.
All figures in this work have been made with {\tt matplotlib} \citep{Hunter:2007}. 
The {\tt gsf} code uses {\tt pynbody} \citep{Pontzen:2013} to load the simulations and  {\tt scikit-learn} \citep{Pedregosa:2011} to run the clustering algorithm, 
as well as the Python libraries {\tt numpy} \citep{Walt:2011} and {\tt scipy} \citep{Jones:2001}. 
{\tt F2PY} \citep{Peterson:2009} has been used to compile the {\tt gsf} Fortran module for Python.
AO and BM have been funded by the Deutsche Forschungsgemeinschaft (DFG, German Research Foundation) -- MO 2979/1-1.
TB acknowledges support from the  Sonderforschungsbereich SFB 881 ``The Milky Way System'' (subproject A1) of the DFG.
GvdV acknowledges funding from the European Research Council (ERC) under the European Union's Horizon 2020 research 
and innovation programme under grant agreement No 724857 (Consolidator Grant ArcheoDyn).

\bibliographystyle{mnras}
\bibliography{paperII}

\appendix

\section{Atlas of the stellar kinematic components}
\label{appendix}

\appendix
\renewcommand{\thefigure}{A\arabic{figure}}
\renewcommand{\thetable}{A\arabic{table}}
\renewcommand{\theequation}{A\arabic{equation}}
\setcounter{figure}{0}
\setcounter{table}{0}
\setcounter{equation}{0}

\begin{figure*}
\includegraphics[width=0.98\textwidth]{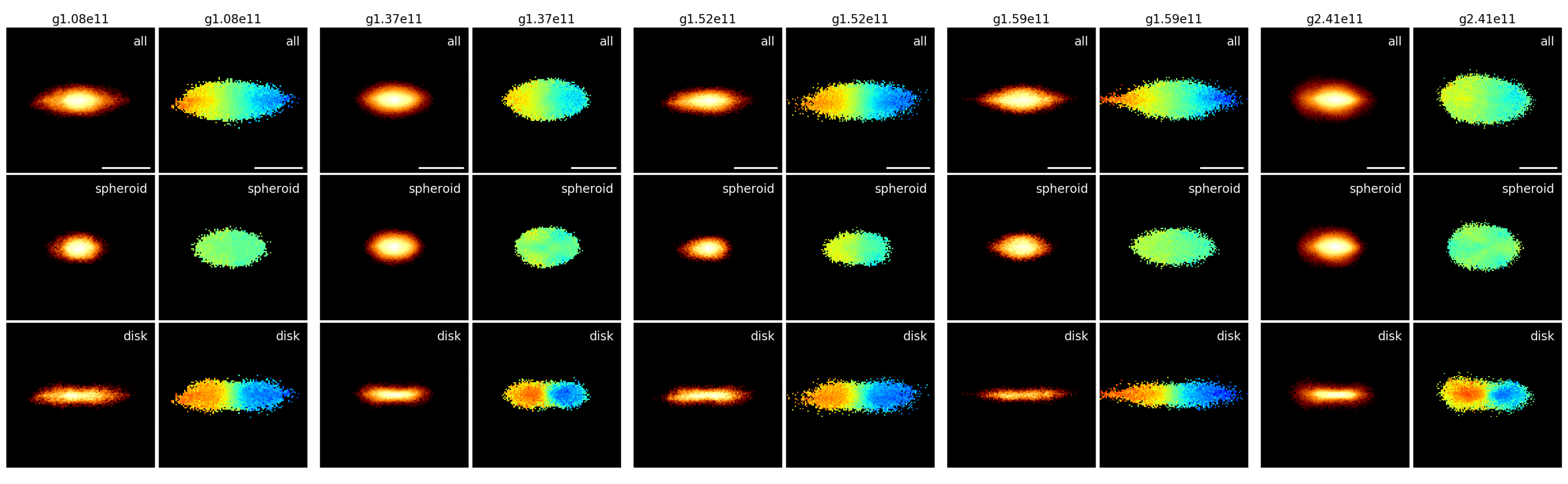}\vspace{0.5cm}
\includegraphics[width=0.98\textwidth]{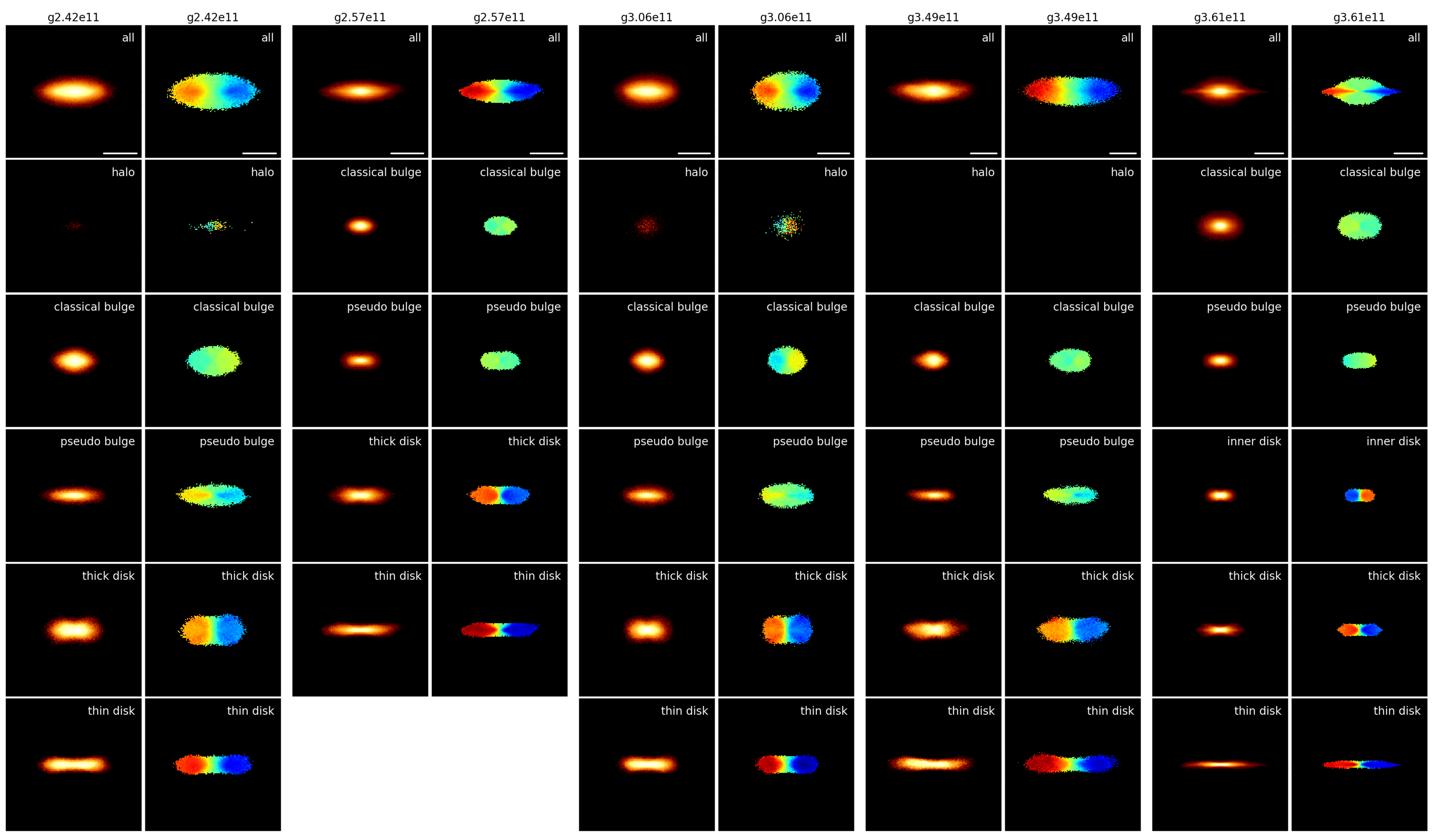}
\caption{The edge-on surface mass density (odd columns) and mass-weighted line-of-sight velocity maps (even columns) for the various kinematic components of the galaxies (names at the top of each column) 
in the simulated sample. The top row shows the whole stellar maps, labeled 'all'. The labels in each panel give the name of the corresponding component. The horizontal lines represent the physical scale of 10 kpc.}
\label{figure_allsampledeco}
\end{figure*}

\renewcommand{\thefigure}{A\arabic{figure} (continued)}
\addtocounter{figure}{-1}

\begin{figure*}
\includegraphics[width=0.98\textwidth]{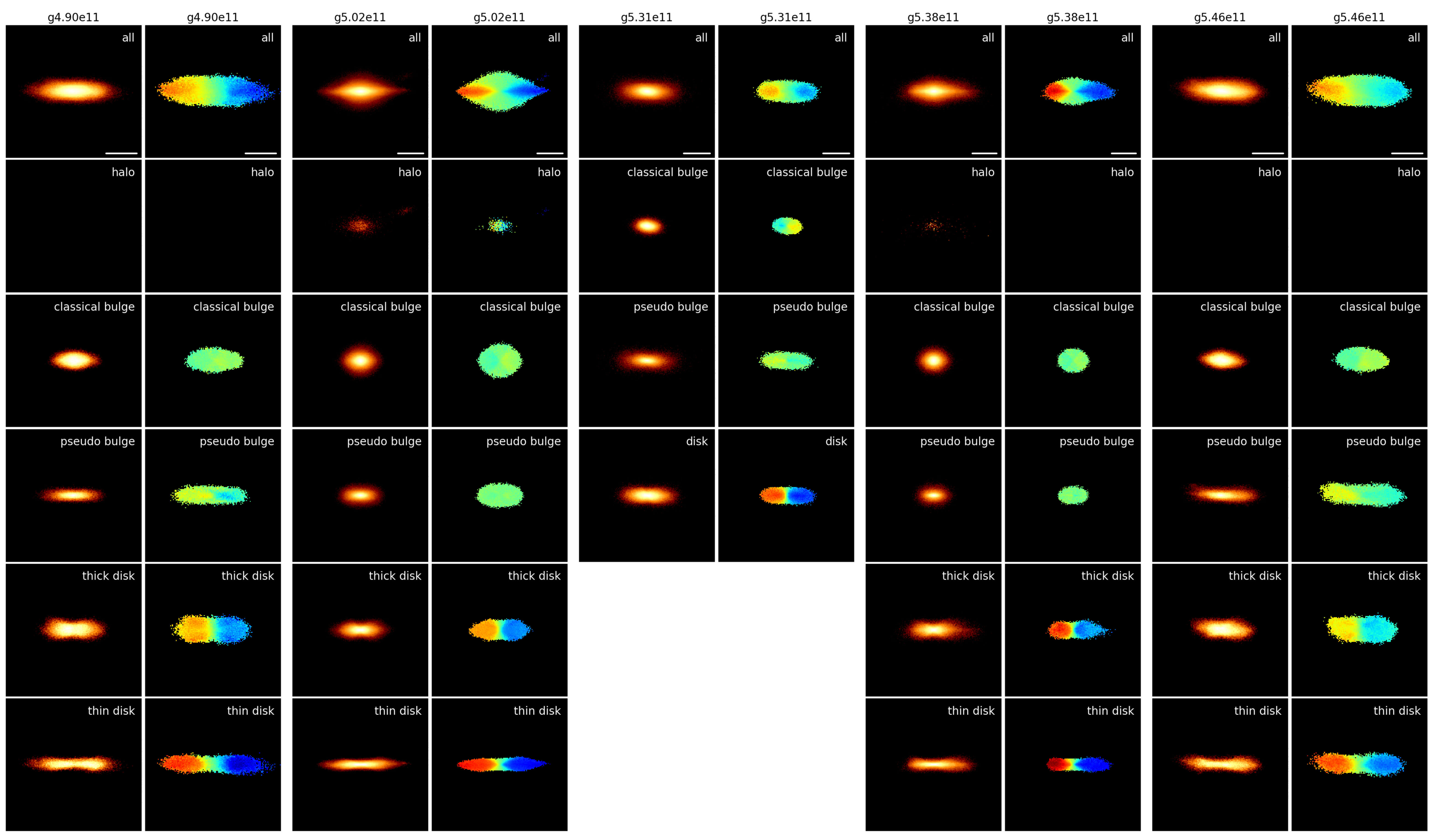}\vspace{0.5cm}
\includegraphics[width=0.98\textwidth]{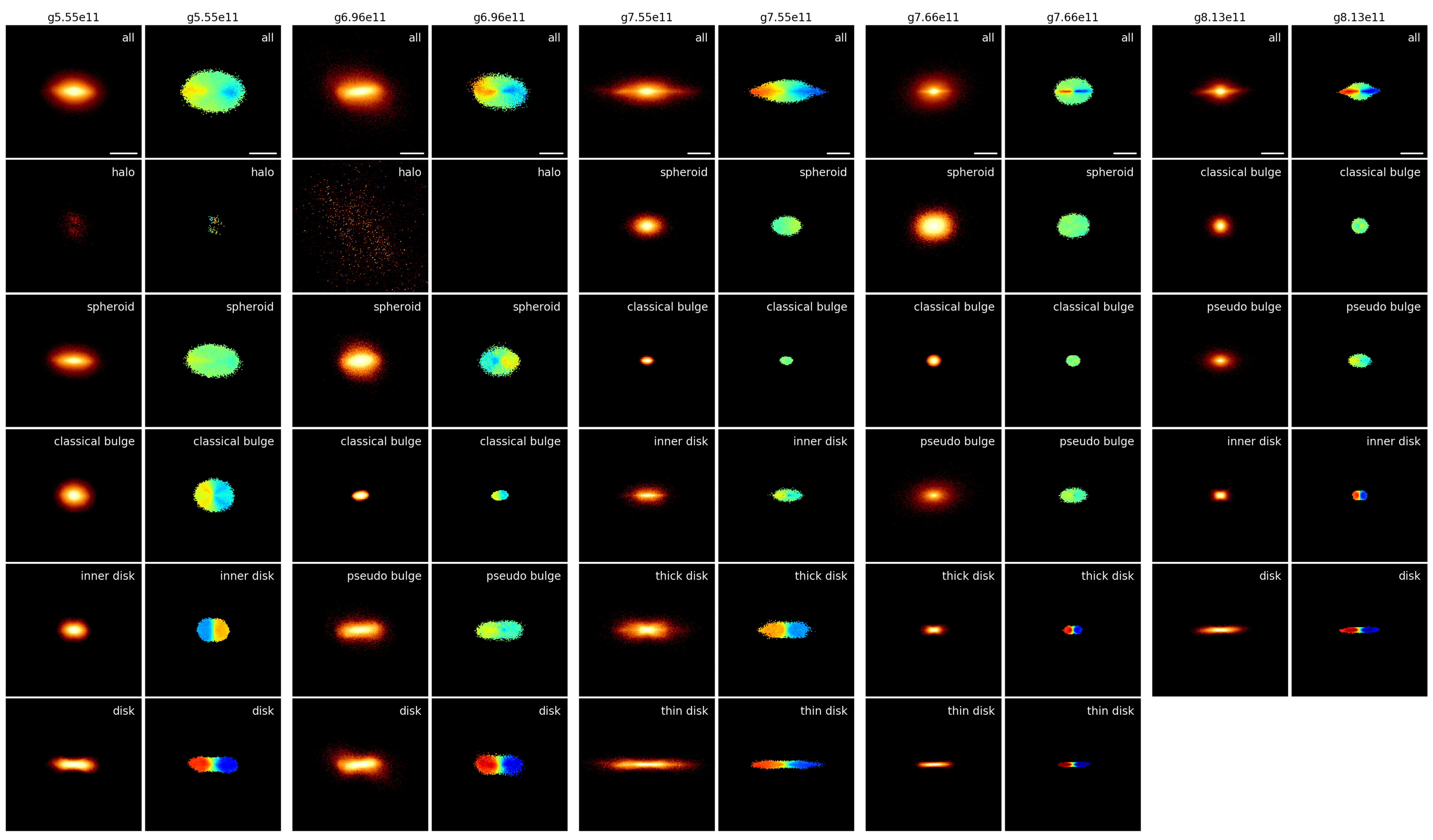}
\caption{}
\end{figure*}

\addtocounter{figure}{-1}

\begin{figure*}
\includegraphics[width=0.98\textwidth]{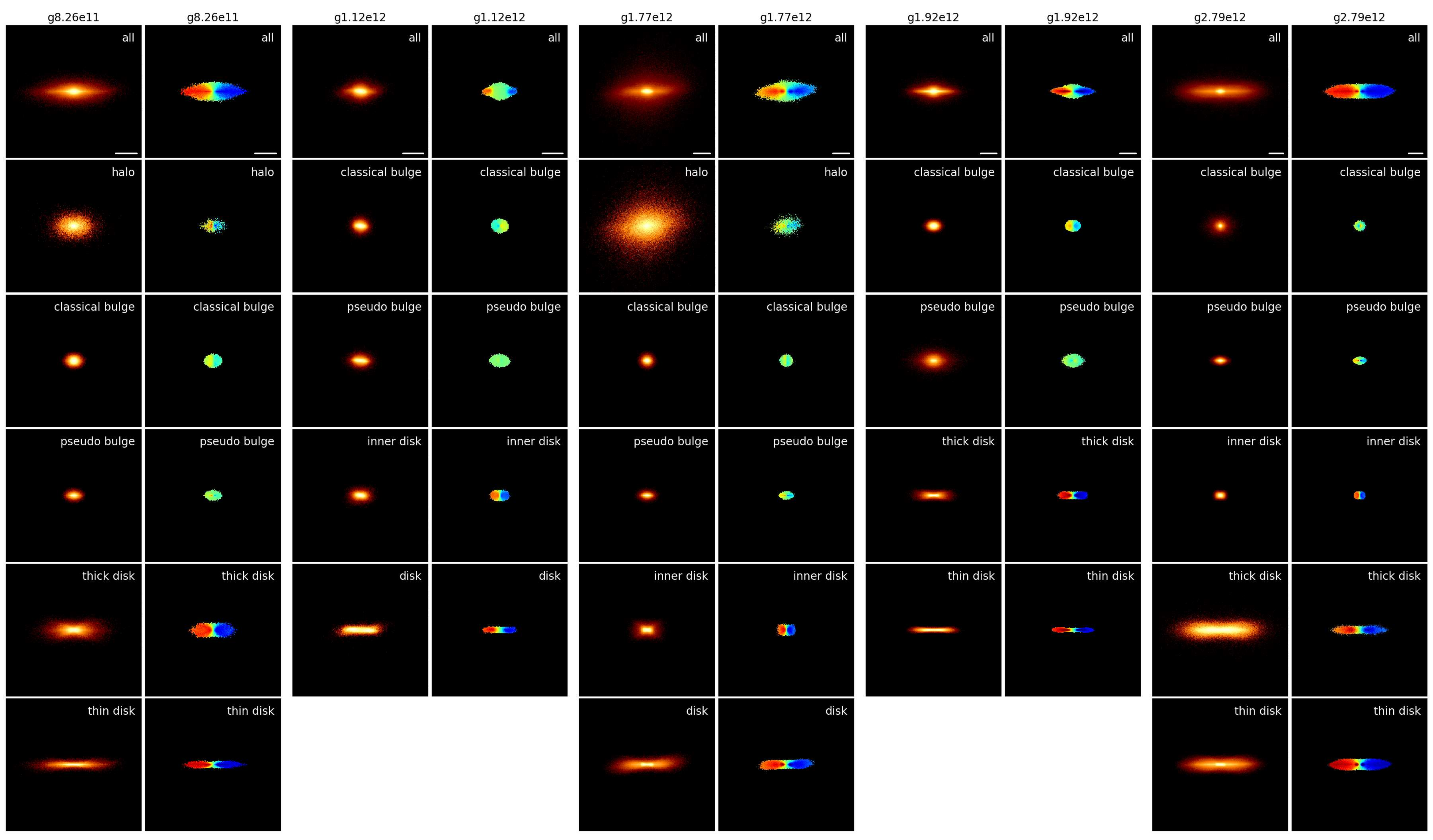}
\caption{}
\end{figure*}

\renewcommand{\thefigure}{A\arabic{figure}}

\begin{figure*}
\includegraphics[width=0.98\textwidth]{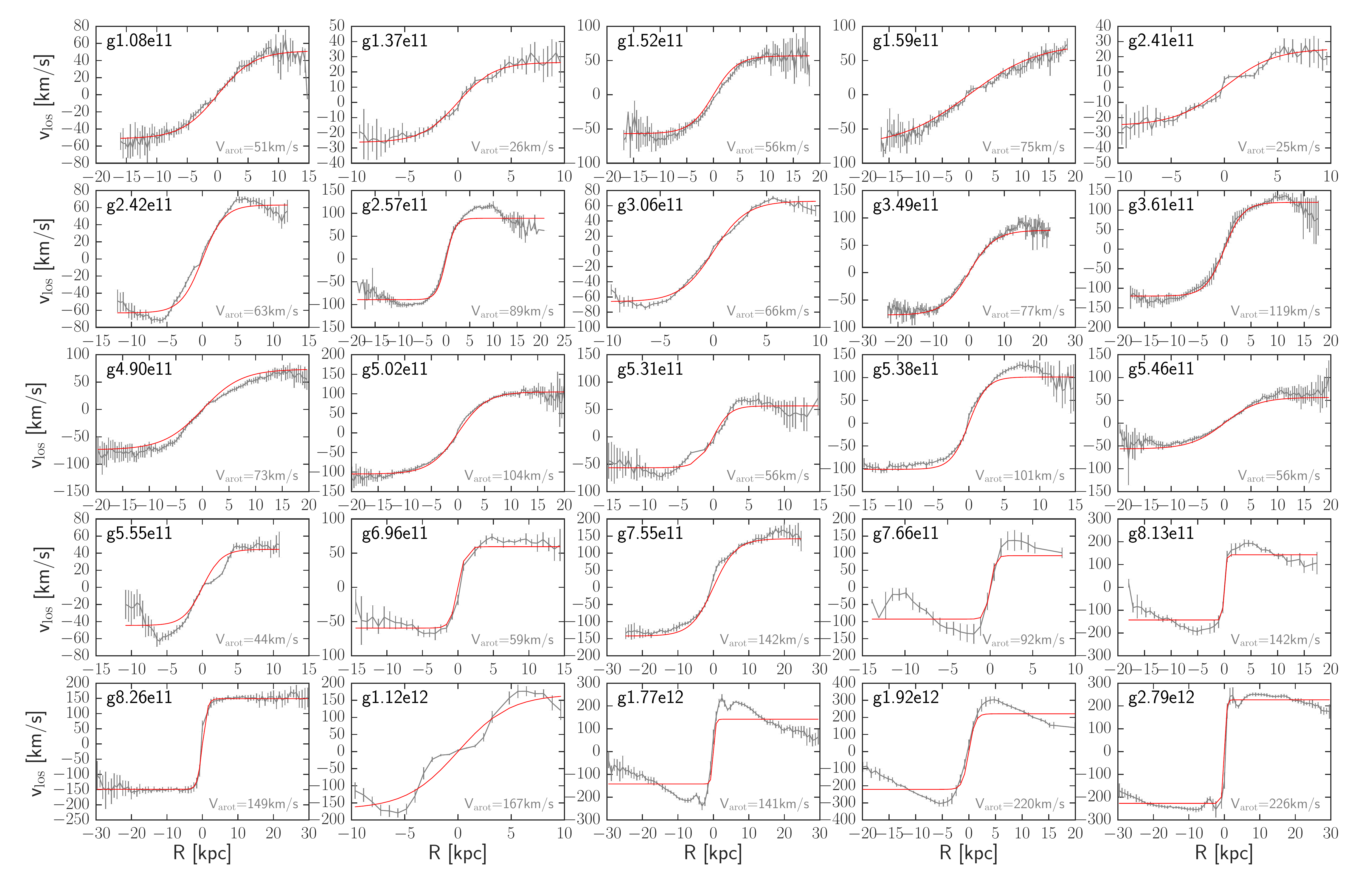}
\includegraphics[width=0.98\textwidth]{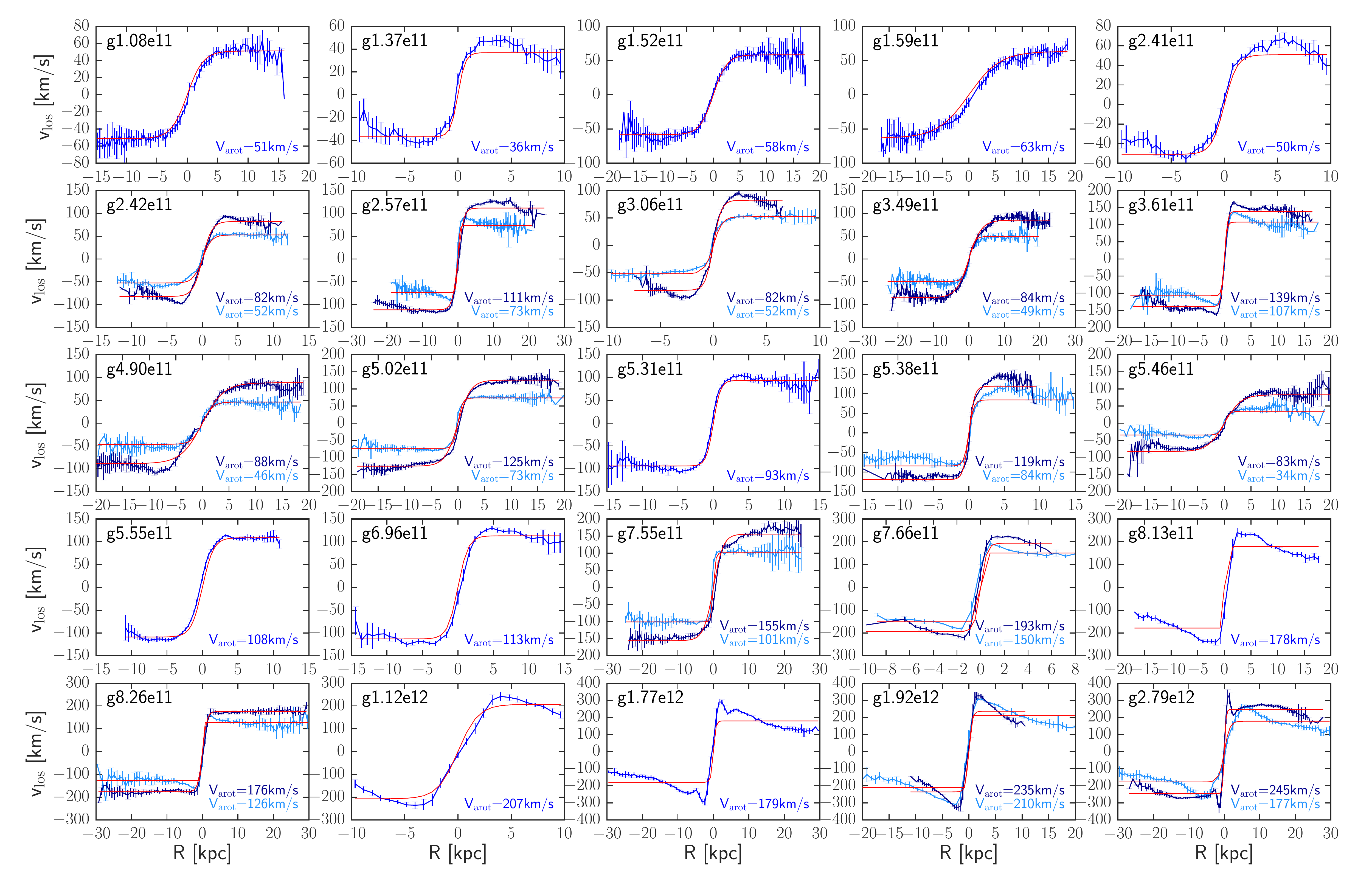}
\caption{The edge-on line-of-sight velocity $v_{\rm los}$ profiles for the complete stellar distributions (top) and 
for the kinematic disc components (bottom) of the galaxies in the simulated sample. 
The red curves give the hyperbolic tangent fits $\rm v_{\rm los}=V_{\rm arot}tanh(R/R_{\rm s})$ \citep[e.g.][]{Martinsson:2013}. 
The values of the fits asymptotic values are shown in each panel as $V_{\rm arot}$.}
\label{figure_appendix_3a}
\end{figure*}

\begin{figure*}
\includegraphics[width=0.98\textwidth]{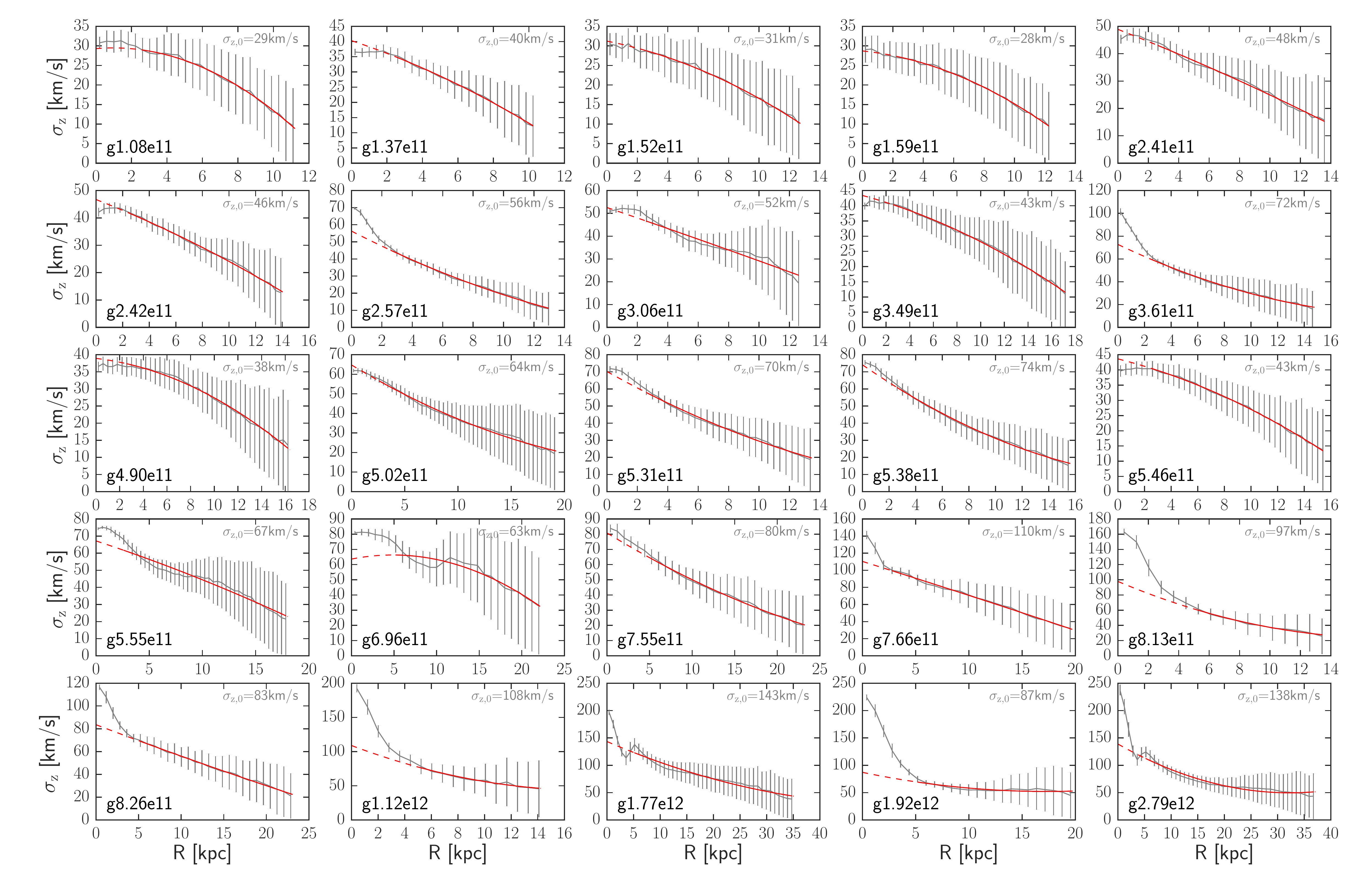}
\includegraphics[width=0.98\textwidth]{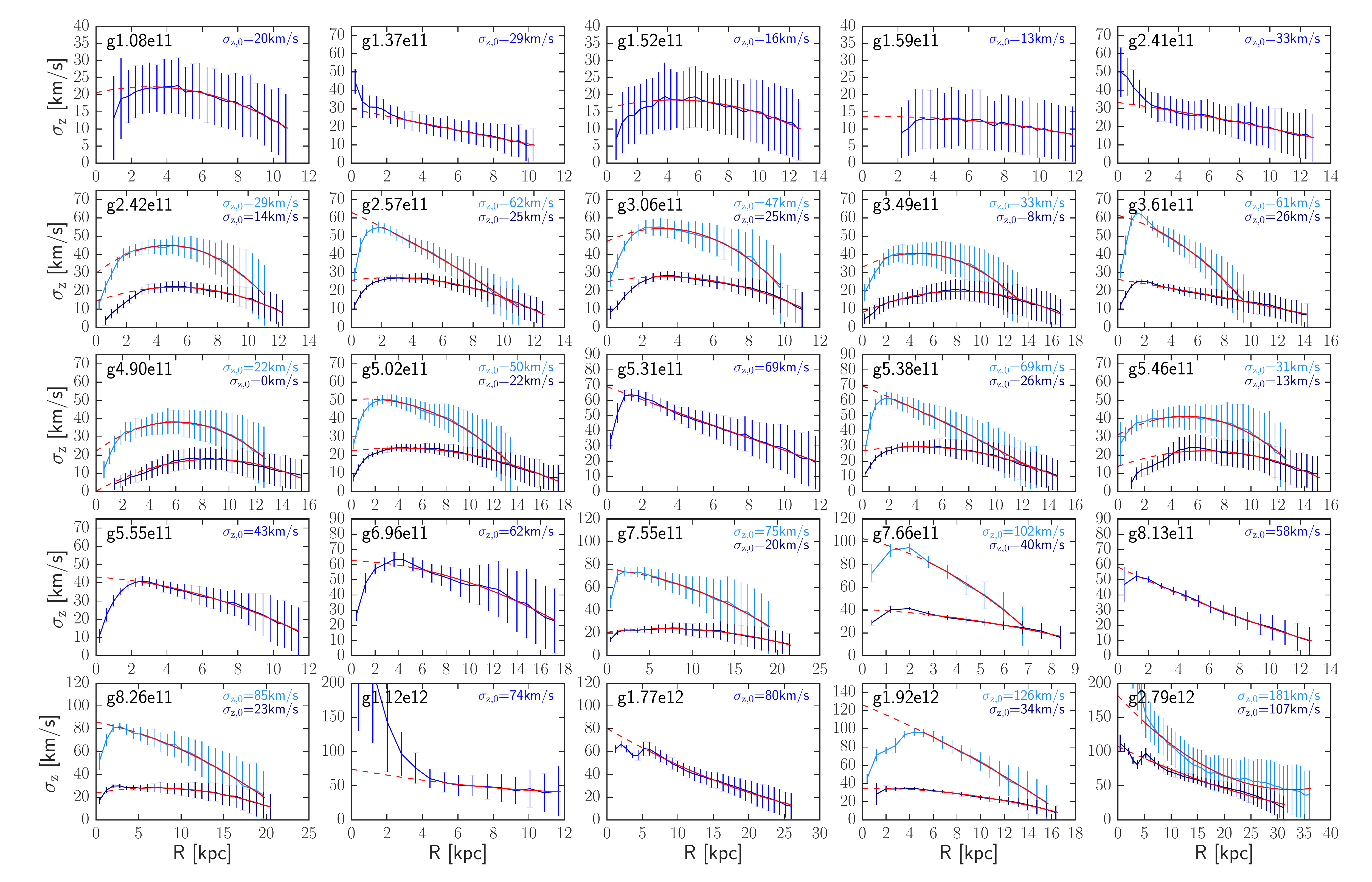}
\caption{The vertical velocity dispersion $\sigma_{\rm z}$ profiles for the complete stellar distributions (top) and 
for the kinematic disc components (bottom) of the galaxies in the simulated sample. 
The red curves give the second order polynomial fits to the profiles. The very inner region where the red curves are dashed has not been used for the fit. 
The values of the extrapolation of the fits to $R=0$ are shown in each panel as $\sigma_{\rm z,0}$.}
\label{figure_appendix_2a}
\end{figure*}

\renewcommand{\tabcolsep}{3pt}
\begin{table*}
\centering
\begin{tabular}{ccccccccccccccccccccc}
\hline
Sim & component & $\langle j_z/j_c\rangle$ & $\langle j_p/j_c\rangle$ & $\langle e/|e|_{\rm max} \rangle$ & $\varepsilon$ & $f_{\sigma}$ & log($J_z$) & $f_j$ & $f_{\rm in-situ}$ & $z_{\rm 1/2(bar/star)}$ & $\tau_{\rm bar/star}$ & $\kappa_{\rm bar/star}$\\ 
\hline
g1.08e11 & all & - & - & - & 0.42 & 0.50 & 11.22 & 0.32 & 0.96 & 3.60/1.47 & 3.45/8.42 & 0.23/0.44\\ 
g1.08e11 & spheroid & 0.08 $\pm$ 0.39 & 0.44 $\pm$ 0.23 & -0.86 $\pm$ 0.07 & 0.28 & 0.14 & 9.93 & 0.05 & 0.99 & 4.02/1.88 & 3.24/6.04 & 0.15/0.40\\ 
g1.08e11 & disk & 0.64 $\pm$ 0.24 & 0.28 $\pm$ 0.16 & -0.71 $\pm$ 0.13 & 0.63 & 0.74 & 11.20 & 0.51 & 0.92 & 2.96/1.00 & 3.67/9.93 & 0.42/0.70\\ 
g1.37e11 & all & - & - & - & 0.42 & 0.44 & 11.24 & 0.30 & 0.98 & 3.60/1.31 & 3.24/9.93 & 0.25/0.44\\ 
g1.37e11 & spheroid & 0.07 $\pm$ 0.41 & 0.44 $\pm$ 0.22 & -0.83 $\pm$ 0.08 & 0.33 & 0.21 & 10.38 & 0.07 & 0.99 & 3.60/1.39 & 3.02/9.28 & 0.27/0.39\\ 
g1.37e11 & disk & 0.44 $\pm$ 0.33 & 0.21 $\pm$ 0.12 & -0.74 $\pm$ 0.12 & 0.64 & 0.77 & 11.17 & 0.71 & 0.95 & 3.60/1.06 & 3.88/10.57 & 0.20/0.58\\ 
g1.52e11 & all & - & - & - & 0.61 & 0.68 & 11.74 & 0.49 & 0.84 & 3.25/0.90 & 6.90/9.93 & 0.14/0.84\\ 
g1.52e11 & spheroid & 0.26 $\pm$ 0.41 & 0.41 $\pm$ 0.22 & -0.85 $\pm$ 0.08 & 0.37 & 0.29 & 10.47 & 0.16 & 0.91 & 3.60/1.47 & 3.67/8.63 & 0.31/0.43\\ 
g1.52e11 & disk & 0.74 $\pm$ 0.25 & 0.26 $\pm$ 0.15 & -0.63 $\pm$ 0.19 & 0.70 & 0.82 & 11.71 & 0.54 & 0.78 & 2.96/0.62 & 8.20/10.14 & 0.15/1.24\\ 
g1.59e11 & all & - & - & - & 0.51 & 0.56 & 11.31 & 0.44 & 0.93 & 2.72/1.39 & 2.37/8.42 & 0.83/0.34\\ 
g1.59e11 & spheroid & 0.14 $\pm$ 0.41 & 0.42 $\pm$ 0.23 & -0.88 $\pm$ 0.06 & 0.38 & 0.18 & 10.36 & 0.14 & 0.95 & 2.96/1.47 & 2.16/6.04 & 0.67/0.56\\ 
g1.59e11 & disk & 0.70 $\pm$ 0.25 & 0.22 $\pm$ 0.13 & -0.75 $\pm$ 0.12 & 0.74 & 0.83 & 11.26 & 0.63 & 0.89 & 2.32/1.06 & 2.81/9.71 & 0.86/0.45\\ 
g2.41e11 & all & - & - & - & 0.40 & 0.44 & 11.56 & 0.09 & 0.94 & 1.56/0.46 & 4.75/9.06 & 1.00/1.21\\ 
g2.41e11 & spheroid & -0.00 $\pm$ 0.44 & 0.41 $\pm$ 0.22 & -0.81 $\pm$ 0.09 & 0.33 & 0.29 & 10.51$^{\rm c}$ & 0.01 & 0.98 & 1.77/0.46 & 4.75/9.06 & 0.69/1.21\\ 
g2.41e11 & disk & 0.38 $\pm$ 0.36 & 0.20 $\pm$ 0.12 & -0.69 $\pm$ 0.16 & 0.60 & 0.74 & 11.60 & 0.29 & 0.85 & 1.24/0.44 & 4.32/8.85 & 1.86/1.05\\ 
g2.42e11 & all & - & - & - & 0.51 & 0.60 & 12.16 & 0.51 & 0.97 & 2.96/0.85 & 4.10/9.28 & 0.27/0.79\\ 
g2.42e11 & halo & -0.06 $\pm$ 0.24 & 0.20 $\pm$ 0.15 & -0.48 $\pm$ 0.16 & 0.59 & 0.62 & 9.74$^{\rm c}$ & 0.07 & 0.38 & 1.18/1.18 & 5.18/3.88 & 3.00/17.00\\ 
g2.42e11 & classical b & -0.09 $\pm$ 0.37 & 0.47 $\pm$ 0.21 & -0.85 $\pm$ 0.08 & 0.28 & 0.10 & 10.91$^{\rm c}$ & 0.12 & 0.99 & 3.60/1.56 & 2.37/6.90 & 0.38/0.45\\ 
g2.42e11 & pseudo b & 0.36 $\pm$ 0.32 & 0.14 $\pm$ 0.07 & -0.77 $\pm$ 0.11 & 0.58 & 0.78 & 11.27 & 0.41 & 0.99 & 2.96/0.85 & 3.88/8.42 & 0.29/0.95\\ 
g2.42e11 & thick disk & 0.58 $\pm$ 0.17 & 0.38 $\pm$ 0.16 & -0.77 $\pm$ 0.10 & 0.49 & 0.46 & 11.65 & 0.61 & 0.99 & 2.96/0.85 & 4.10/8.20 & 0.36/0.90\\ 
g2.42e11 & thin disk & 0.84 $\pm$ 0.09 & 0.21 $\pm$ 0.11 & -0.72 $\pm$ 0.09 & 0.78 & 0.89 & 11.95 & 0.96 & 1.00 & 2.50/0.44 & 4.10/9.06 & 0.46/1.33\\ 
g2.57e11 & all & - & - & - & 0.49 & 0.58 & 12.41 & 0.48 & 0.96 & 2.32/0.85 & 4.53/8.85 & 0.40/0.78\\ 
g2.57e11 & classical b & -0.02 $\pm$ 0.36 & 0.46 $\pm$ 0.20 & -0.78 $\pm$ 0.10 & 0.26 & 0.08 & 10.76$^{\rm c}$ & 0.05 & 0.93 & 2.96/1.31 & 2.81/5.83 & 0.30/0.93\\ 
g2.57e11 & pseudo b & 0.15 $\pm$ 0.36 & 0.16 $\pm$ 0.08 & -0.71 $\pm$ 0.15 & 0.47 & 0.63 & 10.66 & 0.06 & 0.93 & 2.72/1.06 & 3.24/6.69 & 0.36/1.07\\ 
g2.57e11 & thick disk & 0.66 $\pm$ 0.14 & 0.31 $\pm$ 0.15 & -0.63 $\pm$ 0.15 & 0.60 & 0.67 & 11.94 & 0.63 & 0.97 & 2.16/0.77 & 4.10/7.55 & 0.58/1.19\\ 
g2.57e11 & thin disk & 0.89 $\pm$ 0.06 & 0.19 $\pm$ 0.10 & -0.59 $\pm$ 0.15 & 0.81 & 0.91 & 12.23 & 0.69 & 0.99 & 1.47/0.31 & 6.04/6.90 & 0.87/1.46\\ 
g3.06e11 & all & - & - & - & 0.44 & 0.50 & 12.17 & 0.31 & 0.97 & 2.01/0.44 & 4.75/9.49 & 0.69/1.44\\ 
g3.06e11 & halo & -0.10 $\pm$ 0.28 & 0.37 $\pm$ 0.24 & -0.48 $\pm$ 0.17 & 0.30 & 0.16 & 9.20 & 0.11 & 0.48 & 1.24/1.00 & 6.69/3.88 & 1.07/5.00\\ 
g3.06e11 & classical b & -0.18 $\pm$ 0.35 & 0.50 $\pm$ 0.20 & -0.83 $\pm$ 0.09 & 0.25 & 0.03 & 11.12$^{\rm c}$ & 0.13 & 1.00 & 2.50/0.62 & 4.53/8.42 & 0.40/1.79\\ 
g3.06e11 & pseudo b & 0.19 $\pm$ 0.38 & 0.16 $\pm$ 0.08 & -0.76 $\pm$ 0.12 & 0.48 & 0.64 & 11.00 & 0.12 & 0.99 & 2.01/0.49 & 4.75/9.71 & 0.69/1.37\\ 
g3.06e11 & thick disk & 0.55 $\pm$ 0.18 & 0.42 $\pm$ 0.18 & -0.76 $\pm$ 0.11 & 0.41 & 0.33 & 11.66 & 0.34 & 1.00 & 2.01/0.41 & 4.75/9.06 & 0.69/1.63\\ 
g3.06e11 & thin disk & 0.83 $\pm$ 0.10 & 0.24 $\pm$ 0.12 & -0.70 $\pm$ 0.10 & 0.74 & 0.86 & 12.03 & 0.63 & 1.00 & 1.39/0.21 & 4.96/6.26 & 1.56/1.90\\ 
g3.49e11 & all & - & - & - & 0.46 & 0.55 & 12.22 & 0.47 & 0.96 & 1.88/0.65 & 4.75/9.71 & 0.83/0.96\\ 
g3.49e11 & halo & 0.14 $\pm$ 0.30 & 0.61 $\pm$ 0.28 & -0.36 $\pm$ 0.09 & 0.46 & -0.19 & 11.04 & 0.42 & 0.22 & 0.69/0.65 & 4.75/5.39 & 3.40/2.57\\ 
g3.49e11 & classical b & -0.06 $\pm$ 0.37 & 0.51 $\pm$ 0.21 & -0.87 $\pm$ 0.06 & 0.22 & -0.02 & 10.47$^{\rm c}$ & 0.04 & 0.99 & 2.96/1.39 & 3.24/7.55 & 0.50/0.52\\ 
g3.49e11 & pseudo b & 0.26 $\pm$ 0.35 & 0.16 $\pm$ 0.08 & -0.81 $\pm$ 0.09 & 0.54 & 0.68 & 10.79 & 0.12 & 0.99 & 2.01/0.73 & 4.10/9.71 & 0.90/0.88\\ 
g3.49e11 & thick disk & 0.57 $\pm$ 0.19 & 0.38 $\pm$ 0.17 & -0.76 $\pm$ 0.11 & 0.52 & 0.53 & 11.66 & 0.45 & 0.98 & 1.77/0.62 & 4.32/9.06 & 1.22/1.10\\ 
g3.49e11 & thin disk & 0.84 $\pm$ 0.11 & 0.20 $\pm$ 0.10 & -0.67 $\pm$ 0.11 & 0.78 & 0.90 & 12.02 & 0.78 & 0.99 & 1.56/0.25 & 4.96/7.34 & 0.92/1.83\\
g3.61e11 & all & - & - & - & 0.45 & 0.49 & 12.45 & 0.22 & 0.58 & $\rm>$1.4/1.18 & $\rm\lesssim$4.8/$\rm\gtrsim$5.8 & -/$\rm\lesssim$1.0\\ 
g3.61e11 & classical b & 0.11 $\pm$ 0.31 & 0.47 $\pm$ 0.20 & -0.69 $\pm$ 0.16 & 0.22 & -0.00 & 11.34 & 0.14 & 0.39 & $\rm>$1.4/$\rm>$1.4 & $\rm\lesssim$4.8/$\rm\lesssim$6.6 & -/$\rm<$2.1\\ 
g3.61e11 & pseudo b & -0.05 $\pm$ 0.38 & 0.18 $\pm$ 0.09 & -0.72 $\pm$ 0.13 & 0.38 & 0.54 & 11.06$^{\rm c}$ & 0.15 & 0.41 & $\rm>$1.4/$\rm>$1.4 & $\rm\lesssim$4.8/$\rm\lesssim$6.6 & -/$\rm<$2.4\\ 
g3.61e11 & inner disk & -0.61 $\pm$ 0.19 & 0.40 $\pm$ 0.18 & -0.73 $\pm$ 0.11 & 0.51 & 0.50 & 11.91$^{\rm c}$ & 1.00 & 0.47 & $\rm>$1.4/$\rm>$1.4 & $\rm\lesssim$4.8/$\rm\lesssim$6.0 & -/$\rm<$3.0\\ 
g3.61e11 & thick disk & 0.76 $\pm$ 0.11 & 0.30 $\pm$ 0.14 & -0.68 $\pm$ 0.16 & 0.67 & 0.73 & 12.03 & 0.35 & 0.81 & $\rm>$1.4/0.69 & $\rm\lesssim$5.1/4.96 & -/1.30\\ 
g3.61e11 & thin disk & 0.94 $\pm$ 0.04 & 0.13 $\pm$ 0.07 & -0.61 $\pm$ 0.17 & 0.87 & 0.96 & 12.39 & 0.38 & 0.98 & $\rm>$1.4/0.34 & $\rm\lesssim$6.1/5.39 & -/0.79\\ 
g4.90e11 & all & - & - & - & 0.57 & 0.60 & 12.13 & 0.36 & 0.90 & 1.39/0.55 & 5.18/8.42 & 1.18/0.86\\ 
g4.90e11 & halo & 0.15 $\pm$ 0.32 & 0.24 $\pm$ 0.14 & -0.49 $\pm$ 0.18 & 0.54 & 0.27 & 10.32 & 0.21 & 0.28 & 0.81/0.73 & 7.34/6.26 & 1.83/1.64\\ 
g4.90e11 & classical b & -0.04 $\pm$ 0.39 & 0.48 $\pm$ 0.22 & -0.88 $\pm$ 0.06 & 0.42 & 0.06 & 10.44$^{\rm c}$ & 0.03 & 0.93 & 1.47/0.77 & 4.53/7.34 & 1.33/1.00\\ 
g4.90e11 & pseudo b & 0.29 $\pm$ 0.35 & 0.14 $\pm$ 0.07 & -0.81 $\pm$ 0.09 & 0.63 & 0.81 & 11.04 & 0.18 & 0.94 & 1.39/0.49 & 5.18/8.42 & 1.18/1.05\\ 
g4.90e11 & thick disk & 0.59 $\pm$ 0.19 & 0.37 $\pm$ 0.16 & -0.79 $\pm$ 0.09 & 0.59 & 0.51 & 11.59 & 0.39 & 0.93 & 1.47/0.59 & 4.96/7.98 & 1.09/0.85\\ 
g4.90e11 & thin disk & 0.83 $\pm$ 0.11 & 0.19 $\pm$ 0.10 & -0.67 $\pm$ 0.13 & 0.81 & 0.91 & 11.93 & 0.86 & 0.89 & 1.06/0.23 & 5.39/7.77 & 2.12/2.27\\ 
g5.02e11 & all & - & - & - & 0.43 & 0.49 & 12.68 & 0.21 & 0.92 & 1.66/0.73 & 2.16/7.34 & 2.33/0.55\\ 
g5.02e11 & halo & 0.12 $\pm$ 0.28 & 0.27 $\pm$ 0.18 & -0.45 $\pm$ 0.14 & 0.43 & 0.49 & 11.74 & 0.43 & 0.53 & 1.77/1.47 & 3.67/1.94 & 0.70/0.50\\ 
g5.02e11 & classical b & -0.10 $\pm$ 0.34 & 0.53 $\pm$ 0.18 & -0.79 $\pm$ 0.11 & 0.16 & -0.12 & 11.31$^{\rm c}$ & 0.03 & 0.88 & 1.47/0.90 & 2.16/3.45 & 9.00/1.00\\ 
g5.02e11 & pseudo b & 0.07 $\pm$ 0.40 & 0.18 $\pm$ 0.09 & -0.78 $\pm$ 0.10 & 0.40 & 0.55 & 10.20$^{\rm c}$ & 0.00 & 0.93 & 1.66/0.85 & 2.16/4.53 & 2.33/0.75\\ 
g5.02e11 & thick disk & 0.63 $\pm$ 0.16 & 0.35 $\pm$ 0.16 & -0.76 $\pm$ 0.11 & 0.55 & 0.60 & 12.16 & 0.24 & 0.96 & 1.66/0.59 & 1.94/6.04 & 2.00/0.87\\ 
g5.02e11 & thin disk & 0.89 $\pm$ 0.07 & 0.18 $\pm$ 0.10 & -0.68 $\pm$ 0.12 & 0.81 & 0.92 & 12.47 & 0.53 & 0.99 & 1.56/0.27 & 3.67/5.61 & 0.70/1.17\\ 
g5.31e11 & all & - & - & - & 0.46 & 0.50 & 12.41 & 0.17 & 0.83 & $\rm>$1.2/0.52 & $\rm\lesssim$6.7/$\rm\gtrsim$7.6 & -/$\rm\gtrsim$0.7\\ 
g5.31e11 & classical b & -0.19 $\pm$ 0.38 & 0.46 $\pm$ 0.20 & -0.79 $\pm$ 0.10 & 0.35 & 0.19 & 11.72$^{\rm c}$ & 0.08 & 0.87 & $\rm>$1.2/0.46 & $\rm\lesssim$7.1/$\rm\gtrsim$8.0 & -/$\rm\gtrsim$0.8\\ 
g5.31e11 & pseudo b & 0.16 $\pm$ 0.39 & 0.16 $\pm$ 0.09 & -0.66 $\pm$ 0.16 & 0.52 & 0.67 & 11.72 & 0.14 & 0.77 & $\rm>$1.2/0.62 & $\rm\lesssim$6.0/$\rm\gtrsim$6.7 & -/$\rm\gtrsim$0.5\\ 
g5.31e11 & disk & 0.61 $\pm$ 0.23 & 0.35 $\pm$ 0.16 & -0.66 $\pm$ 0.14 & 0.54 & 0.63 & 12.41 & 0.51 & 0.84 & $\rm>$1.2/0.52 & $\rm\lesssim$6.5/$\rm\gtrsim$6.5 & -/$\rm\gtrsim$0.9\\ 
g5.38e11 & all & - & - & - & 0.43 & 0.50 & 12.84 & 0.43 & 0.98 & 1.88/0.81 & 3.24/8.20 & 1.14/0.65\\ 
g5.38e11 & halo & 0.03 $\pm$ 0.34 & 0.23 $\pm$ 0.16 & -0.34 $\pm$ 0.13 & 0.46 & 0.46 & 10.77$^{\rm c}$ & 0.08 & 0.39 & 1.47/1.47 & 8.20/7.55 & 0.36/0.40\\ 
g5.38e11 & classical b & -0.05 $\pm$ 0.34 & 0.54 $\pm$ 0.18 & -0.78 $\pm$ 0.10 & 0.10 & -0.16 & 11.06$^{\rm c}$ & 0.07 & 0.99 & 2.72/1.24 & 2.81/4.10 & 0.44/1.37\\ 
\hline
\end{tabular}
\caption{The properties of the various stellar structures in the sample of 25 galaxies.
$J_z$ is given in M$_{\rm\odot}$~kpc~km~s$^{\rm -1}$, while $\tau_{\rm bar/star}$ are in Gyr. 
A $c$-superscript on the log($J_z$) value denotes counter-rotation.
The simulations g3.61e11 and g5.31e11 do not have information at high redshifts, and therefore their temporal properties $z_{\rm 1/2}$, $\tau$ and $\kappa$ have only upper/lower limits.}
\label{table_aux}
\end{table*}

\renewcommand{\thetable}{A\arabic{table} (continued)}
\addtocounter{table}{-1}

\begin{table*}
\centering
\begin{tabular}{ccccccccccccccccccccc}
\hline
Sim & component & $\langle j_z/j_c\rangle$ & $\langle j_p/j_c\rangle$ & $\langle e/|e|_{\rm max} \rangle$ & $\varepsilon$ & $f_{\sigma}$ & log($J_z$) & $f_j$ & $f_{\rm in-situ}$ & $z_{\rm 1/2(bar/star)}$ & $\tau_{\rm bar/star}$ & $\kappa_{\rm bar/star}$\\ 
\hline
g5.38e11 & pseudo b & 0.10 $\pm$ 0.40 & 0.18 $\pm$ 0.09 & -0.77 $\pm$ 0.10 & 0.35 & 0.48 & 10.61 & 0.03 & 0.99 & 2.32/1.06 & 3.02/4.75 & 0.75/1.20\\ 
g5.38e11 & thick disk & 0.64 $\pm$ 0.15 & 0.33 $\pm$ 0.16 & -0.67 $\pm$ 0.13 & 0.55 & 0.65 & 12.40 & 0.53 & 1.00 & 1.66/0.69 & 3.45/5.83 & 1.67/0.93\\ 
g5.38e11 & thin disk & 0.88 $\pm$ 0.07 & 0.21 $\pm$ 0.11 & -0.66 $\pm$ 0.13 & 0.77 & 0.89 & 12.66 & 0.62 & 1.00 & 1.47/0.36 & 4.75/6.47 & 1.00/0.87\\ 
g5.46e11 & all & - & - & - & 0.53 & 0.57 & 12.06 & 0.31 & 0.92 & 1.47/0.59 & 4.96/8.42 & 1.09/0.86\\ 
g5.46e11 & halo & -0.04 $\pm$ 0.24 & 0.25 $\pm$ 0.13 & -0.31 $\pm$ 0.10 & 0.54 & 0.31 & 9.94$^{\rm c}$ & 0.35 & 0.12 & 0.65/0.62 & 5.18/5.18 & 2.00/2.43\\ 
g5.46e11 & classical b & -0.12 $\pm$ 0.36 & 0.48 $\pm$ 0.21 & -0.86 $\pm$ 0.07 & 0.37 & 0.05 & 10.79$^{\rm c}$ & 0.08 & 0.95 & 1.56/0.77 & 4.75/7.34 & 1.20/0.89\\ 
g5.46e11 & pseudo b & 0.28 $\pm$ 0.34 & 0.14 $\pm$ 0.07 & -0.76 $\pm$ 0.10 & 0.61 & 0.79 & 11.24 & 0.21 & 0.95 & 1.47/0.46 & 4.96/8.20 & 1.09/1.24\\ 
g5.46e11 & thick disk & 0.45 $\pm$ 0.23 & 0.39 $\pm$ 0.15 & -0.79 $\pm$ 0.09 & 0.49 & 0.39 & 11.43 & 0.31 & 0.94 & 1.47/0.65 & 4.75/7.77 & 1.20/0.80\\ 
g5.46e11 & thin disk & 0.78 $\pm$ 0.13 & 0.23 $\pm$ 0.11 & -0.67 $\pm$ 0.14 & 0.75 & 0.86 & 11.89 & 0.77 & 0.88 & 1.31/0.34 & 5.18/8.63 & 1.18/1.50\\ 
g5.55e11 & all & - & - & - & 0.39 & 0.40 & 12.33 & 0.06 & 0.90 & 1.24/0.34 & 5.83/6.69 & 1.08/1.21\\ 
g5.55e11 & halo & -0.02 $\pm$ 0.31 & 0.48 $\pm$ 0.21 & -0.46 $\pm$ 0.20 & 0.19 & -0.11 & 11.44 & 0.69 & 0.47 & 1.39/0.95 & 8.85/4.32 & 0.46/1.50\\ 
g5.55e11 & spheroid & 0.06 $\pm$ 0.33 & 0.17 $\pm$ 0.09 & -0.68 $\pm$ 0.14 & 0.43 & 0.58 & 11.44 & 0.02 & 0.84 & 1.12/0.38 & 5.61/6.26 & 1.36/0.81\\ 
g5.55e11 & classical b & 0.30 $\pm$ 0.27 & 0.49 $\pm$ 0.19 & -0.74 $\pm$ 0.12 & 0.24 & 0.03 & 11.85 & 0.08 & 0.92 & 1.31/0.38 & 6.04/6.69 & 1.00/1.07\\ 
g5.55e11 & inner disk & -0.41 $\pm$ 0.26 & 0.46 $\pm$ 0.20 & -0.78 $\pm$ 0.10 & 0.39 & 0.27 & 11.84$^{\rm c}$ & 0.09 & 0.95 & 1.31/0.34 & 5.83/6.26 & 0.93/1.23\\ 
g5.55e11 & disk & 0.79 $\pm$ 0.12 & 0.26 $\pm$ 0.13 & -0.66 $\pm$ 0.11 & 0.71 & 0.84 & 12.19 & 0.28 & 0.97 & 1.18/0.17 & 5.39/5.39 & 1.08/2.12\\ 
g6.96e11 & all & - & - & - & 0.38 & 0.41 & 13.06 & 0.23 & 0.91 & 1.18/0.23 & 7.12/6.47 & 0.65/1.73\\ 
g6.96e11 & halo & 0.15 $\pm$ 0.29 & 0.48 $\pm$ 0.21 & -0.27 $\pm$ 0.07 & 0.35 & -0.16 & 12.32 & 0.23 & 0.19 & 0.41/0.41 & 5.39/4.32 & 1.08/5.67\\ 
g6.96e11 & spheroid & -0.14 $\pm$ 0.38 & 0.47 $\pm$ 0.20 & -0.68 $\pm$ 0.12 & 0.26 & 0.06 & 12.15$^{\rm c}$ & 0.14 & 0.96 & 1.24/0.44 & 6.04/6.47 & 1.00/1.50\\ 
g6.96e11 & classical b & 0.26 $\pm$ 0.36 & 0.45 $\pm$ 0.21 & -0.89 $\pm$ 0.05 & 0.34 & 0.01 & 11.47 & 0.04 & 1.00 & 1.00/0.11 & 6.90/3.24 & 0.88/1.50\\ 
g6.96e11 & pseudo b & 0.23 $\pm$ 0.37 & 0.17 $\pm$ 0.09 & -0.67 $\pm$ 0.12 & 0.49 & 0.61 & 12.07 & 0.14 & 0.96 & 1.18/0.25 & 6.47/6.26 & 0.76/1.90\\ 
g6.96e11 & disk & 0.70 $\pm$ 0.16 & 0.33 $\pm$ 0.16 & -0.68 $\pm$ 0.14 & 0.61 & 0.68 & 12.97 & 0.55 & 0.93 & 1.00/0.11 & 6.90/5.83 & 0.88/3.50\\ 
g7.55e11 & all & - & - & - & 0.46 & 0.48 & 13.24 & 0.23 & 0.97 & 1.88/0.73 & 3.45/8.20 & 1.00/0.58\\ 
g7.55e11 & spheroid & -0.03 $\pm$ 0.36 & 0.53 $\pm$ 0.20 & -0.75 $\pm$ 0.12 & 0.22 & -0.12 & 11.72$^{\rm c}$ & 0.05 & 0.98 & 2.16/1.00 & 3.24/4.53 & 0.87/1.33\\ 
g7.55e11 & classical b & 0.14 $\pm$ 0.46 & 0.27 $\pm$ 0.13 & -0.87 $\pm$ 0.06 & 0.44 & 0.52 & 10.89 & 0.01 & 1.00 & 1.88/0.73 & 3.45/7.34 & 1.00/0.42\\ 
g7.55e11 & inner disk & 0.32 $\pm$ 0.33 & 0.14 $\pm$ 0.07 & -0.69 $\pm$ 0.14 & 0.49 & 0.62 & 11.79 & 0.08 & 0.96 & 1.88/0.77 & 3.67/8.42 & 1.12/0.70\\ 
g7.55e11 & thick disk & 0.64 $\pm$ 0.18 & 0.37 $\pm$ 0.16 & -0.64 $\pm$ 0.17 & 0.54 & 0.55 & 12.92 & 0.31 & 0.93 & 1.66/0.65 & 3.45/6.90 & 1.67/0.88\\ 
g7.55e11 & thin disk & 0.89 $\pm$ 0.07 & 0.15 $\pm$ 0.08 & -0.56 $\pm$ 0.15 & 0.85 & 0.95 & 12.96 & 0.58 & 0.99 & 1.31/0.23 & 4.53/6.26 & 1.62/1.64\\ 
g7.66e11 & all & - & - & - & 0.38 & 0.38 & 13.09 & 0.12 & 0.93 & 1.24/0.55 & 4.10/5.83 & 2.17/0.69\\ 
g7.66e11 & spheroid & 0.04 $\pm$ 0.39 & 0.45 $\pm$ 0.20 & -0.45 $\pm$ 0.11 & 0.17 & 0.07 & 11.72 & 0.05 & 0.79 & 1.56/0.90 & 4.32/4.10 & 1.22/1.37\\ 
g7.66e11 & classical b & -0.00 $\pm$ 0.36 & 0.49 $\pm$ 0.21 & -0.77 $\pm$ 0.10 & 0.15 & -0.13 & 11.20$^{\rm c}$ & 0.00 & 0.99 & 1.56/0.59 & 3.88/3.02 & 1.00/0.75\\ 
g7.66e11 & pseudo b & 0.10 $\pm$ 0.24 & 0.15 $\pm$ 0.08 & -0.53 $\pm$ 0.21 & 0.32 & 0.41 & 11.82 & 0.05 & 0.74 & 1.00/0.77 & 3.88/3.45 & 17.00/0.60\\ 
g7.66e11 & thick disk & 0.71 $\pm$ 0.14 & 0.34 $\pm$ 0.16 & -0.73 $\pm$ 0.12 & 0.60 & 0.64 & 12.60 & 0.16 & 0.99 & 1.06/0.38 & 4.10/4.53 & 3.75/1.10\\ 
g7.66e11 & thin disk & 0.93 $\pm$ 0.05 & 0.16 $\pm$ 0.09 & -0.62 $\pm$ 0.12 & 0.85 & 0.94 & 12.87 & 0.31 & 1.00 & 1.00/0.21 & 5.61/3.88 & 1.36/1.00\\ 
g8.13e11 & all & - & - & - & 0.40 & 0.37 & 13.31 & 0.26 & 0.99 & 2.96/1.24 & 3.45/9.06 & 0.33/0.35\\ 
g8.13e11 & classical b & -0.04 $\pm$ 0.32 & 0.55 $\pm$ 0.20 & -0.67 $\pm$ 0.18 & 0.12 & -0.25 & 11.96 & 0.27 & 0.96 & 3.25/1.88 & 3.02/5.39 & 0.40/0.32\\ 
g8.13e11 & pseudo b & 0.23 $\pm$ 0.36 & 0.20 $\pm$ 0.10 & -0.65 $\pm$ 0.17 & 0.34 & 0.44 & 12.21 & 0.15 & 0.98 & 3.25/1.77 & 3.24/8.85 & 0.36/0.21\\ 
g8.13e11 & inner disk & 0.59 $\pm$ 0.22 & 0.47 $\pm$ 0.20 & -0.72 $\pm$ 0.12 & 0.47 & 0.36 & 12.80 & 0.18 & 1.00 & 2.72/1.00 & 3.88/9.71 & 0.38/0.45\\ 
g8.13e11 & disk & 0.92 $\pm$ 0.07 & 0.16 $\pm$ 0.09 & -0.49 $\pm$ 0.11 & 0.84 & 0.94 & 13.06 & 0.52 & 1.00 & 1.88/0.69 & 3.67/6.69 & 1.12/0.63\\ 
g8.26e11 & all & - & - & - & 0.50 & 0.51 & 13.48 & 0.23 & 0.94 & 2.96/1.39 & 3.45/8.63 & 0.60/0.38\\ 
g8.26e11 & halo & 0.24 $\pm$ 0.21 & 0.36 $\pm$ 0.22 & -0.45 $\pm$ 0.18 & 0.46 & 0.23 & 12.49 & 0.26 & 0.55 & 2.32/1.56 & 7.77/6.90 & 0.29/0.45\\ 
g8.26e11 & classical b & 0.19 $\pm$ 0.35 & 0.56 $\pm$ 0.18 & -0.79 $\pm$ 0.10 & 0.20 & -0.24 & 11.87 & 0.05 & 1.00 & 3.60/2.16 & 3.02/5.61 & 0.56/0.30\\ 
g8.26e11 & pseudo b & 0.25 $\pm$ 0.43 & 0.20 $\pm$ 0.10 & -0.80 $\pm$ 0.09 & 0.44 & 0.54 & 11.58 & 0.03 & 1.00 & 3.25/1.77 & 3.45/8.42 & 0.45/0.26\\ 
g8.26e11 & thick disk & 0.72 $\pm$ 0.14 & 0.32 $\pm$ 0.15 & -0.63 $\pm$ 0.17 & 0.63 & 0.69 & 13.10 & 0.25 & 0.92 & 2.72/1.31 & 3.24/6.90 & 0.67/0.39\\ 
g8.26e11 & thin disk & 0.91 $\pm$ 0.06 & 0.14 $\pm$ 0.08 & -0.57 $\pm$ 0.16 & 0.86 & 0.96 & 13.14 & 0.37 & 0.98 & 1.77/0.59 & 3.88/8.85 & 2.00/0.86\\ 
g1.12e12 & all & - & - & - & 0.40 & 0.33 & 12.83 & 0.14 & 0.99 & 2.16/0.85 & 5.83/8.63 & 0.35/0.60\\ 
g1.12e12 & classical b & -0.23 $\pm$ 0.34 & 0.51 $\pm$ 0.21 & -0.67 $\pm$ 0.14 & 0.32 & 0.03 & 12.56$^{\rm c}$ & 0.15 & 0.99 & 2.16/0.90 & 6.47/9.06 & 0.30/0.50\\ 
g1.12e12 & pseudo b & 0.05 $\pm$ 0.32 & 0.18 $\pm$ 0.10 & -0.62 $\pm$ 0.13 & 0.48 & 0.56 & 11.66 & 0.03 & 0.99 & 2.01/0.77 & 5.18/7.55 & 0.50/0.84\\ 
g1.12e12 & inner disk & 0.52 $\pm$ 0.25 & 0.47 $\pm$ 0.19 & -0.64 $\pm$ 0.17 & 0.43 & 0.38 & 12.75 & 0.51 & 0.99 & 2.16/0.95 & 6.47/9.06 & 0.30/0.45\\ 
g1.12e12 & disk & 0.66 $\pm$ 0.39 & 0.19 $\pm$ 0.10 & -0.26 $\pm$ 0.09 & 0.57 & 0.82 & 12.64 & 0.99 & 0.82 & 1.88/1.12 & 4.10/6.04 & 0.73/0.40\\ 
g1.77e12 & all & - & - & - & 0.45 & 0.45 & 13.95 & 0.24 & 0.81 & 1.00/0.41 & 3.40/5.18 & 1.03/0.85\\ 
g1.77e12 & halo & 0.13 $\pm$ 0.33 & 0.32 $\pm$ 0.21 & -0.23 $\pm$ 0.08 & 0.14 & 0.12 & 13.04 & 0.16 & 0.61 & 1.12/0.69 & 4.32/3.45 & 1.50/1.67\\ 
g1.77e12 & classical b & 0.15 $\pm$ 0.36 & 0.54 $\pm$ 0.22 & -0.74 $\pm$ 0.13 & 0.28 & -0.17 & 12.26 & 0.02 & 0.76 & 1.00/0.44 & 2.59/4.96 & 1.00/0.77\\ 
g1.77e12 & pseudo b & 0.33 $\pm$ 0.38 & 0.16 $\pm$ 0.08 & -0.69 $\pm$ 0.13 & 0.53 & 0.68 & 12.57 & 0.07 & 0.84 & 1.00/0.36 & 3.18/5.61 & 0.90/0.86\\ 
g1.77e12 & inner disk & 0.69 $\pm$ 0.18 & 0.39 $\pm$ 0.17 & -0.62 $\pm$ 0.15 & 0.57 & 0.56 & 13.30 & 0.23 & 0.85 & 1.00/0.41 & 3.18/5.18 & 0.90/0.71\\ 
g1.77e12 & disk & 0.88 $\pm$ 0.11 & 0.17 $\pm$ 0.09 & -0.42 $\pm$ 0.13 & 0.82 & 0.93 & 13.71 & 0.61 & 0.94 & 1.00/0.34 & 3.40/4.32 & 1.03/1.00\\ 
g1.92e12 & all & - & - & - & 0.47 & 0.47 & 14.05 & 0.34 & 0.98 & 2.01/1.18 & 4.75/9.28 & 0.69/0.43\\ 
g1.92e12 & classical b & 0.30 $\pm$ 0.36 & 0.47 $\pm$ 0.22 & -0.73 $\pm$ 0.11 & 0.30 & 0.05 & 13.17 & 0.10 & 1.00 & 2.72/1.56 & 3.67/7.34 & 0.42/0.31\\ 
g1.92e12 & pseudo b & 0.00 $\pm$ 0.35 & 0.23 $\pm$ 0.14 & -0.56 $\pm$ 0.23 & 0.14 & 0.31 & 12.36 & 0.19 & 0.83 & 3.25/2.32 & 2.81/4.53 & 0.44/0.31\\ 
g1.92e12 & thick disk & 0.84 $\pm$ 0.13 & 0.20 $\pm$ 0.11 & -0.61 $\pm$ 0.15 & 0.80 & 0.89 & 13.57 & 0.44 & 1.00 & 1.88/0.85 & 5.39/9.49 & 0.56/0.57\\ 
g1.92e12 & thin disk & 0.94 $\pm$ 0.07 & 0.09 $\pm$ 0.05 & -0.42 $\pm$ 0.08 & 0.91 & 0.98 & 13.77 & 0.60 & 1.00 & 1.47/0.41 & 5.61/7.12 & 0.62/1.20\\ 
g2.79e12 & all & - & - & - & 0.47 & 0.49 & 14.43 & 0.46 & 0.96 & 2.32/0.81 & 4.96/8.63 & 0.35/0.82\\ 
g2.79e12 & classical b & -0.02 $\pm$ 0.36 & 0.48 $\pm$ 0.18 & -0.61 $\pm$ 0.23 & 0.36 & -0.03 & 12.96 & 0.24 & 0.88 & 2.72/1.47 & 5.18/7.55 & 0.33/0.46\\ 
g2.79e12 & pseudo b & 0.30 $\pm$ 0.38 & 0.17 $\pm$ 0.09 & -0.70 $\pm$ 0.13 & 0.51 & 0.63 & 12.89 & 0.06 & 1.00 & 2.16/0.62 & 4.96/9.49 & 0.44/1.10\\ 
g2.79e12 & inner disk & 0.44 $\pm$ 0.29 & 0.50 $\pm$ 0.21 & -0.71 $\pm$ 0.13 & 0.34 & 0.10 & 13.24 & 0.09 & 1.00 & 2.50/0.95 & 5.18/8.42 & 0.33/0.70\\ 
g2.79e12 & thick disk & 0.58 $\pm$ 0.31 & 0.17 $\pm$ 0.09 & -0.24 $\pm$ 0.06 & 0.51 & 0.73 & 13.79 & 1.00 & 0.76 & 2.32/1.06 & 4.96/5.61 & 0.28/1.00\\ 
g2.79e12 & thin disk & 0.89 $\pm$ 0.08 & 0.20 $\pm$ 0.13 & -0.40 $\pm$ 0.13 & 0.79 & 0.89 & 14.24 & 0.74 & 0.99 & 2.01/0.62 & 4.75/8.63 & 0.47/1.00\\ 
\hline
\end{tabular}
\caption{}
\end{table*}

\renewcommand{\thetable}{A\arabic{table}}

\end{document}